\documentclass[a4paper,12pt]{article}
\usepackage{jheppub} 
\usepackage{tcilatex}
\usepackage{amsfonts}
\usepackage{amssymb}
\usepackage{multicol}
\usepackage{graphicx}
\usepackage{float}
\usepackage{caption}
\usepackage{xcolor}
\usepackage[utf8]{inputenc}
\usepackage{amsmath}
\title{Black Flowers and Real Forms of Higher Spin Symmetries}
\author[1,2]{R. Sammani, }
\author[1,2]{E.H Saidi}

\affiliation[1]{ LPHE-MS, Science Faculty, Mohammed V University in Rabat, Morocco}
\affiliation[2]{Centre of Physics and Mathematics, CPM- Morocco}
\emailAdd{rajae$\_$sammani@um5.ac.ma}
\emailAdd{e.saidi@um5r.ac.ma}
\abstract{ Using Chern-Simons formulation, we investigate higher spin (HS)
black holes in AdS$_{3}$ with soft Heisenberg hair and establish
linkage with the real forms of the underlying complexified gauge symmetries
taken here as $SL(N,\mathbb{C})_{\mathtt{L}}\times SL(N,\mathbb{C})_{\mathtt{%
R}}.$ We study the various conserved currents characterizing the HS
black flowers (HS-BF) and show that they can be formed of layers
indexed by the elements of the centre of the gauge symmetry. This feature
follows from requiring the holonomy of the asymptotic gauge connection
around the thermal cycle to sit in the centre $\mathbb{Z}_{N}$ of the
symmetry group. With regard to the compact subgroups of the real forms of
the complexified gauge symmetry, we calculate the entropies of the
HS-BF and verify that, unless we are considering trivial
holonomies, there are no continuous paths joining the HS-BF to the core spin
2 black holes. As explicit illustrations, we give quantum field
realisations of the soft Heisenberg hair in terms of bosonic and fermionic
primary conformal fields and compute the HS-BF entropy as a
function of the number of fermions occupying the ground state of the
Heisenberg soft hair.}


\begin{document}
\notoc
\maketitle
\flushbottom
\newpage
\tableofcontents
\section{Introduction}

\label{sec:intro} Since their inauguration in \cite{BTZ}, the three
dimensional black hole solutions of AdS$_{3}$ gravity in the Chern-Simons
formulation \cite{AT, W} were built in accordance with the well established
Brown-Henneaux boundary conditions (BC) \cite{2A}. However, over the few
past years, it was evidenced that higher spin AdS$_{3}$ gravity admits
stationary black hole solutions\ that transcend the customary Brown-Henneaux
BC \cite{1A, 1B}. An example of such configuration is the spin $s=2$ gravity
model constructed in \cite{1A}; it features (non) spherical black holes with
$SL(2,\mathbb{R})_{\mathtt{L}}\times SL(2,\mathbb{R})_{\mathtt{R}}$ gauge
symmetry yielding two copies of the asymptotic affine \^{u}(1)$_{\mathrm{k}}$
Heisenberg algebras generated by an infinite number of conserved charges ($%
J_{m}$, $\bar{J}_{\bar{m}}$) labelled by the integers $m$ and $\bar{m}$. The
often anticipated Brown-Henneaux boundary conformal symmetry emerges as a
composite invariance through a twisted Sugawara construction and the
regularity of the gauge configurations at the black hole (BH) horizon was
confirmed to hold independently of the global charges. In addition, these
black holes carry soft hair states \cite{SH, hor} with BH entropy S$_{\text{%
\textsc{bh}}}$ given by the remarkable relation
\begin{equation}
S_{\text{\textsc{bh}}}=2\pi \mathbf{n}\left( J_{0}+\bar{J}_{0}\right)
\end{equation}%
where $\mathbf{n}$ is a positive integer and where the $J_{0}$ and $\bar{J}%
_{0}$ are the zero modes commuting with all of the excited $J_{m},$ $\bar{J}%
_{\bar{m}}$ modes generating the affine $\hat{U}(1)_{\mathrm{k}}\times \hat{U%
}(1)_{\mathrm{\bar{k}}}$ algebra.

Higher spin (HS) extensions \cite{hs6}-\cite{hs9} of the previous spin s=2
AdS$_{3}$\ gravity model were developed in \cite{1B} for the larger $SL(N,%
\mathbb{R})_{\mathtt{L}}\times SL(N,\mathbb{R})_{\mathtt{R}}$ gauge symmetry
with an emphasis on the N=3 case for illustrations. The associated
asymptotic algebra is given by a greater product of affine Heisenberg
symmetries $\dprod_{i}\hat{U}(1)_{\mathrm{k}_{i}}\times \hat{U}(1)_{\mathrm{%
\bar{k}}_{i}}$ \cite{3B}; and similarly to the spin s=2 theory, the boundary
W$_{N}$-invariance \cite{W1}-\cite{W3} can be built out of composites of
Kac-Moody (KM) charges via an extended twisted Sugawara \cite{4B}. The
generalised HS theory comprises black hole solutions dubbed higher spin
black flowers (HS-BF$_{\text{\textsc{n}}}$) \cite{2B} whose entropy S$_{%
\text{\textsc{hs-bh}}}^{\text{\textsc{gpptt}}}$ for the $SL(3)_{\mathtt{L}%
}\times SL(3)_{\mathtt{R}}$ model has the form
\begin{equation}
S_{\text{\textsc{hs-bh}}}^{\text{\textsc{gpptt}}}=2\pi \mathbf{n}\left(
J_{0}+\bar{J}_{0}\right) +3\pi \mathbf{m}\left( K_{0}+\bar{K}_{0}\right)
\end{equation}%
where GPPTT credits the authors Grumiller- Perez- Prohazka- Tempo- Troncoso.
The parameters $\mathbf{n}$ and $\mathbf{m}$ are positive integers, and the
zero mode charges ($J_{0},K_{0};\bar{J}_{0},\bar{K}_{0}$) belong to the
tensorial affine Heisenberg $\dprod_{i=1}^{2}\hat{U}(1$)$_{\mathrm{k}%
_{i}}\times \hat{U}(1)_{\mathrm{\bar{k}}_{i}}$ algebra generated by the
infinite sets ($J_{n},K_{m}$) and ($\bar{J}_{\bar{n}},\bar{K}_{\bar{m}}$).
By defining $\mathtt{I}_{0}^{{\small L}}=\mathbf{n}J_{0}+3\mathbf{m}K_{0}/2$
for the left sector and $\mathtt{I}_{0}^{{\small R}}=\mathbf{n}\bar{J}_{0}+3%
\mathbf{m}\bar{K}_{0}/2$ for the right one, the above entropy formula can be
concisely presented as $2\pi $($\mathtt{I}_{0}^{{\small L}}+\mathtt{I}_{0}^{%
{\small R}}).$ Moreover, we are able to recuperate the fundamental HS black
hole entropy formula of \cite{1C} from the S$_{\text{\textsc{hs-bh}}}^{\text{%
\textsc{gpptt}}}$\ entropy\ of the higher black flower, \textrm{as it is }%
continuously connected to the spin 2 BTZ black hole \cite{1B}. By setting $%
\mathbf{m}=0$ \textrm{and} $\mathbf{n}=1$ in the above spin 3 formula, one
recovers the entropy $2\pi \left( J_{0}+\bar{J}_{0}\right) $ of the spin s=2
BH with gauge symmetry $SL(2)_{\mathtt{L}}\times SL(2)_{\mathtt{R}}$ and
boundary conditions as in \cite{1A}.

In this paper, we\ develop the study of higher spin black flowers (HS-BF$_{%
\text{\textsc{n}}}$) entropy with $SL(N,\mathbb{R})_{\mathtt{L}}\times SL(N,%
\mathbb{R})_{\mathtt{R}}$ gauge symmetry and generalised boundary conditions
as in eqs(\ref{bb}-\ref{bc}). We consider non trivial holonomies around
thermal cycles taking values in the centre $\mathbb{Z}_{N}$ of the gauge
invariance. This discrete group is given by the set $\left\langle \omega
^{p}\right\rangle _{1\leq p<N}$ with $\omega =\exp (2i\pi /N),$ it permits
to extend the trivial holonomies of \cite{1A,1B,1C} beyond the identity
element $\omega ^{0}=I_{id}$ corresponding to $p=0.$ With this enlarged set
of holonomies, we show amongst others that higher spin black flowers possess
N layers labelled by the central group elements $\omega ^{p}$ with $1\leq
p<N $. And for each layer of the HS-BF$_{\text{\textsc{n}}}$, the entropy S$%
_{\text{\textsc{hs-bf}}}^{\text{\textsc{sl}}_{N}}$ is a function of the
rational number p/N (mod N) such that
\begin{equation}
2\pi \left\vert \mathtt{I}_{0}^{\text{\textsc{l}}}(p/N)\right\vert +2\pi
\left\vert \mathtt{I}_{0}^{\text{\textsc{r}}}(p/N)\right\vert
\end{equation}%
where $\mathtt{I}_{0}^{\text{\textsc{l}}}(p/N)$ and $\mathtt{I}_{0}^{\text{%
\textsc{r}}}(p/N)$ are HS gauge invariants given by linear combinations of
the commuting conserved charges like $\sum_{s=2}^{N}\Lambda _{p,\mathbf{n}%
_{s}}^{{\small (s-1)}}J_{0}^{{\small (s)}}$. In this latter expansion, the $%
J_{0}^{{\small (s)}}$ s are the zero modes of the affine Heisenberg-like
symmetry $\dprod \hat{U}(1)_{\mathrm{k}_{i}}\times \hat{U}(1)_{\mathrm{\bar{k%
}}_{i}}$ while the $\Lambda _{p,\mathbf{n}_{s}}^{{\small (s-1)}}$'s
designate quantized Lagrange multipliers determined by solving the
regularity condition of the holonomy of the asymptotic gauge potentials.
Contrary to what was obtained in \cite{1B} for the N=3 theory with trivial
holonomy, there are no continuous paths joining the black flowers entropy S$%
_{\text{\textsc{hs-bf}}}^{\text{\textsc{sl}}_{3}}$ to the core S$_{\text{%
\textsc{hs-bf}}}^{\text{\textsc{sl}}_{2}}$, and more generally no continuous
bridges leading any S$_{\text{\textsc{hs-bf}}}^{\text{\textsc{sl}}_{N}}$
towards S$_{\text{\textsc{hs-bf}}}^{\text{\textsc{sl}}_{M}}$ with $M<N,$
once we consider holonomy layers beyond the identity $\omega ^{0}=I_{id}.$
Therefore, descending from a higher black flower to a lower BF requires
discrete leaps given by Weyl group transformations of the HS gauge algebra.

Besides the disrupted continuity, we uncover additional features. To
rigorously study the thermal properties of higher spin black flowers
including their entropies, one must promote the real gauge symmetry $SL(N,%
\mathbb{R})_{\mathtt{L}}\times SL(N,\mathbb{R})_{\mathtt{R}}$ to\ one of\
the real forms of the complexified $SL(N,\mathbb{C})_{\mathtt{L}}\times SL(N,%
\mathbb{C})_{\mathtt{R}};$ a property previously noted for the Euclidean HS
gravity investigated in \cite{1C} (\textrm{see table I therein}). As a
result, new aspects of HS-BF$_{\text{\textsc{n}}}$\ emerge; in particular
the close-knit relation between the branches of the HS black flowers and the
real forms of the underlying complexified gauge symmetry which is showcased
by the dependency of the entropy on the type of the real form.

In sum, and before delving into the specifics, we list here below some of
our main contributions to the study of HS AdS$_{3}$\ gravities having HS
black flowers solutions with $\left( \mathbf{i}\right) $ generalised
boundary gauge constraints beyond the standard Brown-Henneaux conditions and
$(\mathbf{ii})$ asymptotic gauge potential holonomies sitting in the $%
\mathbb{Z}_{N}$ centre:

\begin{itemize}
\item \textrm{We first describe the extended BC} of \cite{1A,1B} and exploit
this construction to develop a new HS basis\emph{\ }using the roots of the
Lie algebra of the gauge symmetry with illustrative examples. It is founded
on the Cartan-Weyl basis and is equivalent to the more usual one using the
principal $SL_{2}$ representations.\emph{\ }In this new basis, the left ($%
a_{\varphi },a_{t}$) and the right ($\tilde{a}_{\varphi },\tilde{a}_{t}$)
diagonal boundary gauge potentials of \cite{1B} have vanishing commutators $%
[a_{t},a_{\varphi }]$ and $[\tilde{a}_{t},\tilde{a}_{\varphi }]$ as they are
taken in the Cartan subalgebra of the HS gauge invariance. Accordingly, the
associated gauge field strengths reduce to flat Maxwell-like gauge
curvatures given by $\partial _{t}a_{\varphi }-\partial _{\varphi }a_{t}=0.$
We then consider the $a_{t}$ components as Lagrange multipliers constrained
like $\partial _{\varphi }a_{t}=0$ (say $a_{t}=$ constant), giving in result
$\partial _{t}a_{\varphi }=0$ and $\partial _{t}\tilde{a}_{\varphi }=0$.
These boundary potential constraints are to be interpreted as boundary
fields equations following from a 2D variational principle with Lagrange
field parameters. As a result, our approach leads directly to the conserved
total Hamiltonian and the conserved angular momentum of the HS theory with
an infinite asymptotic conserved $\dprod_{i}\hat{U}(1)_{\mathrm{k}%
_{i}}\times \hat{U}(1)_{\mathrm{\bar{k}}_{i}}$ charges while allowing us to
retrieve known results in HS literature.

\item Our construction also enables to interconnect the computation of the
BH entropy, requiring regularity of the asymptotic gauge configurations
around the thermal cycle, with the compact subgroups of the real forms of
the complexified gauge symmetry. In fact in order to derive the thermal BH
properties (\ref{EE}-\ref{28}), one needs to perform a Wick rotation which
leads to a complexification of the $SL(N,\mathbb{R})$ gauge symmetry towards
one of the real forms of $SL(N,\mathbb{C})$. In this regards, we give a
concrete description of the HS branches observed in \cite{1B} using real
forms of $SL(N,\mathbb{C})_{\mathtt{L}}\times SL(N,\mathbb{C})_{\mathtt{R}}$
including the real splits $SL(N,\mathbb{R})$, the compact $SU(N)$ and the
non compact $SU(N_{1},N_{2})$s. Moreover by allowing non trivial gauge
holonomies of the boundary potentials, that is holonomies sitting in the
centre $\mathbb{Z}_{N}$ of the complexified gauge symmetry $SL(N,\mathbb{C})$%
, we find that the HS black flowers have several dissociated layers that
cannot be continuously connected to the spin 2 AdS$_{3}$ black holes. The
crossing between the various layers of the HS-BF$_{\text{\textsc{n}}}$
requires discrete transitions as opposed to the prevailing consensus on HS-BF%
$_{\text{\textsc{n}}}$ with trivial holonomies.

\item Exploiting the conformal symmetry at the boundary of the HS-AdS$_{3}$
gravity, we give bosonic and fermionic oscillator realisations of the BF
particle excitations. We first recall properties on the free fields of
compact and non compact affine KM-symmetries which are known to be closely
related to the sign of the CS level \textrm{k }as first noticed by Dixon,
Peskin, and Lykken for the SU(2)$_{\mathrm{k}}$ and SL(2,$\mathbb{R}$)$_{%
\mathrm{k}}$ conformal models; and successfully used in several applications
like in recent studies regarding string theory on $AdS_{3}\times S^{3}\times
M$ with one unit of NS-NS flux (k = 1) \textrm{\cite{ST1,STT} and refs
therein}. For example, the operator product of the current operator $J_{%
{\small su}_{{\small 2}}}^{0}\left( z\right) $ of the compact SU(2)$_{%
\mathrm{k}}$ displays a positive CS level $+\mathrm{k}$ as shown by $J_{%
{\small su}_{{\small 2}}}^{0}\left( z\right) J_{{\small su}_{{\small 2}%
}}^{0}\left( w\right) \sim +\mathrm{k}/(z-w)^{2}$; unlike its homologue for
the non compact SL(2,$\mathbb{R}$)$_{\mathrm{k}}$, it exhibits a negative CS
level $\mathrm{\tilde{k}=}-\mathrm{k}$ as shown by the corresponding OPE
namely $J_{{\small sl}_{{\small 2},\mathbb{R}}}^{0}\left( z\right) J_{%
{\small sl}_{{\small 2},\mathbb{R}}}^{0}\left( w\right) \sim -\mathrm{k}%
/(z-w)^{2}.$ Focusing on the Cartan sectors of the real forms of SL(N,$%
\mathbb{C}$)$_{\mathrm{k}}$ with emphasis on the compact SU(N,$\mathbb{C}$)$%
_{\mathrm{k}}$, we give the quantum numbers characterising the A$_{{\small %
N-1}}$ soft hair BH, describe the ground state of this black flowers and
comment on their critical properties. For the bosonic realisation of the
spin s=3 black flower, we show that the zero modes of the primary scalar
fields sit in the $\mathbb{Z}^{4}$ lattice as given by \textrm{eqs(\ref{525})%
}. As for the fermionic oscillator realisation, we distinguish two pictures
given by periodic and anti-periodic boundary fermions. For both of these
(anti-) periodicities, we find amongst others that the entropy of the black
flowers (\ref{NE}) can be interpreted in terms of the number of fermionic
particles living in the ground state of the HS-BF soft hair.
\end{itemize}

The organisation of this paper is as follows: \textrm{In \autoref{sec2}}, we
describe the generalised boundary conditions of \cite{1A,1B} and\emph{\ }set
up the basic tools of the CS formulation in preparation for the upcoming
study of black holes solutions. \textrm{In \autoref{sec3}}, we give new
results on higher spin black flowers physics. In particular, we establish an
affiliation between the thermodynamics of black hole solutions and the real
forms of the complexified gauge symmetry of the HS AdS$_{3}$ gravity. We%
\emph{\ }utilise\emph{\ }field theory techniques on constrained hamiltonian
to develop a method that allows to build the boundary field action. All for
the purpose of deriving the HS black flowers entropies for the various real
forms of the complex gauge group. \textrm{In \autoref{sec4}}, we derive\emph{%
\ }the affine Heisenberg soft hair algebra with regard to the considered
real forms and explicitly compute the entropy formulas for the various black
flowers branches. \textrm{In \autoref{sec5}}, we realise the charge
potentials in terms of bosonic and fermionic primary conformal fields on the
horizon of the black hole flowers as concrete illustrations of our analysis.
Last \textrm{\autoref{conc}} is devoted to conclusion and discussions. We
accompany our work with three appendices \textrm{A, B and C. In \autoref%
{appA}, }we build HS AdS$_{3}$ theories using real forms of complex Lie
algebras with the help of Tits-Satake diagrams. \textrm{In \autoref{appB}, }%
we compute the asymptotic current algebra of the boundary invariants
operators. And finally in \textrm{\autoref{appC},} we study the physical
significance of the $\mathbb{Z}_{N}$ centre symmetry behind the non trivial
holonomies.

\section{Asymptotic potentials in AdS$_{3}$}

\label{sec2} In this section, we lay the groundwork for our study by
describing useful tools on higher spin (HS) AdS$_{3}$ gravity formulated in
terms of two copies of Chern-Simons potentials A$_{\mu }\left( r,\varphi
,t\right) $ and \~{A}$_{\mu }\left( r,\varphi ,t\right) $ with opposite CS\
couplings ($\mathrm{\tilde{k}}=-\mathrm{k}$). These bulk potentials are
valued in the Lie algebra of the HS gauge symmetry with some boundary
conditions%
\begin{equation}
\begin{tabular}{lllll}
$A_{r}$ & $\rightarrow $ & $A_{r}^{{\small asym}}$ & $\simeq $ & $%
g^{-1}\partial _{r}g$ \\
$A_{\pm }$ & $\rightarrow $ & $A_{\pm }^{{\small asym}}$ & $\simeq $ & $%
g^{-1}a_{\pm }g+g^{-1}\partial _{\pm }g$%
\end{tabular}
\label{20}
\end{equation}%
characterised by the gauge transformation $g$ taken as in \cite{1A,1B}. The
asymptotic $A_{\pm }^{{\small asym}}$ and $\tilde{A}_{\pm }^{{\small asym}}$
sit in the near horizon of the HS black holes (BH); they are generated by
the boundary gauge potentials $a$ and $\tilde{a}$ investigated hereafter.

\subsection{Structure of boundary potentials}

We start by introducing the structure of the connections $a$ and $\tilde{a}$
living in the 2D boundary surface $\Sigma _{{\small 2D}}=\partial \mathcal{M}%
_{{\small 3D}}$ of the higher spin AdS$_{3}$ gravity expressed in terms of
two Chern-Simons potentials. These boundary fields $a\left( \varphi
,t\right) $ and $\tilde{a}\left( \varphi ,t\right) $ are local 1-forms
valued in the Lie algebra of the gauge symmetry $G\times \tilde{G}.$ They
allow to build the three dimensional CS gauge potentials $A$ and $\tilde{A}$
(\ref{20}) of the bulk $\mathcal{M}_{{\small 3D}}$ through the following%
\textrm{\ }boundary condition%
\begin{equation}
A=\mathfrak{b}^{-1}\left( d+a\right) \mathfrak{b}\qquad ,\qquad \tilde{A}=%
\mathfrak{\tilde{b}}^{-1}\left( d+\tilde{a}\right) \mathfrak{\tilde{b}}
\label{bb}
\end{equation}%
The expressions of $\mathfrak{b}$ and $\mathfrak{\tilde{b}}$ are given by $%
e^{-L_{+}/\zeta _{-}}e^{-\rho L_{-}/2}$ and $e^{+L_{+}/\zeta _{-}}e^{+\rho
L_{-}/2}$ \cite{1A,1B}. They have the typical structure $g=e^{\theta
_{-}L_{+}}e^{\theta _{0}L_{0}}e^{\theta _{+}L_{-}}$ with real gauge
parameters $\theta _{0,\pm }$ and the pure gauge 1-form $g^{-1}dg$ expanding
like%
\begin{equation}
g^{-1}dg=L_{-}d\theta _{+}+\left( L_{0}+\theta _{+}L_{-}\right) d\theta
_{0}+\left( L_{+}+2\theta _{+}L_{0}+2\theta _{+}\theta _{+}L_{-}\right)
d\theta _{-}  \label{bc}
\end{equation}%
where the $L_{0,\pm }$ generate the principal SL(2,$\mathbb{R}$).
Interesting examples of such (non) spherical boundary conditions include the
three following:\textrm{\ }$(\mathbf{i})$ the $\left( \theta _{+},\theta
_{0},\theta _{-}\right) \sim \left( 0,\rho ,0\right) $ used in \cite{1D}, $%
\left( \mathbf{ii}\right) $\ the $\left( \theta _{+},\theta _{0},\theta
_{-}\right) \sim \left( 0,\rho ,\zeta _{-}\right) $ with constant $\zeta
_{-} $ of \cite{1E}, and $\left( \mathbf{iii}\right) $ the $\left( \theta
_{+},\theta _{0},\theta _{-}\right) \sim \left( \rho ,0,\zeta _{-}\right) $
considered in \cite{1A,1B} as well as\textrm{\ }in the present study.

From this description, it is clear that the connections $a$ and $\tilde{a}$
as well as the transformations ($\mathfrak{b},\mathfrak{\tilde{b}}$)
constitute fundamental\ ingredients in the formulation of higher spin AdS$%
_{3}$ gravity required for the study of the (non) spherical higher spin
black holes properties with soft Heisenberg hair.

Recall that the 3D field action of this HS gauge theory is given by the
difference of two Chern-Simons (CS) actions as follows \cite{AT,W}%
\begin{equation}
\mathcal{S}_{{\small 3D}}=\frac{\mathrm{k}}{4\pi }\int tr(AdA\mathbf{+}\frac{%
2}{3}A^{3}\mathbf{)}-\frac{\mathrm{k}}{4\pi }\int tr(\tilde{A}d\tilde{A}%
\mathbf{+}\frac{2}{3}\tilde{A}^{3}\mathbf{)}  \label{action}
\end{equation}%
where the CS coupling is equal to $\mathrm{k}=l_{\mathrm{AdS}}/(4G_{3})$
with $G_{3}$ being the 3D Newton constant and\emph{\ }$l_{\mathrm{AdS}}$ the
AdS$_{3}$ radius related to the negative cosmological constant \textrm{as} $%
\Lambda =-1/l_{\mathrm{AdS}}^{2}.$ The gravity/gauge duality can be
recognized through the CS fields relation to the HS-dreibein $e_{\mu }$ and
the HS-spin $\omega _{\mu }$ potentials of the 3D gravity given by%
\begin{equation}
A_{\mu }=\omega _{\mu }+\frac{1}{l_{\mathrm{AdS}}}e_{\mu }\qquad ,\qquad
\tilde{A}_{\mu }=\omega _{\mu }-\frac{1}{l_{\mathrm{AdS}}}e_{\mu }
\end{equation}%
from which we acquire the 3D space time metric $g_{\mu \nu }$ and the higher
spin fields $\Phi _{\mu _{1}...\mu _{s}}$ as follows \cite{hs6}%
\begin{equation}
g_{\mu \nu }\sim Tr\left( e_{\mu }e_{\nu }\right) \qquad ,\qquad \Phi _{\mu
_{1}...\mu _{s}}\sim Tr\left( e_{(\mu _{1}}...e_{\mu _{s})}\right)
\end{equation}%
The field equations of motion of (\ref{action}) are given by the 2-form
field strengths $F=dA+A\wedge A=0$ and $\tilde{F}=d\tilde{A}+\tilde{A}\wedge
\tilde{A}=0$. With the substitution (\ref{bb}), they factorise as $F=g^{-1}%
\boldsymbol{f}$ $g=0$ and $\tilde{F}=\tilde{g}^{-1}\boldsymbol{\tilde{f}}$ $%
\tilde{g}=0$ with vanishing
\begin{equation}
\boldsymbol{f}=da+a\wedge a\qquad ,\qquad \boldsymbol{\tilde{f}}=d\tilde{a}+%
\tilde{a}\wedge \tilde{a}
\end{equation}%
These\ constraints will prove to be instrumental\textrm{\ }in this
investigation.

Recall also that here the two 1-forms $a$ and $\tilde{a}$ live on the
boundary cylinder $\mathcal{C}_{\varphi }\times \mathbb{R}_{t}$ coordinated
by the variables ($\varphi ,t$) with $0\leq \varphi <2\pi $ and $t\in
\mathbb{R}_{t}.$\ In practice, one might use\ instead\textrm{\ }the complex $%
e^{t\pm i\varphi }$ (with periodicity $\varphi \equiv \varphi +2\pi $) which
can be associated with the compactification of the real $\sigma $- variable
in the real $e^{t\pm \sigma }$ parameterising $\mathbb{R}^{1,1}$ (or
Euclidean $\mathbb{R}^{2}$)$.$ Given that, the boundary $a$ and $\tilde{a}$
are generated by the differential forms $d\varphi $ and $dt$ as follows%
\begin{equation}
\begin{tabular}{lll}
$a$ & $=$ & $a_{\varphi }d\varphi +a_{t}dt$ \\
$\tilde{a}$ & $=$ & $\tilde{a}_{\varphi }d\varphi +\tilde{a}_{t}dt$%
\end{tabular}
\label{aa}
\end{equation}%
with components ($a_{\varphi },a_{t}$) and ($\tilde{a}_{\varphi },\tilde{a}%
_{t}$) valued in the Lie algebra of the boundary gauge symmetry $G\times
\tilde{G}$. They obey the following differential equations%
\begin{equation}
\begin{tabular}{lll}
$\partial _{t}a_{\varphi }-\partial _{\varphi }a_{t}+\left[ a_{t},a_{\varphi
}\right] $ & $=$ & $0$ \\
$\partial _{t}\tilde{a}_{\varphi }-\partial _{\varphi }\tilde{a}_{t}+\left[
\tilde{a}_{t},\tilde{a}_{\varphi }\right] $ & $=$ & $0$%
\end{tabular}
\label{ct}
\end{equation}%
which are well-known\ in conformal (and affine) Toda theories as
integrability conditions \textrm{\cite{1H}} for building conformal invariant
models such as the Liouville and sine-Gordon equations \textrm{\cite{2H,2HH}}%
.\textrm{\ }They will be explored further below in the context of higher
spin black flowers.

Following the interesting studies \textrm{\cite{1A,1B,1C}}, one
distinguishes \emph{two kinds} of equivalent matrix representations of the
connections $a$ and $\tilde{a}$ namely: $\left( \mathbf{1}\right) $ the off
diagonal highest weight (HW) representation ($a^{\text{\textsc{hw}}},\tilde{a%
}^{\text{\textsc{hw}}}$) \textrm{\cite{2A, spin3, 1J, 1J2, 1J3}}; and $%
\left( \mathbf{2}\right) $ the diagonal realisation ($a^{\text{\textsc{diag}}%
},\tilde{a}^{\text{\textsc{diag}}}$) \cite{1B, 1C} thought of as%
\begin{equation}
\begin{tabular}{lll}
$a_{\varphi }^{\text{\textsc{diag}}}$ & $=$ & $\dsum\limits_{s}\mathcal{J}_{%
{\small (s)}}^{0}W_{0}^{{\small (s)}}$ \\
$a_{t}^{\text{\textsc{diag}}}$ & $=$ & $\dsum\limits_{s}\lambda _{{\small (s)%
}}^{0}W_{0}^{{\small (s)}}$%
\end{tabular}%
\end{equation}%
Pertaining to the purpose of our inquiry, we will use the diagonal
representation as it permits $\left( \mathbf{i}\right) $ a refined
formulation of the entropy S$_{\text{\textsc{hs-bh}}}$ for (non) spherical
HS black holes; and $\left( \mathbf{ii}\right) $ the computation of
conserved charges of the HS black flowers such as the classical gauge
invariants
\begin{equation}
\text{\texttt{I}}_{n}=\int_{0}^{\mathrm{\tau }}dt\left[ \frac{1}{2\pi }%
\int_{0}^{2\pi }d\varphi e^{in\varphi }Tr\left( a_{t}^{\text{\textsc{diag}}%
}a_{\varphi }^{\text{\textsc{diag}}}\right) \right]  \label{CS}
\end{equation}%
and their twild homologue $\mathtt{\tilde{I}}_{n}$. Considering time
independent $a_{t}^{\text{\textsc{diag}}}$ with constraints (\ref{ct}) on
the velocity $\partial _{t}a_{\varphi }^{\text{\textsc{diag}}}$, we can
perform the trace $Tr(a_{t}^{\text{\textsc{diag}}}a_{\varphi }^{\text{%
\textsc{diag}}})$ leading to the form $\sum G^{{\small s\sigma }}\lambda _{%
{\small (s)}}^{0}\mathcal{J}_{{\small (\sigma )}}^{0}$ with metric $G^{%
{\small s\sigma }}=Tr(W_{0}^{{\small (s)}}W_{0}^{{\small (\sigma )}}).$ The
integration with respect to time is therefore%
\begin{equation}
\text{\texttt{I}}_{n}=2\pi \sum_{s}\mathtt{d}_{s}\lambda _{{\small (s)}}^{%
{\small 0}}J_{n}^{{\small 0(s)}}\qquad ,\qquad G^{{\small s\sigma }}=\frac{%
\mathtt{d}_{s}}{2\pi }\delta ^{{\small s\sigma }}
\end{equation}%
with $\mathtt{d}_{s}$ to be determined. The higher spin black hole entropy
\textsc{S}$_{\text{\textsc{hs-bh}}}^{\text{\textsc{spin s}}}$ (see later
analysis for details) \textrm{is} given by the sum of zero modes like $2\pi
\sum_{s}\mathtt{d}_{s}\lambda _{{\small (s)}}^{{\small 0}}(J_{0}^{{\small %
0(s)}}+\tilde{J}_{0}^{{\small 0(s)}})$ while assuming $\lambda _{{\small (s)}%
}^{{\small 0}}>0$ and $J_{0}^{{\small 0(s)}},$ $\tilde{J}_{0{\small (s)}}^{%
{\small 0(s)}}>0$ otherwise one uses absolute values. This sum reads for
spin 2 AdS$_{3}$ gravity like%
\begin{equation}
\text{\textsc{S}}_{\text{\textsc{hs-bh}}}^{\text{\textsc{spin 2}}}=2\pi
\mathtt{d}_{2}\lambda _{{\small (2)}}^{0}\left( J_{0}^{{\small 0(2)}}+\tilde{%
J}_{0}^{{\small 0(2)}}\right)
\end{equation}%
For completeness, as well as later comparisons between our\textrm{\ }%
findings and results in higher spin black holes literature, we briefly
describe the two matrix representations here below along with some useful
comments.

\subsection{Matrix potentials: realisations and examples}

We first describe the highest weight representation (HWR) \cite{2A, spin3,
1J, 1J2, 1J3}; then we\ attend to the diagonal realisation of the boundary
gauge potentials.

\subsubsection{ Highest weight representation}

We give the HWR for the simplest higher spin models\emph{\ }with small
ranked \texttt{r}$_{{\small G}}$ gauge symmetries$.$ This is the case of $%
\left( \mathbf{a}\right) $ the basic spin s=2 gravity with symmetry $%
SL\left( 2,\mathbb{R}\right) _{\mathtt{L}}\times SL\left( 2,\mathbb{R}%
\right) _{\mathtt{R}}$ having a rank $\mathtt{r}_{sl_{2}}=1,$ and $\left(
\mathbf{b}\right) $ the two spins gravities with rank $\mathtt{r}_{G}=2.$
For instance, we cite the familiar HS model with $SL\left( 3,\mathbb{R}%
\right) _{\mathtt{L}}\times SL\left( 3,\mathbb{R}\right) _{\mathtt{R}}$
algebra\textrm{\ }having the spins\textrm{\ }\{$s_{1}=2,s_{2}=3$\}.

For the basic spin s=2 model, the HWR of the boundary gauge connection in
terms of the $SL\left( 2,\mathbb{R}\right) $ generators $L_{0,\pm }$ is as
follows%
\begin{equation}
\begin{tabular}{lll}
$a_{\varphi }^{\text{\textsc{hw}}}$ & $=$ & $a_{\varphi
}^{+}L_{+}+a_{\varphi }^{-}L_{-}$ \\
$a_{t}^{\text{\textsc{hw}}}$ & $=$ & $%
a_{t}^{+}L_{+}+a_{t}^{0}L_{0}+a_{t}^{-}L_{-}$%
\end{tabular}%
\end{equation}%
where $\left( \mathbf{i}\right) $ the $a_{\varphi }^{\pm }$\ are the charge
potentials with free $a_{\varphi }^{-}$ and fixed $a_{\varphi }^{+}=1$ that
can be thought of as $\delta _{\varphi }^{+}$; while $\left( \mathbf{ii}%
\right) \ $the time components $a_{t}^{0}$ and $a_{t}^{\pm }$ are
constrained by the eqs(\ref{ct}), they are functionals of $a_{\varphi }^{-}$%
. Quite similar expressions can be written down for ($\tilde{a}_{\varphi }^{%
\text{\textsc{hw}}},\tilde{a}_{t}^{\text{\textsc{hw}}}),$ they are given by%
\begin{equation}
\begin{tabular}{lll}
$\tilde{a}_{\varphi }^{\text{\textsc{hw}}}$ & $=$ & $\tilde{a}_{\varphi
}^{-}L_{-}+\tilde{a}_{\varphi }^{+}L_{+}$ \\
$\tilde{a}_{t}^{\text{\textsc{hw}}}$ & $=$ & $\tilde{a}_{t}^{-}L_{-}+\tilde{a%
}_{t}^{0}L_{0}+\tilde{a}_{t}^{+}L_{+}$%
\end{tabular}%
\end{equation}%
with $\tilde{a}_{\varphi }^{-}=1$.

For the generic higher spin theory with $SL\left( N,\mathbb{R}\right) _{%
\mathtt{L}}\times SL\left( N,\mathbb{R}\right) _{\mathtt{R}}$ symmetry, we
have the following generalised expansions%
\begin{equation}
\begin{tabular}{lll}
$a_{\varphi }^{\text{\textsc{hw}}}$ & $=$ & $a_{\varphi
}^{+}L_{+}+a_{\varphi }^{-}L_{-}+\dsum\limits_{s=3}^{N}a_{\varphi }^{\left(
1-s\right) }W_{1-s}^{(s)}$ \\
$a_{t}^{\text{\textsc{hw}}}$ & $=$ & $\dsum\limits_{s=2}^{N}\dsum%
\limits_{m_{s}=1-s}^{s-1}a_{t}^{(m_{s})}W_{m_{s}}^{(s)}$%
\end{tabular}%
\end{equation}%
with $a_{\varphi }^{+}=1$\textrm{. }The\textrm{\ }set \{$W_{m_{s}}^{(s)}$\}
designate the$\ N^{2}-1$ generators of $SL\left( N,\mathbb{R}\right) $
labelled by two integers $\left( s,m_{s}\right) $ taking values as $2\leq
s\leq N$ and $1-s\leq m_{s}\leq s-1.$ Similar relations can be written as
well for the potentials $\tilde{a}_{\varphi }$ and $\tilde{a}_{t}$ of the
right sector.

Before proceeding, take note of the\emph{\ }following:

\begin{description}
\item[$\left( \mathbf{1}\right) $] The three leading $W_{m_{2}}^{(s=2)}$ are
just the three generators $L_{0,\pm }$ of the principal $SL\left( 2,\mathbb{R%
}\right) $ satisfying the commutation relations $[L_{n},L_{m}]=\left(
n-m\right) L_{n+m}$. In the isospin $j=1/2$ representation, we have the 2$%
\times $2 matrix realisation%
\begin{equation}
L_{+}=\left(
\begin{array}{cc}
{\small 0} & {\small 0} \\
{\small 1} & {\small 0}%
\end{array}%
\right) ,\quad L_{0}=\left(
\begin{array}{cc}
\frac{{\small 1}}{2} & {\small 0} \\
{\small 0} & {\small -}\frac{{\small 1}}{2}%
\end{array}%
\right) ,\quad L_{-}=\left(
\begin{array}{cc}
{\small 0} & -{\small 1} \\
{\small 0} & {\small 0}%
\end{array}%
\right)  \label{22}
\end{equation}%
acting on the two eigenvectors $L_{0}\left\vert q\right\rangle =q\left\vert
q\right\rangle $ with $q=\pm 1/2.$

\item[$\left( \mathbf{2}\right) $] The angular components \{$a_{\varphi }^{%
{\small (1-s)}},\tilde{a}_{\varphi }^{{\small (s-1)}}$\} termed as the
charge potentials are commonly scaled by a (2$\pi /\mathrm{k}$) \textrm{%
factor.} For the spin s=2, they are often denoted like $a_{\varphi
}^{-}:=a_{++}$ and $\tilde{a}_{\varphi }^{+}:=\tilde{a}_{--}$ with%
\begin{equation}
a_{++}=-\frac{2\pi }{\mathrm{k}}\mathcal{L}_{++}\qquad ,\qquad \tilde{a}%
_{--}=-\frac{2\pi }{\mathrm{k}}\mathcal{\tilde{L}}_{--}
\end{equation}%
More generally, we have $a_{\varphi }^{{\small (1-s)}}=-\frac{2\pi \eta _{s}%
}{\mathrm{k}}\mathcal{W}_{+s}$ and $\tilde{a}_{\varphi }^{{\small (s-1)}}=-%
\frac{2\pi \eta _{s}}{\mathrm{k}}\mathcal{\tilde{W}}_{-s}$\ with some
normalisation numbers $\eta _{s}$ required by the canonical commutation
relations of the W-symmetry \textrm{\cite{W,W3}}. For the spin s=3 example,
we get $a_{\varphi }^{--}=-\frac{\pi }{2\mathrm{k}}\mathcal{W}_{+3}$ and $%
\tilde{a}_{\varphi }^{--}=-\frac{\pi }{2\mathrm{k}}\mathcal{\tilde{W}}_{-3}.$
The charge potentials ($\mathcal{L}_{++},\mathcal{W}_{+s})$ and ($\mathcal{%
\tilde{L}}_{--},\mathcal{\tilde{W}}_{-s})$ are conserved quantities
(constant of motion), they will be further discussed later on (\textrm{see
subsection 3.2}).

\item[$\left( \mathbf{3}\right) $] The time components $a_{t}^{{\small (m}%
_{s}{\small )}}$ will fulfill the role of Lagrange multipliers, they are
constrained by the field equations (\ref{ct}) and interpreted as chemical
potentials in the thermodynamics of HS- black holes \textrm{\cite{spin3, t1,
t2, t3}}.
\end{description}

\ \newline
To truly understand the usefulness of\ the HW representation,\ we take on
the simple HS model with $SL\left( 3,\mathbb{R}\right) $ symmetry. This
group has eight traceless generators split in two subsets as $3+5:$ $\left(
\mathbf{i}\right) $ the isotriplet $W_{0,\pm }^{(s=2)}=L_{0,\pm }$ generate
the principal SL$\left( 2,\mathbb{R}\right) $ represented as%
\begin{equation}
L_{+}=\sqrt{2}\left(
\begin{array}{ccc}
{\small 0} & {\small 0} & {\small 0} \\
{\small 1} & {\small 0} & {\small 0} \\
{\small 0} & {\small 1} & {\small 0}%
\end{array}%
\right) ,\quad L_{0}=\left(
\begin{array}{ccc}
{\small 1} & {\small 0} & {\small 0} \\
{\small 0} & {\small 0} & {\small 0} \\
{\small 0} & {\small 0} & {\small -1}%
\end{array}%
\right) ,\quad L_{-}=\sqrt{2}\left(
\begin{array}{ccc}
{\small 0} & -{\small 1} & {\small 0} \\
{\small 0} & {\small 0} & -{\small 1} \\
{\small 0} & {\small 0} & {\small 0}%
\end{array}%
\right)  \label{at}
\end{equation}%
with real $\left( L_{n}\right) ^{\ast }=L_{n}$ and $\left( L_{n}\right)
^{\dagger }=\left( -\right) ^{n}L_{-n}$ as well as%
\begin{equation}
Tr\left( L_{0}L_{0}\right) =2,\qquad Tr\left( L_{-}L_{+}\right) =-4
\end{equation}%
$\left( \mathbf{ii}\right) $ the 5-uplet $W_{0,\pm 1,\pm 2}^{(s=3)}$ forming
a 5-dim representation of SL$\left( 2,\mathbb{R}\right) $ are given by%
\begin{equation}
\begin{tabular}{lllll}
$W_{+2}$ & $=\left(
\begin{array}{ccc}
{\small 0} & {\small 0} & {\small 0} \\
{\small 0} & {\small 0} & {\small 0} \\
{\small 4} & {\small 0} & {\small 0}%
\end{array}%
\right) $ & , & $W_{+}$ & $=\left(
\begin{array}{ccc}
{\small 0} & {\small 0} & {\small 0} \\
\sqrt{{\small 2}} & {\small 0} & {\small 0} \\
{\small 0} & {\small -}\sqrt{{\small 2}} & {\small 0}%
\end{array}%
\right) $ \\
$W_{0}$ & $=\left(
\begin{array}{ccc}
\frac{2}{3} & {\small 0} & {\small 0} \\
{\small 0} & {\small -}\frac{4}{3} & {\small 0} \\
{\small 0} & {\small 0} & \frac{2}{3}%
\end{array}%
\right) $ &  &  &  \\
$W_{-2}$ & $=\left(
\begin{array}{ccc}
{\small 0} & {\small 0} & {\small 4} \\
{\small 0} & {\small 0} & {\small 0} \\
{\small 0} & {\small 0} & {\small 0}%
\end{array}%
\right) $ & , & $W_{-}$ & $=\left(
\begin{array}{ccc}
{\small 0} & {\small -}\sqrt{{\small 2}} & {\small 0} \\
{\small 0} & {\small 0} & \sqrt{{\small 2}} \\
{\small 0} & {\small 0} & {\small 0}%
\end{array}%
\right) $%
\end{tabular}
\label{mat}
\end{equation}%
\textrm{where} $\left( W_{n}\right) ^{\ast }=W_{n}$ and $\left( W_{n}\right)
^{\dagger }=\left( -\right) ^{n}W_{-n}$ as well as
\begin{equation}
Tr\left( W_{0}W_{0}\right) =\frac{8}{3},\qquad Tr\left( W_{-}W_{+}\right)
=-4,\qquad Tr\left( W_{-2}W_{+2}\right) =16
\end{equation}%
\textrm{with} the remarkable orthogonality relation $Tr\left(
L_{0}W_{0}\right) =0.$ In this particular higher spin model, the potentials
read as follows:%
\begin{equation}
\begin{tabular}{lll}
$a_{\varphi }^{\text{\textsc{hw}}}$ & $=$ & $L_{+}+a_{\varphi
}^{-}L_{-}+a_{\varphi }^{--}W_{--}$ \\
$a_{t}^{\text{\textsc{hw}}}$ & $=$ & $\dsum%
\limits_{m_{2}=-1}^{+1}a_{t}^{m_{2}}L_{m_{2}}+\dsum%
\limits_{m_{3}=-2}^{+2}a_{t}^{m_{3}}W_{m_{3}}$%
\end{tabular}%
\end{equation}%
with the constraint relations (\ref{ct}). As noticed before, this
realisation can be recast like%
\begin{equation}
\begin{tabular}{lll}
$a_{+}^{\text{\textsc{hw}}}$ & $=$ & $L_{+}-\frac{2\pi }{\mathrm{k}}\mathcal{%
L}_{++}L_{-}-\frac{\pi }{2\mathrm{k}}\mathcal{W}_{+3}W_{--}$ \\
$a_{-}^{\text{\textsc{hw}}}$ & $=$ & $\dsum%
\limits_{m_{2}=-1}^{+1}a_{-}^{m_{2}}L_{m_{2}}+\dsum%
\limits_{m_{3}=-2}^{+2}a_{-}^{m_{3}}W_{m_{3}}$%
\end{tabular}
\label{car}
\end{equation}%
where the components $a_{\varphi }^{\text{\textsc{hw}}}$ and $a_{t}^{\text{%
\textsc{hw}}}$ can be interpreted as $a_{+}^{\text{\textsc{hw}}}$ (charges)
and $a_{-}^{\text{\textsc{hw}}}$ (Lagrange multipliers).

\subsubsection{ Diagonal representation}

For the rest of the analysis, we settle on this type of representation and
set our boundary fields to obey the diagonal form. For the spin 3 case, the
boundary potentials are valued in the Cartan subalgebra of $SL\left( 3,%
\mathbb{R}\right) $ generated by the two commuting $L_{0}$ and $W_{0}$ with
the orthogonal property $Tr(L_{0}W_{0})=0.$ The 1-form connections read as
follows%
\begin{equation}
\begin{tabular}{lll}
$a^{\text{\textsc{diag}}}$ & $=$ & $\left( \Upsilon ^{0}+\mathcal{J}%
^{0}\right) L_{0}+\left( \Gamma ^{0}+\mathcal{K}^{0}\right) W_{0}$ \\
$\tilde{a}^{\text{\textsc{diag}}}$ & $=$ & $(\tilde{\Upsilon}^{0}-\mathcal{%
\tilde{J}}^{0})L_{0}+(\tilde{\Gamma}^{0}-\mathcal{\tilde{K}}_{\varphi
}^{0})W_{0}$%
\end{tabular}
\label{dg}
\end{equation}%
with $L_{0}$ and $W_{0}$ as in (\ref{mat}). They are real potentials as they
verify:
\begin{eqnarray}
(a^{\text{\textsc{diag}}})^{\ast } &=&a^{\text{\textsc{diag}}}\qquad ,\qquad
(\tilde{a}^{\text{\textsc{diag}}})^{\ast }=\tilde{a}^{\text{\textsc{diag}}}
\notag \\
(a^{\text{\textsc{diag}}})^{\dagger } &=&a^{\text{\textsc{diag}}}\qquad
,\qquad (\tilde{a}^{\text{\textsc{diag}}})^{\dagger }=\tilde{a}^{\text{%
\textsc{diag}}}  \label{ca}
\end{eqnarray}%
Notice that for the particular case where $\Gamma ^{0}=\mathcal{K}^{0}=0$
and $\tilde{\Gamma}^{0}=\mathcal{\tilde{K}}_{\varphi }^{0}=0$, the boundary
potentials reduc\textrm{e t}o $a^{\text{\textsc{diag}}}=\left( \Upsilon ^{0}+%
\mathcal{J}^{0}\right) L_{0}$ and $\tilde{a}^{\text{\textsc{diag}}}=(\tilde{%
\Upsilon}^{0}-\mathcal{\tilde{J}}^{0})L_{0}$ meaning that they are in fact
valued in the principal subset $sl\left( 2,\mathbb{R}\right) $ of $sl\left(
3,\mathbb{R}\right) $.

In the above (\ref{dg}), the partial connections are given by%
\begin{equation}
\begin{tabular}{lllll}
$\mathcal{J}^{0}$ & $=\mathcal{J}_{\varphi }^{0}d\varphi $ & $,\qquad $ & $%
\mathcal{K}^{0}$ & $=\mathcal{K}_{\varphi }^{0}d\varphi $ \\
$\Upsilon ^{0}$ & $=\Upsilon _{t}^{0}dt$ & $,\qquad $ & $\Gamma ^{0}$ & $%
=\Gamma _{t}^{0}dt$%
\end{tabular}
\label{ppt}
\end{equation}%
and%
\begin{equation}
\begin{tabular}{lllll}
$\mathcal{\tilde{J}}^{0}$ & $=\mathcal{\tilde{J}}_{\varphi }^{0}d\varphi $ &
$,\qquad $ & $\mathcal{\tilde{K}}^{0}$ & $=\mathcal{\tilde{K}}_{\varphi
}^{0}d\varphi $ \\
$\tilde{\Upsilon}^{0}$ & $=\tilde{\Upsilon}_{t}^{0}dt$ & $,\qquad $ & $%
\tilde{\Gamma}^{0}$ & $=\tilde{\Gamma}_{t}^{0}dt$%
\end{tabular}
\label{pps}
\end{equation}%
with the reality property%
\begin{equation}
(\mathcal{J}^{0})^{\ast }=\mathcal{J}^{0},\qquad (\mathcal{K}^{0})^{\ast }=%
\mathcal{K}^{0},\qquad (\mathcal{\tilde{J}}^{0})^{\ast }=\mathcal{\tilde{J}}%
^{0},\qquad (\mathcal{\tilde{K}}^{0})^{\ast }=\mathcal{\tilde{K}}^{0}
\end{equation}%
The use of partial 1-form potentials is \textrm{essential as} $\left(
\mathbf{i}\right) $ it \textrm{distinguishes} the charges ($\mathcal{J},%
\mathcal{K};\mathcal{\tilde{J}},\mathcal{\tilde{K}}$) \textrm{from} the
Lagrange multipliers ($\Upsilon ,\Gamma ;\tilde{\Upsilon},\tilde{\Gamma}$);
and $\left( \mathbf{ii}\right) $ \textrm{differentiates} between the two
diagonal gl(1,$\mathbb{R}$)$_{\mathcal{J}}$ and gl(1,$\mathbb{R}$)$_{%
\mathcal{K}}$ \textrm{for} \textrm{each copy of }sl(3,$\mathbb{R}$). Hence,
the components of $a^{\text{\textsc{diag}}}$ and $\tilde{a}^{\text{\textsc{%
diag}}}$ are defined as%
\begin{equation}
\begin{tabular}{lll}
$a_{\varphi }^{\text{\textsc{diag}}}$ & $=$ & $\mathcal{J}_{\varphi
}^{0}L_{0}+\mathcal{K}_{\varphi }^{0}W_{0}$ \\
$a_{t}^{\text{\textsc{diag}}}$ & $=$ & $\Upsilon _{t}^{0}L_{0}+\Gamma
_{t}^{0}W_{0}$%
\end{tabular}%
\qquad ,\qquad
\begin{tabular}{lll}
$\tilde{a}_{\varphi }^{\text{\textsc{diag}}}$ & $=$ & $-\mathcal{\tilde{J}}%
_{\varphi }^{0}L_{0}-\mathcal{\tilde{K}}_{\varphi }^{0}W_{0}$ \\
$\tilde{a}_{t}^{\text{\textsc{diag}}}$ & $=$ & $\tilde{\Upsilon}%
_{t}^{0}L_{0}+\tilde{\Gamma}_{t}^{0}W_{0}$%
\end{tabular}
\label{da}
\end{equation}%
Note that the diagonal potentials $(a^{\text{\textsc{diag}}},\tilde{a}^{%
\text{\textsc{diag}}})$ are related to \textrm{the highest weight gauge
potentials} $(a^{\text{\textsc{hw}}},\tilde{a}^{\text{\textsc{hw}}})$
through $SL\left( 3,\mathbb{R}\right) $ gauge transformations $a^{\text{%
\textsc{hw}}}=g_{\text{\textsc{bridge}}}^{-1}a^{\text{\textsc{diag}}}g_{%
\text{\textsc{bridge}}}$ and $\tilde{a}^{\text{\textsc{hw}}}=\tilde{g}_{%
\text{\textsc{bridge}}}^{-1}\tilde{a}^{\text{\textsc{diag}}}\tilde{g}_{\text{%
\textsc{bridge}}}$ with commutative diagram%
\begin{equation}
\begin{tabular}{ccccc}
\textbf{Bulk} & :$\quad $ & $A^{\text{\textsc{hw}}}$ & $\quad -\boldsymbol{G}%
_{I}-\rightarrow \quad $ & $A^{\text{\textsc{diag}}}$ \\
$\downarrow $ \emph{asymp}{\small .} &  & $\mathbf{\downarrow }\boldsymbol{G}%
_{II}^{\prime }$ &  & $\mathbf{\downarrow }\boldsymbol{G}_{II}$ \\
\textbf{Boundary} & :$\quad $ & $a^{\text{\textsc{hw}}}$ & $\quad -%
\boldsymbol{G}_{I}^{\prime }-\rightarrow \quad $ & $a^{\text{\textsc{diag}}}$%
\end{tabular}%
\end{equation}%
where ($\boldsymbol{G}_{I},\boldsymbol{G}_{II}$) and ($\boldsymbol{G}%
_{I}^{\prime },\boldsymbol{G}_{II}^{\prime }$) are some gauge
transformations. The bridge transformations between $a^{\text{\textsc{hw}}}$
and $a^{\text{\textsc{diag}}}$\ have the structure $g_{\text{\textsc{bridge}}%
}=g_{1}g_{2}$ with $g_{1}=\exp (\theta _{-}L_{+}+\zeta _{-}W_{+}+\zeta
_{-2}W_{+2})$ and $g_{2}=\exp (\mathcal{F}_{+}L_{-}+\mathcal{G}_{+}W_{-}+%
\mathcal{G}_{+2}W_{-2})$. They were computed in \cite{1B} where it was shown
that the left charges ($\mathcal{L}_{++},\mathcal{W}_{+3})\equiv (\mathcal{L}%
_{\varphi \varphi },\mathcal{W}_{\varphi \varphi \varphi })$ and their twild
homologue ($\mathcal{\tilde{L}}_{--},\mathcal{\tilde{W}}_{-3})\equiv (%
\mathcal{\tilde{L}}_{\varphi \varphi },\mathcal{\tilde{W}}_{\varphi \varphi
\varphi })$ are related to ($\mathcal{J}_{\varphi }^{0},\mathcal{K}_{\varphi
}^{0})$ and ($\mathcal{\tilde{J}}_{\varphi }^{0},\mathcal{\tilde{K}}%
_{\varphi }^{0}$) as follows%
\begin{equation}
\begin{tabular}{lll}
$\mathcal{L}_{\varphi \varphi }$ & $=$ & $+\frac{\mathrm{k}}{4\pi }\left[
\frac{1}{2}\left( \mathcal{J}_{\varphi }^{0}\right) ^{2}+\frac{\partial
\mathcal{J}_{\varphi }^{0}}{\partial \varphi }+\frac{2}{3}\left( \mathcal{K}%
_{\varphi }^{0}\right) ^{2}\right] $ \\
$\mathcal{W}_{\varphi \varphi \varphi }$ & $=$ & $-\frac{\mathrm{k}}{6\pi }%
\left[ -\frac{8}{9}\left( \mathcal{K}_{\varphi }^{0}\right) ^{3}+\mathcal{K}%
_{\varphi }^{0}\left( 2\left( \mathcal{J}_{\varphi }^{0}\right)
^{2}+\partial _{\varphi }\mathcal{J}_{\varphi }^{0}\right) +3\mathcal{J}%
_{\varphi }^{0}\partial _{\varphi }\mathcal{K}_{\varphi }^{0}+\partial
_{\varphi }^{2}\mathcal{K}_{\varphi }^{0}\right] $%
\end{tabular}
\label{L1}
\end{equation}%
and%
\begin{equation}
\begin{tabular}{lll}
$\mathcal{\tilde{L}}_{\varphi \varphi }$ & $=$ & $-\frac{\mathrm{k}}{4\pi }%
\left[ \frac{1}{2}(\mathcal{\tilde{J}}_{\varphi }^{0})^{2}+\frac{\partial
\mathcal{\tilde{J}}_{\varphi }^{0}}{\partial \varphi }+\frac{2}{3}(\mathcal{%
\tilde{K}}_{\varphi }^{0})^{2}\right] $ \\
$\mathcal{\tilde{W}}_{\varphi \varphi \varphi }$ & $=$ & $+\frac{\mathrm{k}}{%
6\pi }\left[ -\frac{8}{9}(\mathcal{\tilde{K}}_{\varphi }^{0})^{3}+\mathcal{%
\tilde{K}}_{\varphi }\left( 2(\mathcal{\tilde{J}}_{\varphi
}^{0})^{2}+\partial _{\varphi }\mathcal{\tilde{J}}_{\varphi }^{0}\right) +3%
\mathcal{\tilde{J}}_{\varphi }^{0}\partial _{\varphi }\mathcal{\tilde{K}}%
_{\varphi }^{0}+\partial _{\varphi }^{2}\mathcal{\tilde{K}}_{\varphi }^{0}%
\right] $%
\end{tabular}
\label{L2}
\end{equation}%
\begin{equation*}
\end{equation*}%
These Sugawara like relationships between ($\mathcal{L}_{\varphi \varphi },%
\mathcal{W}_{\varphi \varphi \varphi }$) and ($\mathcal{J}_{\varphi }^{0},%
\mathcal{K}_{\varphi }^{0}$)\ as well as the ones between their right
counterparts ($\mathcal{\tilde{L}}_{\varphi \varphi },\mathcal{\tilde{W}}%
_{\varphi \varphi \varphi }$) and ($\mathcal{\tilde{J}}_{\varphi }^{0},%
\mathcal{\tilde{K}}_{\varphi }^{0}$) are well known in the Literature of the%
\textrm{\ }W-algebras extending the 2D boundary conformal invariance; they
are generated by the so-called Miura transformations \textrm{\cite{miura}}
of the W$_{3}$-symmetry.

\section{Real forms and boundary configurations}

\label{sec3} \qquad In this section, we \textrm{develop} our approach to
deal with the equations of motion (\ref{ct}) in order to investigate the
linkage between HS black flowers and the conserved charges of the real forms
of the complexified gauge symmetry $SL\left( 3,\mathbb{C}\right) _{L}\times
SL\left( 3,\mathbb{C}\right) _{R}$. First, we \textrm{shed}\emph{\ }light on
the role of the real forms of the SL(N,$\mathbb{C}$) algebra in the
computation of the entropy \textsc{S}$_{\text{\textsc{hs-bf}}}$ of the
thermal HS black flowers. Then, we explore various possible solutions of the
equations of motion (\ref{ct}) with potentials as in (\ref{dg}).

For these purposes, we start by the equations of motion
\begin{equation}
\begin{tabular}{lll}
$\partial _{t}a_{\varphi }^{\text{\textsc{diag}}}-\partial _{\varphi }a_{t}^{%
\text{\textsc{diag}}}$ & $=$ & $0$ \\
$\partial _{t}\tilde{a}_{\varphi }^{\text{\textsc{diag}}}-\partial _{\varphi
}\tilde{a}_{t}^{\text{\textsc{diag}}}$ & $=$ & $0$%
\end{tabular}
\label{tc}
\end{equation}%
having no quadratic term because the diagonal gauge connections commute;
i.e: $[a_{\varphi }^{\text{\textsc{diag}}},a_{t}^{\text{\textsc{diag}}}]=[%
\tilde{a}_{\varphi }^{\text{\textsc{diag}}},\tilde{a}_{t}^{\text{\textsc{diag%
}}}]=0$ due to the trivial commutation relation $\left[ L_{0},W_{0}\right]
=0.$ Using (\ref{da}), the equations (\ref{tc}) split as follows%
\begin{equation}
\begin{tabular}{lll}
$\partial _{t}\mathcal{J}_{\varphi }^{{\small 0}}-\partial _{\varphi
}\Upsilon _{t}^{{\small 0}}$ & $=$ & $0$ \\
$\partial _{t}\mathcal{\tilde{J}}_{\varphi }^{{\small 0}}+\partial _{\varphi
}\tilde{\Upsilon}_{t}^{{\small 0}}$ & $=$ & $0$%
\end{tabular}%
\qquad ,\qquad
\begin{tabular}{lll}
$\partial _{t}\mathcal{K}_{\varphi }^{{\small 0}}-\partial _{\varphi }\Gamma
_{t}^{{\small 0}}$ & $=$ & $0$ \\
$\partial _{t}\mathcal{\tilde{K}}_{\varphi }^{{\small 0}}+\partial _{\varphi
}\tilde{\Gamma}_{t}^{{\small 0}}$ & $=$ & $0$%
\end{tabular}
\label{fem}
\end{equation}%
These differential equations give dynamical relationships between the
boundary charges ($\mathcal{J}_{\varphi }^{{\small 0}},\mathcal{K}_{\varphi
}^{{\small 0}}$) and ($\mathcal{\tilde{J}}_{\varphi }^{{\small 0}},\mathcal{%
\tilde{K}}_{\varphi }^{{\small 0}}$) of the HS\ theory $SL\left( 3,\mathbb{R}%
\right) _{L}\times SL\left( 3,\mathbb{R}\right) _{R}$ and the associated
Lagrange multipliers ($\Upsilon _{t}^{{\small 0}},\Gamma _{t}^{{\small 0}}$)
and ($\tilde{\Upsilon}_{t}^{{\small 0}},\tilde{\Gamma}_{t}^{{\small 0}}$).

There are various methods to solve and work out the properties of the eqs(%
\ref{tc}), among which we cite\textrm{:}

\begin{description}
\item[$\left( \mathbf{a}\right) $] the topological boundary gauge
configuration. Here, the gauge curvatures $f^{\text{\textsc{diag}}}=da^{%
\text{\textsc{diag}}}$ and $\tilde{f}^{\text{\textsc{diag}}}=d\tilde{a}^{%
\text{\textsc{diag}}}$ vanish identically ($da^{\text{\textsc{diag}}}=d%
\tilde{a}^{\text{\textsc{diag}}}=0);$ they correspond to exact boundary
potential forms namely $a^{\text{\textsc{diag}}}=d\vartheta $ and $\tilde{a}%
^{\text{\textsc{diag}}}=d\tilde{\vartheta}$. This asymptotic image of the
potentials does not suit the goal of our study and will be accordingly
dropped.

\item[$\left( \mathbf{b}\right) $] the fixed Lagrange multipliers. This
solution is based on boundary potentials having constrained Lagrange
multipliers like $\partial _{\varphi }\Upsilon _{t}=\partial _{\varphi
}\Gamma _{t}=0.$ These conditions can be effectively fulfilled by just taking%
\textrm{\ }$\Upsilon _{t}=$\textsc{cte}$_{2}$ and $\Gamma _{t}=$\textsc{cte}$%
_{3}$; they allow explicit calculations and permit to extract interesting
results without need of involved computations. It will be thoroughly
examined for the remainder of the paper.

\item[$\left( \mathbf{c}\right) $] the primary scalar and/or fermionic
fields. In this approach, we consider bosonic and fermionic conformal fields
fulfilling boundary field equations of motion given by (\ref{tc}). This
two-dimensional QFT model can be viewed as a boundary CFT$_{2}$ solution
that permits interesting physical interpretations.\textrm{\ }For example, it
allows to realise\textrm{\ }the Heisenberg soft hair states $\left\vert N_{%
\text{\textsc{soft-hair}}}\right\rangle $ at the horizon of the HS- black
flowers \textrm{\cite{SH}} in terms of particles of the HS black flower.
Later on this section, we delve into the key idea behind this proposal\
which we will further develop \textrm{in section 5}.
\end{description}

\subsection{Real forms of complex Lie symmetries}

Before going \textrm{into further details}, we\emph{\ }\textrm{should first
shed}\emph{\ }light on the \textrm{significant} role played by the real
forms of the complex SL(N,$\mathbb{C}$) in the derivation of the entropy
\textsc{S}$_{\text{\textsc{hs-bf}}}$ of the thermal HS black flowers. Indeed,%
\textrm{\ }in order to calculate the entropies \textsc{S}$_{\text{\textsc{%
hs-bf}}}$ of the thermal HS- black holes, one needs to $\left( \mathbf{i}%
\right) $ compactify the time variable by performing a Wick rotation $%
t=it_{E}$ parameterising the thermal cycle $\mathcal{C}_{t_{{\small E}}}=%
\left[ 0,\mathrm{\beta }\right] $ with $\mathrm{\beta }$ referring to the
inverse of temperature; and $\left( \mathbf{ii}\right) $ demand regularity
of the holonomy%
\begin{equation}
\exp \left( i\dint\nolimits_{\mathcal{C}_{t_{{\small E}}}}a_{t_{{\small E}%
}}dt_{E}\right)  \label{reg}
\end{equation}%
that must be equal to $(-)^{N+1}I_{id}$ due to the reality of the gauge
symmetry SL(N,$\mathbb{R}).$ In this regard, notice that for odd values of
N, this holonomy must be equal to the identity $I_{id}$ (termed below as
trivial holonomy or neutral holonomy because $I_{id}$ is the neutral element
of $\mathbb{Z}_{N}$) while for even $N$, it is equal to $-I_{id}.$ More
generally, for the complexified SL(N,$\mathbb{C})$, one may replace this
condition by $zI_{id}$ where the complex number z is constrained like $%
z^{N}=1$ having N solutions given by $\omega ^{p}=e^{2i\pi p/N}$
corresponding to the group elements of the $\mathbb{Z}_{N}$ centre of SL(N,$%
\mathbb{C}).$ For odd N, we have only one real solution $I_{id}$ while for
even N we get both $\pm I_{id}$. As for the complex solutions, they depend
on the rational $p/N$ numbers associated to the phase $2p\pi /N$ beyond the
values 0 and $\pi $ $\func{mod}2\pi $.\newline
Moreover, the Wick rotation $t=it_{E}$ maps the boundary cylinder $\mathcal{C%
}_{\varphi }\times \mathbb{R}_{t}$ parameterised by ($\varphi ,t$) to the
torus $\mathbb{T}^{2}=\mathcal{C}_{\varphi }\times \mathcal{C}_{t_{E}}$
coordinated by the periodic variables ($\varphi ,t_{E}$). Furthermore, under
time compactification we have the following features:

\begin{description}
\item[$\left( \mathbf{1}\right) $] the \emph{real} partial connections $%
a_{t}^{\text{\textsc{diag}}}dt=\Upsilon +\Gamma $ and $\tilde{a}_{t}^{\text{%
\textsc{diag}}}dt=\tilde{\Upsilon}+\Gamma $ valued in sl(3,$\mathbb{R}$)
with L$_{0}$ and W$_{0}$ components as
\begin{equation}
\begin{tabular}{ll}
$\Upsilon $ & $=\left( \Upsilon _{t}^{0}L_{0}\right) dt$ \\
$\tilde{\Upsilon}$ & $=(\tilde{\Upsilon}_{t}^{0}L_{0})dt$%
\end{tabular}%
\qquad ,\qquad
\begin{tabular}{ll}
$\Gamma $ & $=\left( \Gamma _{t}^{0}W_{0}\right) dt$ \\
$\tilde{\Gamma}$ & $=(\tilde{\Gamma}_{t}^{0}W_{0})dt$%
\end{tabular}
\label{34}
\end{equation}%
get mapped to the \emph{pure imaginary }ones%
\begin{equation}
\begin{tabular}{ll}
$i\Upsilon _{E}$ & $=\Upsilon _{t_{E}}^{0}\left( iL_{0}\right) dt_{E}$ \\
$i\tilde{\Upsilon}_{E}$ & $=\tilde{\Upsilon}_{t_{E}}^{0}(iL_{0})dt_{E}$%
\end{tabular}%
\qquad ,\qquad
\begin{tabular}{ll}
$i\Gamma _{E}$ & $=\Gamma _{t_{E}}^{0}\left( iW_{0}\right) dt_{E}$ \\
$i\tilde{\Gamma}_{E}$ & $=\tilde{\Gamma}_{t_{E}}^{0}(iW_{0})dt_{E}$%
\end{tabular}%
\end{equation}%
which no longer sit in the real sl(3,$\mathbb{R}$).

\item[$\left( \mathbf{2}\right) $] \textrm{the }real exponentiations of (\ref%
{34}) valued in the real group SL(3,$\mathbb{R}$) namely
\begin{equation}
\begin{tabular}{ll}
$\exp (\int \Upsilon )$ & $=\exp \left( \int \left( \Upsilon
_{t}^{0}L_{0}\right) dt\right) $ \\
$\exp (\int \tilde{\Upsilon})$ & $=\exp \left( \int (\tilde{\Upsilon}%
_{t}^{0}L_{0})dt\right) $%
\end{tabular}%
\qquad ,\qquad
\begin{tabular}{ll}
$\exp (\int \Gamma )$ & $=\exp \left( \int \left( \Gamma
_{t}^{0}W_{0}\right) dt\right) $ \\
$\exp (\int \tilde{\Gamma})$ & $=\exp \left( \int (\tilde{\Gamma}%
_{t}^{0}W_{0})dt\right) $%
\end{tabular}%
,
\end{equation}%
get in turn mapped into the following complex U(1) phases%
\begin{equation}
\begin{tabular}{ll}
$\exp (i\int \Upsilon _{E})$ & $=\exp \left( i\int \left( \Upsilon
_{t_{E}}^{0}L_{0}\right) dt_{E}\right) $ \\
$\exp (i\int \tilde{\Upsilon}_{E})$ & $=\exp \left( i\int (\tilde{\Upsilon}%
_{t_{E}}^{0}L_{0})dt_{E}\right) $%
\end{tabular}%
,\quad
\begin{tabular}{ll}
$\exp (i\int \Gamma _{E})$ & $=\exp \left( i\int \left( \Gamma
_{t_{E}}^{0}W_{0}\right) dt_{E}\right) $ \\
$\exp (i\int \tilde{\Gamma}_{E})$ & $=\exp \left( i\int (\tilde{\Gamma}%
_{t_{E}}^{0}W_{0})dt_{E}\right) $%
\end{tabular}
\label{38}
\end{equation}
\end{description}

These mappings show that under Wick rotation, the real matrices ($%
L_{0},W_{0} $) are replaced by the \emph{pure imaginary} ($iL_{0},iW_{0}$).
Notice that the real ($L_{0},W_{0}$) generates the real abelian subgroup $S%
\left[ GL(1,\mathbb{R})\times GL(1,\mathbb{R})\right] $ of SL(3,$\mathbb{R}$%
) with group elements as
\begin{equation}
g_{_{{\small SL}_{{\small 3}}}}=\left(
\begin{array}{ccc}
e^{\zeta ^{0}L_{0}} & 0 & 0 \\
0 & e^{\eta ^{0}W_{0}} & 0 \\
0 & 0 & e^{-\zeta ^{0}L_{0}}e^{-\eta ^{0}W_{0}}%
\end{array}%
\right) ,\qquad \zeta ^{0},\eta ^{0}\in \mathbb{R},\qquad g_{_{{\small SL}_{%
{\small 3}}}}^{\ast }=g_{_{{\small SL}_{{\small 3}}}}^{\dagger }=g_{_{%
{\small SL}_{{\small 3}}}},
\end{equation}%
However, the pure imaginary ($iL_{0},iW_{0}$) engender the complex unitary $S%
\left[ U(1)\times U(1)\right] $ subgroup of SU(3) with elements as%
\begin{equation}
g_{_{{\small SU}_{{\small 3}}}}=\left(
\begin{array}{ccc}
e^{i\zeta ^{0}L_{0}} & 0 & 0 \\
0 & e^{i\eta ^{0}W_{0}} & 0 \\
0 & 0 & e^{-i\zeta ^{0}L_{0}}e^{-i\eta ^{0}W_{0}}%
\end{array}%
\right) ,\qquad g_{_{{\small SU}_{{\small 3}}}}^{\ast }=g_{_{{\small SU}_{%
{\small 3}}}}^{\dagger }=g_{_{{\small SU}_{{\small 3}}}}^{-1}
\end{equation}%
So, under the Wick rotation, the real SL(3,$\mathbb{R}$) gets mapped into
the complex unitary SU(3). This relationship between SL(3,$\mathbb{R}$) and
SU(3) is not unusual considering that they are, respectively, the split and
compact real forms of SL(3,$\mathbb{C}$). This feature indicates that the
real forms of SL(3,$\mathbb{C}$) should play an important role\textrm{\ }in
the investigation\textrm{\ }of black hole thermodynamics. As such, the
various real forms of SL(N,$\mathbb{C}$) have to be monitored while
computing the HS black holes entropies especially while descending towards
the core spin 2 black holes of AdS$_{\mathrm{3}}$\ gravity. For these
reasons, we briefly overview the real forms of the\textrm{\ }$SL(N,\mathbb{C}%
)$\textrm{\ }series with concrete examples and graphical illustrations given
by Tits-Satake diagrams \textrm{\cite{ST, ST2} }in appendix A\textrm{.}

\subsection{Solving the equations of motion}

Since the equations of motion (\ref{fem}) play an important role in the
study of HS black holes, we find it imperative to explore the various
possible solutions. Strictly speaking, the boundary ($a^{\text{\textsc{diag}}%
},\tilde{a}^{\text{\textsc{diag}}}$) involved in these relations are not
arbitrary fields seeing that\ they must obey (\ref{fem}) and undergo certain
gauge transformations. Thus, the fundamental variables are a priori obtained
by solving the field equations (\ref{fem}); which can be achieved in various
ways depending on the targeted features. Below, we give three types of
solutions: the first two are described in details here after while the third
is briefly addressed as it will be further investigated \textrm{in section 5}%
.

\subsubsection{Topological gauge configuration}

The first type of\emph{\ }\textrm{(\ref{fem})'s }solutions is given by the
trivial realisation with boundary potentials taken as follows%
\begin{equation}
\begin{tabular}{lll}
$\mathcal{J}_{\varphi }^{{\small 0}}$ & $=$ & $\partial _{\varphi }\vartheta
_{{\small 1}}^{{\small 0}}$ \\
$\mathcal{K}_{\varphi }^{{\small 0}}$ & $=$ & $\partial _{\varphi }\vartheta
_{{\small 2}}^{{\small 0}}$%
\end{tabular}%
,\quad
\begin{tabular}{lll}
$\Upsilon _{t}^{{\small 0}}$ & $=$ & $\partial _{t}\vartheta _{{\small 1}}^{%
{\small 0}}$ \\
$\Gamma _{t}^{{\small 0}}$ & $=$ & $\partial _{t}\vartheta _{{\small 2}}^{%
{\small 0}}$%
\end{tabular}%
,\quad
\begin{tabular}{lll}
$\mathcal{\tilde{J}}_{\varphi }^{{\small 0}}$ & $=$ & $\partial _{\varphi }%
\tilde{\vartheta}_{{\small 1}}^{{\small 0}}$ \\
$\mathcal{\tilde{K}}_{\varphi }^{{\small 0}}$ & $=$ & $\partial _{\varphi }%
\tilde{\vartheta}_{{\small 2}}^{{\small 0}}$%
\end{tabular}%
,\quad
\begin{tabular}{lll}
$\tilde{\Upsilon}_{t}^{{\small 0}}$ & $=$ & $\partial _{t}\tilde{\vartheta}_{%
{\small 1}}^{{\small 0}}$ \\
$\tilde{\Gamma}_{t}^{{\small 0}}$ & $=$ & $\partial _{t}\tilde{\vartheta}_{%
{\small 2}}^{{\small 0}}$%
\end{tabular}%
\end{equation}%
where the fields $\vartheta _{i}^{{\small 0}}\left( \varphi ,t\right) $ and $%
\tilde{\vartheta}_{i}^{{\small 0}}\left( \varphi ,t\right) $ are arbitrary
scalar functions. This solution reflects\emph{\ }the\textrm{\ }topological
nature of the abelian potentials $a^{\text{\textsc{diag}}}$ and $\tilde{a}^{%
\text{\textsc{diag}}}$ where\textrm{\ }the vanishing of the curvatures $%
f=da^{\text{\textsc{diag}}}$ and $\tilde{f}=d\tilde{a}^{\text{\textsc{diag}}%
} $ is assured by
\begin{equation}
\begin{tabular}{lllll}
$a^{\text{\textsc{diag}}}$ & $=$ & $d\vartheta _{1}+d\vartheta _{2}$ & $=$ &
$\left( d\vartheta _{{\small 1}}^{{\small 0}}\right) L_{{\small 0}}+\left(
d\vartheta _{{\small 2}}^{{\small 0}}\right) W_{{\small 0}}$ \\
$\tilde{a}^{\text{\textsc{diag}}}$ & $=$ & $d\tilde{\vartheta}_{1}+d\tilde{%
\vartheta}_{2}$ & $=$ & $(d\tilde{\vartheta}_{{\small 1}}^{{\small 0}})L_{%
{\small 0}}+(d\tilde{\vartheta}_{{\small 2}}^{{\small 0}})W_{{\small 0}}$%
\end{tabular}%
\end{equation}%
with $d^{2}\vartheta _{i}^{{\small 0}}=0$ and $d^{2}\tilde{\vartheta}_{i}^{%
{\small 0}}=0.$ Here, gauge invariant objects are given by topological
defects \cite{wilson, wilson2,y}. This realisation treats ($\mathcal{J}%
_{\varphi }^{{\small 0}},\mathcal{K}_{\varphi }^{{\small 0}};\mathcal{\tilde{%
J}}_{\varphi }^{{\small 0}},\mathcal{\tilde{K}}_{\varphi }^{{\small 0}}$)
and ($\Upsilon _{t}^{{\small 0}},\Gamma _{t}^{{\small 0}};\tilde{\Upsilon}%
_{t}^{{\small 0}},\tilde{\Gamma}_{t}^{{\small 0}}$) on a somehow equal
footing which makes it less significant for our study of higher spin black
holes and will be dropped for the rest of the analysis\textrm{.}

\subsubsection{Fixing Lagrange multipliers}

The second type of solutions fulfilling the constraint (\ref{fem}) is given
by the interesting realisation where the angular gradient of Lagrange
multipliers namely $(\partial _{\varphi }\Upsilon _{t}^{0},\partial
_{\varphi }\Gamma _{t}^{0})$ and ($\partial _{\varphi }\tilde{\Upsilon}%
_{t}^{0},\partial _{\varphi }\tilde{\Gamma}_{t}^{0})$ are assumed to be
vanishing quantities; that is
\begin{equation}
\begin{tabular}{lll}
$\partial _{\varphi }\Upsilon _{t}^{0}$ & $=$ & $0$ \\
$\partial _{\varphi }\Gamma _{t}^{0}$ & $=$ & $0$%
\end{tabular}%
\qquad ,\qquad
\begin{tabular}{lll}
$\partial _{\varphi }\tilde{\Upsilon}_{t}^{0}$ & $=$ & $0$ \\
$\partial _{\varphi }\tilde{\Gamma}_{t}^{0}$ & $=$ & $0$%
\end{tabular}
\label{cst}
\end{equation}%
requiring $(\Upsilon _{t}^{0},\Gamma _{t}^{0})$ and ($\tilde{\Upsilon}%
_{t}^{0},\tilde{\Gamma}_{t}^{0})$ $\varphi $-independent functions but they
may still be time dependent. Below, we consider the simple case where both%
\textrm{\ }$\Upsilon _{t}^{0}$\textrm{\ and }$\Gamma _{t}^{0}$\textrm{\ are }%
constant and study the arising features.

\paragraph{\textbf{A)} \textbf{Case of constant Lagrange multipliers:}}

\ \ \newline
We satisfy the requirements of the eqs(\ref{cst}) by setting the Lagrange
multipliers to take\emph{\ }constant real numbers as follows%
\begin{eqnarray}
\Upsilon _{t}^{0} &=&\lambda ^{0}\qquad ,\qquad \Gamma _{t}^{0}=\Lambda ^{0}
\notag \\
\tilde{\Upsilon}_{t}^{0} &=&\tilde{\lambda}^{0}\qquad ,\qquad \tilde{\Gamma}%
_{t}^{0}=\tilde{\Lambda}^{0}
\end{eqnarray}%
which by putting into (\ref{da}), the real line functionals%
\begin{equation}
\boldsymbol{W}_{a_{\tau }}=\exp \left( \int_{0}^{\mathrm{\tau }}a_{t}^{\text{%
\textsc{diag}}}dt\right) \qquad ,\qquad \boldsymbol{\tilde{W}}_{a_{\tau
}}=\exp \left( \int_{0}^{\mathrm{\tau }}\tilde{a}_{t}^{\text{\textsc{diag}}%
}dt\right)  \label{1W}
\end{equation}%
\textrm{read }explicitly as follows%
\begin{eqnarray}
\boldsymbol{W}_{a_{\tau }} &=&\exp \left[ \mathrm{\tau }\left(
\begin{array}{ccc}
{\small \frac{2}{3}\Lambda ^{0}+\lambda }^{0} & {\small 0} & {\small 0} \\
{\small 0} & -\frac{4}{3}{\small \Lambda }^{0} & {\small 0} \\
{\small 0} & {\small 0} & {\small \frac{2}{3}\Lambda ^{0}-\lambda }^{0}%
\end{array}%
\right) \right]  \label{W1} \\
&&  \notag \\
\boldsymbol{\tilde{W}}_{a_{\tau }} &=&\exp \left[ \mathrm{\tau }\left(
\begin{array}{ccc}
{\small \frac{2}{3}\tilde{\Lambda}^{0}+\tilde{\lambda}}^{0} & {\small 0} &
{\small 0} \\
{\small 0} & -\frac{4}{3}{\small \tilde{\Lambda}}^{0} & {\small 0} \\
{\small 0} & {\small 0} & {\small \frac{2}{3}\tilde{\Lambda}^{0}-\tilde{%
\lambda}}^{0}%
\end{array}%
\right) \right]  \notag
\end{eqnarray}%
with $\det \boldsymbol{W}_{a_{\tau }}=\det \boldsymbol{\tilde{W}}_{a_{\tau
}}=1$ and $\boldsymbol{W}_{a_{\tau }}^{\ast }=\boldsymbol{W}_{a_{\tau }}$ as
well as $\boldsymbol{\tilde{W}}_{a_{\tau }}^{\ast }=\boldsymbol{\tilde{W}}%
_{a_{\tau }}.$

\

\textbf{A.1) Spin 2 branches in higher spin 3 theory }\newline
The matrices $T=\lambda ^{0}L_{0}+\frac{2}{3}\Lambda ^{0}W_{0}$ and $\tilde{T%
}=\tilde{\lambda}^{0}L_{0}+\frac{2}{3}\tilde{\Lambda}^{0}W_{0}$ in the above
exponentials belong to the Cartan subalgebra of sl(3,$\mathbb{R}$); they
read explicitly as%
\begin{equation}
T=\left(
\begin{array}{ccc}
\frac{2}{3}\Lambda ^{0}+\lambda ^{0} & 0 & 0 \\
0 & -\frac{4}{3}\Lambda ^{0} & 0 \\
0 & 0 & \frac{2}{3}\Lambda ^{0}-\lambda ^{0}%
\end{array}%
\right)  \label{M}
\end{equation}%
A similar formula is also available for $\tilde{T}.$ These matrices capture
important data on the black holes entropies \textrm{provided by} \textrm{the}
constraints on the values of the chemical potentials ($\lambda ^{0},\Lambda
^{0}$). Because sl(3) contains three basic sl(2)s \textrm{in one} to one
correspondence with its 3 positive roots namely%
\begin{equation}
\begin{tabular}{ccccc}
positive roots & : & $\alpha _{1}$ & $\alpha _{2}$ & $\ \ \alpha _{3}=\alpha
_{1}+\alpha _{2}$ \\
$sl(2)$s & : & $\ \ sl(2)_{\alpha _{1}}$ \ \ \  & $\ \ \ sl(2)_{\alpha _{2}}$
\ \ \  & $sl(2)_{\alpha _{3}}$ \\
Cartans & : & $H_{\alpha _{1}}$ & $H_{\alpha _{2}}$ & $\ \ \ \ H_{\alpha
_{3}}=H_{\alpha _{1}}+H_{\alpha _{2}}$ \ \ \ \
\end{tabular}%
\end{equation}%
we can\emph{\ }identify several 2-dim chambers in the parameter space of ($%
\lambda ^{0},\Lambda ^{0}$); see Figure \textbf{\ref{sl2} }\textrm{where we
set} the identification $(\lambda ^{0},\Lambda ^{0})=(\zeta ,\eta ).$\
Below, we give \textrm{the different boundary lines chambers} \textrm{of}
the Cartan subspace:

\begin{itemize}
\item \textbf{Boundary line} $\Lambda ^{0}=0$ \textbf{and} $\lambda ^{0}\in
\mathbb{R}:$ In this case, the matrix (\ref{M}) reduces to
\begin{equation}
T_{j=1}=\lambda ^{0}\left(
\begin{array}{ccc}
1 & 0 & 0 \\
0 & 0 & 0 \\
0 & 0 & -1%
\end{array}%
\right)
\end{equation}%
with $L_{0}$ corresponding to the Cartan of $sl(2)_{\alpha _{3}}$ in the
isospin representation $j=1.$ This line corresponds to the horizontal axis
in Figure \textbf{\ref{sl2}}.

\item \textbf{Boundary line }$\lambda ^{0}=0$ \textbf{and} $\Lambda ^{0}\in
\mathbb{R}:$ Here, the matrix (\ref{M}) becomes%
\begin{equation}
T_{j=2}=\frac{2}{3}\Lambda ^{0}\left(
\begin{array}{ccc}
1 & 0 & 0 \\
0 & -2 & 0 \\
0 & 0 & 1%
\end{array}%
\right)
\end{equation}%
where the $W_{0}$ is the Cartan of $sl(2)_{\alpha _{3}}$ in the isospin
representation $j=2$. This line corresponds to the vertical line in Figure
\textbf{\ref{sl2}}

\item \textbf{Boundary line} $\lambda ^{0}=+\frac{2}{3}\Lambda ^{0}:$ the
matrix (\ref{M}) is given by
\begin{equation}
T_{\alpha _{1}}=\lambda ^{0}\left(
\begin{array}{ccc}
2 & 0 & 0 \\
0 & -2 & 0 \\
0 & 0 & 0%
\end{array}%
\right)
\end{equation}%
$T_{\alpha _{1}}$ generate the Cartan of $sl(2)_{\alpha _{1}}$ in the
isospin j=2 representation.

\item \textbf{Line} $\lambda ^{0}=-\frac{2}{3}\Lambda ^{0};$ the matrix (\ref%
{M}) reads as
\begin{equation}
T_{\alpha _{2}}=\lambda ^{0}\left(
\begin{array}{ccc}
0 & 0 & 0 \\
0 & 2 & 0 \\
0 & 0 & -2%
\end{array}%
\right)
\end{equation}%
Similarly, $T_{\alpha _{2}}$ generate the Cartan of $sl(2)_{\alpha _{2}}$ in
isospin j=2 representation.
\end{itemize}

Using the boundary sl(2) straight lines, the 4+4 chambers having $\Lambda
^{0}\geq 0$ are as follows%
\begin{equation}
\begin{tabular}{ccccccccc}
I$_{\mathbf{++}}$ & : & $0$ & $\leq $ & $\Lambda ^{0}$ & $\leq $ & $+\frac{2%
}{3}\lambda ^{0}$ & $,$ & $\lambda ^{0}\geq 0$ \\
II$_{\mathbf{++}}$ & : & $+\frac{2}{3}\lambda ^{0}$ & $\leq $ & $\Lambda
^{0} $ & $\leq $ & $\infty $ & $,$ & $\lambda ^{0}\geq 0$ \\
I$_{\mathbf{-+}}$ & : & $-\frac{3}{2}\lambda ^{0}$ & $\leq $ & $\Lambda ^{0}$
& $\leq $ & $\infty $ & $,$ & $\lambda ^{0}\leq 0$ \\
II$_{\mathbf{-+}}$ & : & $0$ & $\leq $ & $\Lambda ^{0}$ & $\leq $ & $-\frac{2%
}{3}\lambda ^{0}$ & $,$ & $\lambda ^{0}\leq 0$%
\end{tabular}%
\end{equation}

\textbf{A.2) regularity around thermal cycle}\newline
Using the quantities (\ref{W1}) and the real forms of SL(3,$\mathbb{C}$)
(see \ref{geo},\ref{T1},\ref{316}-\ref{321}), we have the values%
\begin{equation}
\begin{tabular}{|c|c|c|}
\hline
$sl(3,\mathbb{C})${\small \ real forms} & $\boldsymbol{W}_{a_{\mathrm{\beta }%
}}$ & $\boldsymbol{\tilde{W}}_{a_{\mathrm{\beta }}}$ \\ \hline
$su\left( 3\right) $ & $\ \ \ e^{i\mathrm{\beta }\lambda L_{0}}e^{i\mathrm{%
\beta }\Lambda W_{0}}$ \ \ \  & $\ \ \ e^{-i\mathrm{\beta }\lambda
L_{0}}e^{-i\mathrm{\beta }\Lambda W_{0}}$ \ \ \  \\ \hline
$su(2,1)_{{\small 12}}$ & $e^{i\mathrm{\beta }\lambda L_{0}}e^{\mathrm{\beta
}\Lambda W_{0}}$ & $e^{-i\mathrm{\beta }\lambda L_{0}}e^{\mathrm{\beta }%
\Lambda W_{0}}$ \\ \hline
$su(2,1)_{{\small 21}}$ & $e^{\mathrm{\beta }\lambda L_{0}}e^{i\mathrm{\beta
}\Lambda W_{0}}$ & $e^{\mathrm{\beta }\lambda L_{0}}e^{-i\mathrm{\beta }%
\Lambda W_{0}}$ \\ \hline
$sl\left( 3,\mathbb{R}\right) $ & $e^{\mathrm{\beta }\lambda L_{0}}e^{%
\mathrm{\beta }\Lambda W_{0}}$ & $e^{\mathrm{\beta }\lambda L_{0}}e^{\mathrm{%
\beta }\Lambda W_{0}}$ \\ \hline
\end{tabular}
\label{hol}
\end{equation}%
\begin{equation*}
\end{equation*}%
From these relations, it is evident that the regularity of the holonomies
around the thermal cycle can be only imposed on the compact $su\left(
3\right) $ and $su\left( 2\right) $ forms inside the real forms $su(2,1)_{%
{\small 12}}$ and $su(2,1)_{{\small 21}}$. Accordingly, we demand the
following regularity conditions
\begin{equation}
\begin{tabular}{|c|c|c|c|}
\hline
$su\left( 3\right) $ & $e^{i\mathrm{\beta }\lambda L_{0}}e^{i\mathrm{\beta }%
\Lambda W_{0}}$ & $=$ & z$I_{id}$ \\ \hline
$su(2,1)_{{\small 12}}$ & $e^{i\mathrm{\beta }\lambda L_{0}}$ & $=$ & z$%
I_{id}$ \\ \hline
$su(2,1)_{{\small 21}}$ & $e^{i\mathrm{\beta }\Lambda W_{0}}$ & $=$ & z$%
I_{id}$ \\ \hline
\end{tabular}%
\end{equation}%
where we have set $\mathrm{\tau }=i\mathrm{\beta }$ and where a priori $z=1$
fo\textrm{r a}\emph{\ }trivial holonomy.\textrm{\ }However, it will be taken
below like $z^{N}=1$, that is in the centre of SU($N$) and SU($N_{1},N_{2}$)
with\textrm{\ }$N_{1}+N_{2}=N$\textrm{\ to }prevent any loss of generality.
For the compact case where $N=N_{1}=3,$ the condition $z^{3}=1$ is\textrm{\ }%
fulfilled by the complex numbers $z=\omega ^{p}$ with $\omega =\exp (i2\pi
/3)$ and p an integer mod 3.
\begin{figure}[tbph]
\begin{center}
\includegraphics[width=10cm]{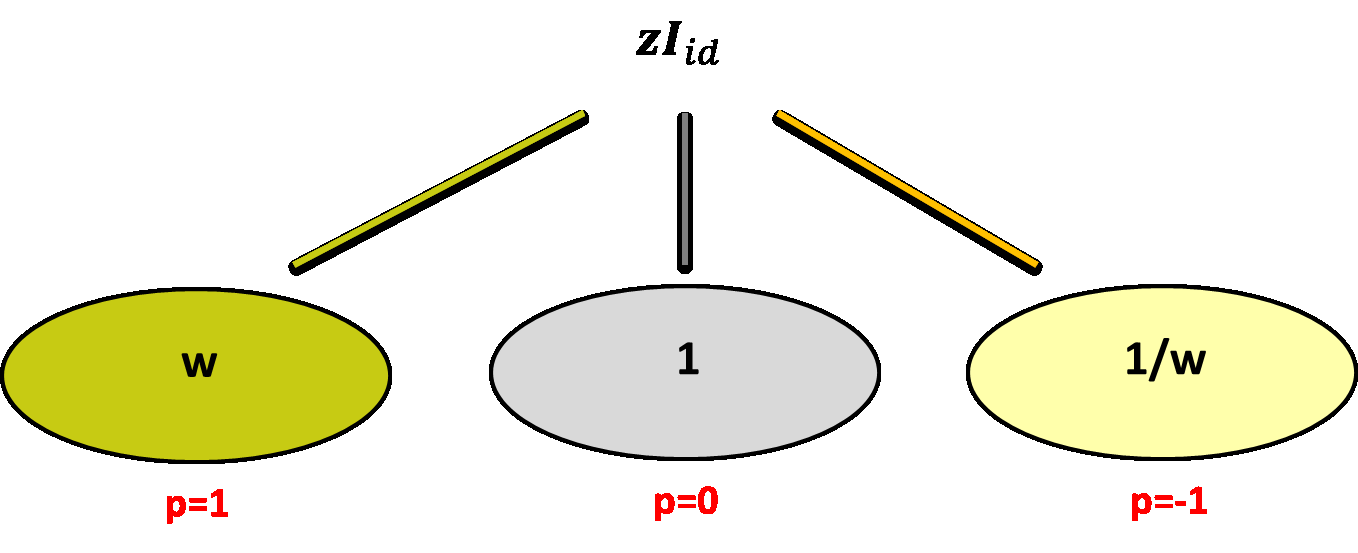}
\end{center}
\par
\vspace{0.0cm}
\caption{The three layers of the $\mathbb{Z}_{3}$ group centre of the SL(3).}
\label{z3}
\end{figure}
These three $\omega ^{p}$ elements form the discrete abelian group $\mathbb{Z%
}_{3}=\left\{ I_{id},\omega ,\omega ^{-1}\right\} $ with three layers as
depicted by Figure \textbf{\ref{z3}}. Notice that in \textrm{\cite{1B}}, the
holonomies (\ref{hol}) took values in the subgroup $\{I_{id}\}$ of the
centre $\mathbb{Z}_{3}$ (neutral holonomy). However, below we shall consider
all three layers of\textrm{\ }\textbf{Figure}\emph{\ }\textbf{\ref{z3}}.
Note that for the compact part of $su(2,1),$ we have $\Lambda =0$ for $%
su(2,1)_{{\small 12}}$ and $\lambda ^{0}=0$ for $su(2,1)_{{\small 21}}.$

$\bullet $ \textbf{Compact real form} $su\left( 3\right) :$\newline
In the case of the compact real form $su\left( 3\right) $, the associated
constraint $e^{i\mathrm{\beta }\lambda L_{0}}e^{i\mathrm{\beta }\Lambda
W_{0}}=zI_{id}$ is given by the matrix equation:%
\begin{equation}
\left(
\begin{array}{ccc}
e^{i\mathrm{\beta }(\frac{2}{3}\Lambda ^{0}+\lambda ^{0})} & 0 & 0 \\
0 & e^{-i\frac{4}{3}\mathrm{\beta }\Lambda ^{0}} & 0 \\
0 & 0 & e^{i\mathrm{\beta }(\frac{2}{3}\Lambda ^{0}-\lambda ^{0})}%
\end{array}%
\right) =e^{i2\pi p/3}\left(
\begin{array}{ccc}
e^{i2\pi N_{+}} & 0 & 0 \\
0 & e^{-i2\pi M} & 0 \\
0 & 0 & e^{i2\pi N_{-}}%
\end{array}%
\right)
\end{equation}%
where $N_{\pm }$\ and $M$ are integers and where we have used $p\equiv p+3q$
indicating that $p\equiv -2p$ mod 3. This leads to
\begin{equation}
\left\{
\begin{tabular}{ccc}
$\frac{2}{3}\Lambda ^{0}+\lambda ^{0}$ & $=$ & $\frac{2\pi }{\mathrm{\beta }}%
\left( N_{+}+\frac{1}{3}p\right) $ \\
$-\frac{4}{3}\Lambda ^{0}$ & $=$ & $\frac{2{\small \pi }}{\mathrm{\beta }}%
(-M-\frac{2}{3}p)$ \\
$\frac{2}{3}\Lambda ^{0}-\lambda ^{0}$ & $=$ & $\frac{2\pi }{\mathrm{\beta }}%
\left( N_{-}+\frac{1}{3}p\right) $%
\end{tabular}%
\right.  \label{LM}
\end{equation}%
solved by ${\small M}=N_{+}+N_{-}$ and the quantized
\begin{equation}
\Lambda ^{0}=\frac{{\small 3\pi }}{2\mathrm{\beta }}\left( M+\frac{2}{3}%
p\right) \qquad ,\qquad \lambda ^{0}=\frac{\pi }{\mathrm{\beta }}(2N+M)
\label{RF}
\end{equation}%
labelled by three integers ($N,M,p$) \textrm{such that} $\Lambda ^{0}\equiv
\Lambda _{0,M,p}^{0}$ and $\lambda ^{0}\equiv \lambda _{N,M,0}^{0}$. This
discrete solution indicates that the dependence \textrm{on} the central
parameter z is captured \textrm{only by the spin} $s=3$ \textrm{component} $%
\Lambda ^{0}.$ Because p can take three values 0,1,2 [mod 3], we then
distinguish three sectors for $\Lambda ^{0}=\Lambda ^{0}\left( p\right) $
namely%
\begin{equation}
\begin{tabular}{|c|c||c|c|c|}
\hline
\multicolumn{2}{|c||}{$p$ \ \ $\func{mod}3$} & $0$ & $1$ & $2$ \\ \hline
{\small spin s=2} & $\lambda _{N,M,0}^{0}$ & $\ \frac{\pi }{\mathrm{\beta }}%
(2N+M)$ & $\frac{\pi }{\mathrm{\beta }}(2N+M)$ & $\frac{\pi }{\mathrm{\beta }%
}(2N+M)$ \\ \hline
{\small spin s=3} & $\ \Lambda _{0,M,p}^{0}$ \  & $\ \frac{\pi }{2\mathrm{%
\beta }}\left( 3M\right) $ & $\ \frac{\pi }{2\mathrm{\beta }}\left(
3M+1\right) $ & $\ \frac{\pi }{2\mathrm{\beta }}\left( 3M+2\right) $ \\
\hline
\end{tabular}
\label{I1}
\end{equation}%
\begin{equation*}
\end{equation*}%
Notice that $\Lambda ^{0}\left( p\right) $ \textrm{can vanish only} when $%
p=0 $ corresponding to \textrm{trivial holonomies in the }subgroup $%
\{I_{id}\}$ of the centre $\mathbb{Z}_{3}$. Notice also that for the branch $%
M=0,$ we have%
\begin{equation}
\begin{tabular}{|c|c||c|c|c|}
\hline
\multicolumn{2}{|c||}{$p$ \ \ $\func{mod}3$} & $0$ & $1$ & $2$ \\ \hline
{\small spin s=2} & $\lambda ^{0}$ & $\ \frac{\pi }{\mathrm{\beta }}2N$ & $%
\frac{\pi }{\mathrm{\beta }}2N$ & $\frac{\pi }{\mathrm{\beta }}2N$ \\ \hline
{\small spin s=3} & $\ \Lambda ^{0}\left( p\right) $ \  & $0$ & $\ \frac{\pi
}{2\mathrm{\beta }}$ & $\ \frac{\pi }{\mathrm{\beta }}$ \\ \hline
\end{tabular}%
\end{equation}%
\begin{equation*}
\end{equation*}%
showing that $\left( \mathbf{i}\right) $ the spin $s=3$ has non trivial
contributions $\frac{\pi }{2\mathrm{\beta }}$ and $\frac{\pi }{\mathrm{\beta
}}$ \textrm{stemming from} the central values 2$\pi p/3$; and $\left(
\mathbf{ii}\right) $ the crossing of the line $\Lambda ^{0}=0$ [ the SU(2)$%
_{\alpha _{3}}$ direction] requires a discrete jump by an amount\emph{\ }%
\textrm{of }2$\pi p/3.$

$\bullet $ \textbf{Real form} $su\left( 2,1\right) :$\newline
Quite similar relations hold for the real form $su\left( 2,1\right) \simeq
su\left( 1,2\right) $. \newline
For the case $\Lambda ^{0}=0$, we have%
\begin{equation}
\left(
\begin{array}{ccc}
e^{i\mathrm{\beta }\lambda ^{0}} & 0 & 0 \\
0 & 1 & 0 \\
0 & 0 & e^{-i\mathrm{\beta }\lambda ^{0}}%
\end{array}%
\right) =e^{i2\pi p/3}\left(
\begin{array}{ccc}
e^{i2\pi N_{0}} & 0 & 0 \\
0 & 1 & 0 \\
0 & 0 & e^{-i2\pi N_{0}}%
\end{array}%
\right)
\end{equation}%
giving%
\begin{equation}
\left\{
\begin{tabular}{ccc}
$\lambda ^{0}$ & $=$ & $\frac{2\pi }{\mathrm{\beta }}\left( N_{0}+\frac{p}{3}%
\right) $ \\
$0$ & $=$ & $\frac{2{\small \pi }}{3\mathrm{\beta }}p$ \\
$\lambda ^{0}$ & $=$ & $\frac{2\pi }{\mathrm{\beta }}\left( N_{0}-\frac{p}{3}%
\right) $%
\end{tabular}%
\right. \qquad \Rightarrow \qquad \lambda ^{0}=\frac{2\pi }{\mathrm{\beta }}N
\label{I2}
\end{equation}%
solved by $p=0$ mod 3, and the quantized $\lambda ^{0}=\frac{2\pi }{\mathrm{%
\beta }}N$ \textrm{can be }denoted like $\lambda _{N}^{0}=\frac{2\pi }{%
\mathrm{\beta }}N.$ \newline
For the case $\lambda ^{0}=0$, we have%
\begin{equation}
\left(
\begin{array}{ccc}
e^{i\mathrm{\beta }\frac{2}{3}\Lambda ^{0}} & 0 & 0 \\
0 & e^{-i\frac{4}{3}\mathrm{\beta }\Lambda ^{0}} & 0 \\
0 & 0 & e^{i\mathrm{\beta }\frac{2}{3}\Lambda ^{0}}%
\end{array}%
\right) =e^{i2\pi p/3}\left(
\begin{array}{ccc}
e^{i2\pi N} & 0 & 0 \\
0 & e^{-i2\pi {\small M}} & 0 \\
0 & 0 & e^{i2\pi N}%
\end{array}%
\right)
\end{equation}%
giving
\begin{equation}
\left\{
\begin{tabular}{ccc}
$\frac{2}{3}\Lambda ^{0}$ & $=$ & $\frac{2\pi }{\mathrm{\beta }}\left( N+%
\frac{1}{3}p\right) $ \\
$\frac{4}{3}\Lambda ^{0}$ & $=$ & $\frac{2{\small \pi }}{\mathrm{\beta }}%
\left( M+\frac{2}{3}p\right) $%
\end{tabular}%
\right. \qquad \Rightarrow \qquad \Lambda ^{0}=\frac{3\pi }{\mathrm{\beta }}%
\left( N+\frac{1}{3}p\right)  \label{I3}
\end{equation}%
solved by $M=2N$ and then $\Lambda ^{0}=\frac{3\pi }{\mathrm{\beta }}\left(
N+\frac{1}{3}p\right) $ can be written like $\Lambda _{N,p}^{0}=\frac{3\pi }{%
\mathrm{\beta }}\left( N+\frac{1}{3}p\right) .$

\paragraph{\textbf{B)} \textbf{Eqs(\protect\ref{fem}) from a variational
principle and BH entropy}}

\ \ \newline
Here, we go over the variational principle leading to the relations (\ref%
{fem}) in preparation for the derivation of our\textrm{\ }\textsc{S}$_{\text{%
\textsc{hs-bh}}}$ formula for the entropy of HS black flowers. As a follow
up, we compare the \textrm{well- known} \textsc{S}$_{\text{\textsc{hs-bh}}}^{%
\text{\textsc{gpptt}}},\ $calculated by GPPTT (\emph{Grumiller et al})
\textrm{in \cite{1B} [see eq(4.5) there]}, with our\textrm{\ S}$_{\text{%
\textsc{hs-bh}}}$\textrm{\ }and comment on the conserved quantities for this
family of HS\ black holes.

\ \

\textbf{B.1)} \textbf{Variational principle}\newline
Using the hypothesis (\ref{cst}), the field equations (\ref{fem}) reduce
down to the remarkable relations%
\begin{equation}
\begin{tabular}{lllll}
$\partial _{t}\mathcal{J}_{\varphi }$ & $=$ & $\left( \partial _{t}\mathcal{J%
}_{\varphi }^{0}\right) L_{0}$ & $=$ & $0$ \\
$\partial _{t}\mathcal{K}_{\varphi }$ & $=$ & $\left( \partial _{t}\mathcal{K%
}_{\varphi }^{0}\right) W_{0}$ & $=$ & $0$%
\end{tabular}%
\qquad ,\qquad
\begin{tabular}{lllll}
$\partial _{t}\mathcal{\tilde{J}}_{\varphi }$ & $=$ & $(\partial _{t}%
\mathcal{\tilde{J}}_{\varphi }^{0})L_{0}$ & $=$ & $0$ \\
$\partial _{t}\mathcal{\tilde{K}}_{\varphi }$ & $=$ & $(\partial _{t}%
\mathcal{\tilde{K}}_{\varphi }^{0})W_{0}$ & $=$ & $0$%
\end{tabular}
\label{eq}
\end{equation}%
These relations can be interpreted as 2D field equations of motion
descending from a 2D action principle as follows%
\begin{eqnarray}
\frac{\delta \mathcal{H}}{\delta \lambda } &=&\partial _{t}\mathcal{J}%
_{\varphi }\qquad ,\qquad \frac{\delta \mathcal{H}}{\delta \tilde{\lambda}}%
=\partial _{t}\mathcal{\tilde{J}}_{\varphi }  \notag \\
\frac{\delta \mathcal{H}}{\delta \Lambda } &=&\partial _{t}\mathcal{K}%
_{\varphi }\qquad ,\qquad \frac{\delta \mathcal{H}}{\delta \tilde{\Lambda}}%
=\partial _{t}\mathcal{\tilde{K}}_{\varphi }
\end{eqnarray}%
\textrm{\ }here the ($\mathcal{J}_{\varphi },\mathcal{K}_{\varphi }$) and ($%
\mathcal{\tilde{J}}_{\varphi },\mathcal{\tilde{K}}_{\varphi }$)\textrm{\ }%
are functions of the variables ($\varphi ,t$); i.e:%
\begin{equation}
\mathcal{J}_{\varphi }\left( \varphi ,t\right) \quad ,\quad \mathcal{K}%
_{\varphi }\left( \varphi ,t\right) \quad ,\quad \mathcal{\tilde{J}}%
_{\varphi }\left( \varphi ,t\right) \quad ,\quad \mathcal{\tilde{K}}%
_{\varphi }\left( \varphi ,t\right)  \label{td}
\end{equation}%
we find that the action $\mathcal{H}$ is a linear combination of two objects
$\mathtt{I}$ and $\mathtt{\tilde{I}}$ such that $\mathrm{\beta }\mathcal{H}=%
\mathtt{I}+\mathtt{\tilde{I}}$ \textrm{where}%
\begin{eqnarray}
2\pi \mathtt{I} &=&\frac{\mathrm{k}}{4}\int_{0}^{\mathrm{\beta }}dt\left(
\int_{0}^{2\pi }d\varphi Tr\left[ \lambda \partial _{t}\mathcal{J}_{\varphi
}+\Lambda \partial _{t}\mathcal{K}_{\varphi }\right] \right)  \notag \\
2\pi \mathtt{\tilde{I}} &=&\frac{\mathrm{k}}{4}\int_{0}^{\mathrm{\beta }%
}dt\left( \int_{0}^{2\pi }d\varphi Tr\left[ \tilde{\lambda}\partial _{t}%
\mathcal{\tilde{J}}_{\varphi }+\tilde{\Lambda}\partial _{t}\mathcal{\tilde{K}%
}_{\varphi }\right] \right)  \label{int}
\end{eqnarray}%
with Lagrange multipliers ($\lambda ,\Lambda $) and ($\tilde{\lambda},\tilde{%
\Lambda}$) being constant parameters as previously chosen\textrm{.} An
alternative approach to derive (\ref{eq}) is to consider instead of $H$\ the
quantity $\mathrm{\beta }\mathcal{P}=\mathtt{I}-\mathtt{\tilde{I}}$ defining
the angular momentum.\

\ \

\textbf{B.2)} \textbf{Deriving BH entropy}\newline
We show below that the absolute values $\left\vert \mathtt{I}\right\vert $
and $|\mathtt{\tilde{I}}|$ are\emph{\ }closely related to the GPPTT entropy $%
\QTR{sc}{S}_{\text{\textsc{hs-bh}}}^{\text{\textsc{gpptt}}}$ given by,%
\begin{equation}
\text{\textsc{S}}_{\text{\textsc{hs-bh}}}^{\text{\textsc{gpptt}}}=\pi \left(
2n+m\right) \left( J_{0}+\bar{J}_{0}\right) +3\pi m\left( K_{0}+\bar{K}%
_{0}\right)  \label{GPP}
\end{equation}%
with an affiliation\ to the boundary hamiltonian given by the variation $%
\delta \text{\textsc{S}}_{\text{\textsc{hs-bh}}}^{\text{\textsc{gpptt}}}=%
\mathrm{\beta }\left( \delta \mathcal{H}^{\text{\textsc{gpptt}}}\right) $.
In what follows, we establish that:

\begin{description}
\item[$\left( \mathbf{i}\right) $] \textrm{the entropy} $\QTR{sc}{S}_{\text{%
\textsc{hs-bh}}}^{\text{\textsc{gpptt}}}$ denotes a particular layer of the
black flower solution of higher spin gravity with symmetry $SU(3)_{{\small L}%
}\times SU(3)_{{\small R}}.$ It is associated with the neutral holonomy $%
\{I_{id}\}$ sitting in the $\mathbb{Z}_{3}=\{I_{id},\omega ,\omega ^{2}\}$
\textrm{centre}\emph{\ }of $SU(3)$.

\item[$\left( \mathbf{ii}\right) $] \textrm{the eq} (\ref{GPP}) corresponds
to the maximal value of\emph{\ }two gauge invariants
\begin{equation}
\mathtt{I}^{\pm }=2\pi \left\vert \mathtt{I}\pm \mathtt{\tilde{I}}\right\vert
\label{pm}
\end{equation}%
where the left sector charge (and equivalently the right) is given by $%
\mathtt{I}=2\pi \mathfrak{N}J_{0}+3\pi \mathfrak{M}K_{0}$ \textrm{with} $%
\mathfrak{N}$ and $\mathfrak{M}$ to be determined. Due to the inequality $|%
\mathtt{I}\pm \mathtt{\tilde{I}}|$ $\leq $ $|\mathtt{I}|+|\mathtt{\tilde{I}}%
|,$ the invariants (\ref{pm}) are upper bounded by the maximal value%
\begin{equation}
\text{\textsc{\textsc{S}}}_{\text{\textsc{hs-bh}}}=2\pi \left( \left\vert
\mathtt{I}\right\vert +|\mathtt{\tilde{I}}|\right)
\end{equation}%
defining the entropy of the HS-BF.
\end{description}

To prove that \textsc{S}$_{\text{\textsc{hs-bh}}}$ \textrm{comprises}
\textsc{S}$_{\text{\textsc{hs-bh}}}^{\text{\textsc{gpptt}}},$ we must
calculate the integrals in eq(\ref{int}). However, this requires a certain
knowledge of the dependencies of\textrm{\ (}$\mathcal{J}_{\varphi },\mathcal{%
K}_{\varphi }$\textrm{) }an\textrm{d (}$\mathcal{\tilde{J}}_{\varphi },%
\mathcal{\tilde{K}}_{\varphi }$) \textrm{on} the Euclidean time $t=it_{%
{\small E}}$ and the angular variable $\varphi .$ In fact, in addition to
the Lagrange multipliers being constants in our description, we can impose a
reliance on the Euclidean time in\textrm{\ (\ref{td}) }as follows%
\begin{equation}
\begin{tabular}{lll}
$\mathcal{J}_{\varphi }\left( \varphi ,t_{{\small E}}\right) $ & $=$ & $-%
\frac{\mathrm{\beta }}{2}\mathcal{J}_{\varphi }\left( \varphi \right) \cos
(\pi \frac{t_{{\small E}}}{\mathrm{\beta }})$ \\
$\mathcal{K}_{\varphi }\left( \varphi ,t_{{\small E}}\right) $ & $=$ & $-%
\frac{\mathrm{\beta }}{2}\mathcal{K}_{\varphi }\left( \varphi \right) \cos
(\pi \frac{t_{{\small E}}}{\mathrm{\beta }})$ \\
&  &  \\
$\mathcal{\tilde{J}}_{\varphi }\left( \varphi ,t_{{\small E}}\right) $ & $=$
& $-\frac{\mathrm{\beta }}{2}\mathcal{\tilde{J}}_{\varphi }\left( \varphi
\right) \cos (\pi \frac{t_{{\small E}}}{\mathrm{\beta }})$ \\
$\mathcal{\tilde{K}}_{\varphi }\left( \varphi ,t_{{\small E}}\right) $ & $=$
& $-\frac{\mathrm{\beta }}{2}\mathcal{\tilde{K}}_{\varphi }\left( \varphi
\right) \cos (\pi \frac{t_{{\small E}}}{\mathrm{\beta }})$%
\end{tabular}%
\end{equation}%
with $t_{{\small E}}$ varying in the 1-cycle $\left[ 0,\mathrm{\beta }\right]
.$ Notice that the above factorisation of the variables ($\varphi ,t_{%
{\small E}}$) follows from the equations of motion (\ref{eq}) demanding the
equality of the time derivatives (velocities) of these relations at $t_{%
{\small E}}=0$ and $t_{{\small E}}=\mathrm{\beta }.$ For example, the $%
\mathcal{J}_{\varphi }\left( \varphi ,t_{{\small E}}\right) $ \textrm{obey} $%
\partial _{t_{{\small E}}}\mathcal{J}_{\varphi }\left( \varphi ,t_{{\small E}%
}\right) =-\frac{\pi }{2}\mathcal{J}_{\varphi }\left( \varphi \right) \sin
(\pi \frac{t_{{\small E}}}{\mathrm{\beta }})$ which vanishes at $t_{{\small E%
}}=n\mathrm{\beta }.$ The integration \textrm{in} $\mathtt{I}$\textrm{\ (as
well as in }$\mathtt{\tilde{I}}$\textrm{) }with respect to the Euclidean
time coordinate gives%
\begin{equation}
2\pi \mathtt{I}=\left( \frac{\mathrm{k\beta }}{4}\dint\nolimits_{0}^{2\pi
}d\varphi Tr\left[ \lambda \mathcal{J}_{\varphi }+\Lambda \mathcal{K}%
_{\varphi }\right] \right) \left[ -\frac{1}{2}\dint\nolimits_{0}^{\mathrm{%
\beta }}dt_{{\small E}}\partial _{t_{{\small E}}}\left( \cos (\pi \frac{t_{%
{\small E}}}{\mathrm{\beta }})\right) \right]
\end{equation}%
leading to
\begin{equation}
\begin{tabular}{lll}
$2\pi \mathtt{I}$ & $=$ & $\frac{\mathrm{k\beta }}{4}\dint\nolimits_{0}^{2%
\pi }d\varphi Tr\left[ \lambda \mathcal{J}_{\varphi }+\Lambda \mathcal{K}%
_{\varphi }\right] $ \\
$2\pi \mathtt{\tilde{I}}$ & $=$ & $\frac{\mathrm{k\beta }}{4}%
\dint\nolimits_{0}^{2\pi }d\varphi Tr\left[ \tilde{\lambda}\mathcal{\tilde{J}%
}_{\varphi }+\tilde{\Lambda}\mathcal{\tilde{K}}_{\varphi }\right] $%
\end{tabular}
\label{II}
\end{equation}%
By substituting $\lambda =\lambda ^{0}L_{0},$ $\Lambda =\Lambda ^{0}W_{0}$
and $\mathcal{J}_{\varphi }=\mathcal{J}_{\varphi }^{0}L_{0},$ $\mathcal{K}%
_{\varphi }=\mathcal{K}_{\varphi }^{0}W_{0}$ along with their twild
homologue as well as using the properties $Tr\left( L_{0}L_{0}\right) =2$
and $Tr\left( W_{0}W_{0}\right) =\frac{8}{3},$ we bring the above relation
to
\begin{eqnarray}
\mathtt{I} &=&\mathrm{\beta }\frac{\mathrm{k}}{4\pi }\int_{0}^{2\pi
}d\varphi \left[ \lambda ^{0}\mathcal{J}_{\varphi }^{0}+\frac{4}{3}\Lambda
^{0}\mathcal{K}_{\varphi }^{0}\right]  \notag \\
\mathtt{\tilde{I}} &=&\mathrm{\beta }\frac{\mathrm{k}}{4\pi }\int_{0}^{2\pi
}d\varphi \left[ \tilde{\lambda}^{0}\mathcal{\tilde{J}}_{\varphi }^{0}+\frac{%
4}{3}\tilde{\Lambda}^{0}\mathcal{\tilde{K}}_{\varphi }^{0}\right]
\end{eqnarray}%
\textrm{with}\emph{\ }the periodicities $\mathcal{J}_{\varphi }^{0}\left(
\varphi +2\pi \right) =\mathcal{J}_{\varphi }^{0}\left( \varphi \right) $
and $\mathcal{K}_{\varphi }^{0}\left( \varphi +2\pi \right) =\mathcal{K}%
_{\varphi }^{0}\left( \varphi \right) $\emph{. }The integration with respect
to the angular variable gives%
\begin{eqnarray}
\mathtt{I} &=&\mathrm{\beta }\left[ \lambda ^{0}J_{0}^{0}+\Lambda
^{0}K_{0}^{0}\right]  \notag \\
\mathtt{\tilde{I}} &=&\mathrm{\beta }\left[ \tilde{\lambda}^{0}\tilde{J}%
_{0}^{0}+\tilde{\Lambda}^{0}\tilde{K}_{0}^{0}\right]  \label{IPM}
\end{eqnarray}%
where we have used%
\begin{equation}
\frac{1}{2\pi }\int_{0}^{2\pi }d\varphi \mathcal{J}_{\varphi }=\frac{2}{%
\mathrm{k}}J_{0}\qquad ,\qquad \frac{1}{2\pi }\int_{0}^{2\pi }d\varphi
\mathcal{K}_{\varphi }=\frac{3}{2\mathrm{k}}K_{0}
\end{equation}%
Putting (\ref{RF}) into (\ref{IPM}), we have for\emph{\ }($\lambda ,\Lambda $%
)%
\begin{equation}
\begin{tabular}{|c|c|}
\hline
{\small real forms} & invariant $\mathtt{I}$ \\ \hline\hline
$su\left( 3\right) $ & $\ \ \ 2\pi \mathfrak{N}J_{0}^{0}+3\pi \mathfrak{M}%
K_{0}^{0}$ \ \ \ \ \ \  \\ \hline\hline
$\ \ \ \ su(2,1)_{{\small 12}}$ \ \ \ \  & $\left( 2\pi N\right) J_{0}^{0}$
\\ \hline
$su(1,2)_{{\small 21}}$ & $\left( 3\pi \mathfrak{M}\right) K_{0}^{0}$ \\
\hline\hline
$sl\left( 3,\mathbb{R}\right) $ & $\mathrm{\beta }\left[ \lambda
^{0}J_{0}^{0}+\Lambda ^{0}K_{0}^{0}\right] $ \\ \hline
\end{tabular}
\label{GF}
\end{equation}%
with
\begin{equation}
\mathfrak{N}=n+\frac{1}{2}m\qquad ,\qquad \mathfrak{M}=\frac{1}{2}m+\frac{1}{%
3}p  \label{I4}
\end{equation}%
where $\mathfrak{M}$ evidently depends on the group elements of the $\mathbb{%
Z}_{3}=\{I_{id},\omega ,\omega ^{2}\}$ \textrm{centre} of $SU(3)$. \textrm{%
The }equivalent expressions for $\mathtt{\tilde{I}}$ are generated by ($%
\tilde{J}_{0}^{0},\tilde{K}_{0}^{0}$).

\ \ \

\textbf{B.3)} \textbf{Comments on }\textsc{S}$_{\text{\textsc{hs-bh}}}$\emph{%
\ }\textbf{and}\emph{\ }\textsc{S}$_{\text{\textsc{hs-bh}}}^{\text{\textsc{%
gpptt}}}$\emph{\ }\textbf{entropies}\newline
Our generalised solution is given\textrm{\ }by the maximal value \textsc{S}$%
_{\text{\textsc{hs-bh}}}=2\pi (\left\vert \mathtt{I}\right\vert +|\mathtt{%
\tilde{I}}|)$; it reads for the compact real form $SU(3)$ of the
complexified gauge symmetry as follows%
\begin{equation}
\text{\textsc{S}}_{\text{\textsc{hs-bh}}}=2\pi \left\vert \mathfrak{N}%
\right\vert \left( \left\vert J_{{\small 0}}^{{\small 0}}\right\vert +|%
\tilde{J}_{{\small 0}}^{{\small 0}}|\right) +3\pi \left\vert \mathfrak{M}%
\right\vert \left( \left\vert K_{{\small 0}}^{{\small 0}}\right\vert +|%
\tilde{K}_{{\small 0}}^{{\small 0}}|\right)
\end{equation}%
where $(\left\vert J_{{\small 0}}^{{\small 0}}\right\vert ,|K_{{\small 0}}^{%
{\small 0}}|)$ and $(|\tilde{J}_{{\small 0}}^{{\small 0}}|,|\tilde{K}_{%
{\small 0}}^{{\small 0}})$ [ for short ($\left\vert J_{0}\right\vert
,|K_{0}| $) and ($|\tilde{J}_{0}|,|\tilde{K}_{0}$)] are the absolute value
of the zero modes of conserved $\dprod $\^{U}(1)$_{\mathrm{k}}$ currents.
This entropy can be also presented as%
\begin{equation}
\begin{tabular}{lll}
\textsc{S}$_{\text{\textsc{hs-bh}}}$ & $=$ & $2\pi \left\vert n+\frac{1}{2}%
m\right\vert \left( J_{0}+\tilde{J}_{0}\right) +$ \\
&  & $3\pi \left\vert \frac{1}{2}m+\frac{1}{3}p\right\vert \left( K_{0}+%
\tilde{K}_{0}\right) $%
\end{tabular}
\label{max}
\end{equation}%
The number n, m are integers ($n,m\in \mathbb{Z}$) with $p$ belonging to $%
\mathbb{Z}/3\mathbb{Z}.$ This reformulation of the expression helps to
expose several features:

\begin{itemize}
\item \textbf{Case} $p=0$\textrm{\ }\textbf{mod} \textbf{3}:\newline
With this parametrisation, the higher spin black hole entropy relation
coincides with the entropy \textsc{S}$_{\text{\textsc{hs-bh}}}^{\text{%
\textsc{gpptt}}}$ given by $\pi \left( 2n+m\right) \left( J_{0}+\bar{J}%
_{0}\right) +$ $3\pi m\left( K_{0}+\bar{K}_{0}\right) .$ Because \textsc{S}$%
_{\text{\textsc{hs-bh}}}^{\text{\textsc{gpptt}}}>0,$ we can assure\ the
fulfillment of this condition by taking $n$ and $m$ to be positive integers
with\textrm{\ }$J_{0}+\bar{J}_{0}>0;$ that is $\func{Re}(J_{0})>0$ and $%
\func{Re}(K_{0})>0$. Additionally, by setting $m=0,$ the GPPTT reduces to
\begin{equation}
\left. \text{\textsc{S}}_{\text{\textsc{hs-bh}}}^{\text{\textsc{gpptt}}%
}\right\vert _{m=0}=2n\pi \left( J_{0}+\bar{J}_{0}\right)
\end{equation}%
recuperating therefore the spin 2 black holes characterised by the even
integer $2n$. As we will see below [see eq(\ref{82})], this even parity is
closely related to the identity element in the $\mathbb{Z}%
_{2}=\{I_{id},-I_{id}\}$ group centre of $SU(2).$

\item \textbf{Case} $p\neq 0$ \textbf{mod 3 and m=0}:\newline
By setting $m=0$, the BH entropy (\ref{max}) reads as follows%
\begin{equation}
\text{\textsc{S}}_{\text{\textsc{hs-bh}}}=2\pi n\left( J_{0}+\bar{J}%
_{0}\right) +\pi |p|\left( K_{0}+\bar{K}_{0}\right)
\end{equation}%
we therefore have%
\begin{equation}
\text{\textsc{S}}_{\text{\textsc{hs-bh}}}=\left. \text{\textsc{S}}_{\text{%
\textsc{hs-bh}}}^{\text{\textsc{gpptt}}}\right\vert _{m=0}+\pi |p|\left(
K_{0}+\bar{K}_{0}\right)
\end{equation}%
indicating that for non trivial holonomies, the branch m=0 has a
contribution stemming from the spin s=3 fields. As such, for $p\neq 0$ mod
3, one cannot continuously\ go from the entropy \textsc{S}$_{\text{\textsc{%
hs-bh}}}$ of the HS black flowers towards the entropy of the spin 2 black
hole; a \textrm{leap} is required as depicted by Figure \textbf{\ref{32}}.
\begin{figure}[tbph]
\begin{center}
\includegraphics[width=12cm]{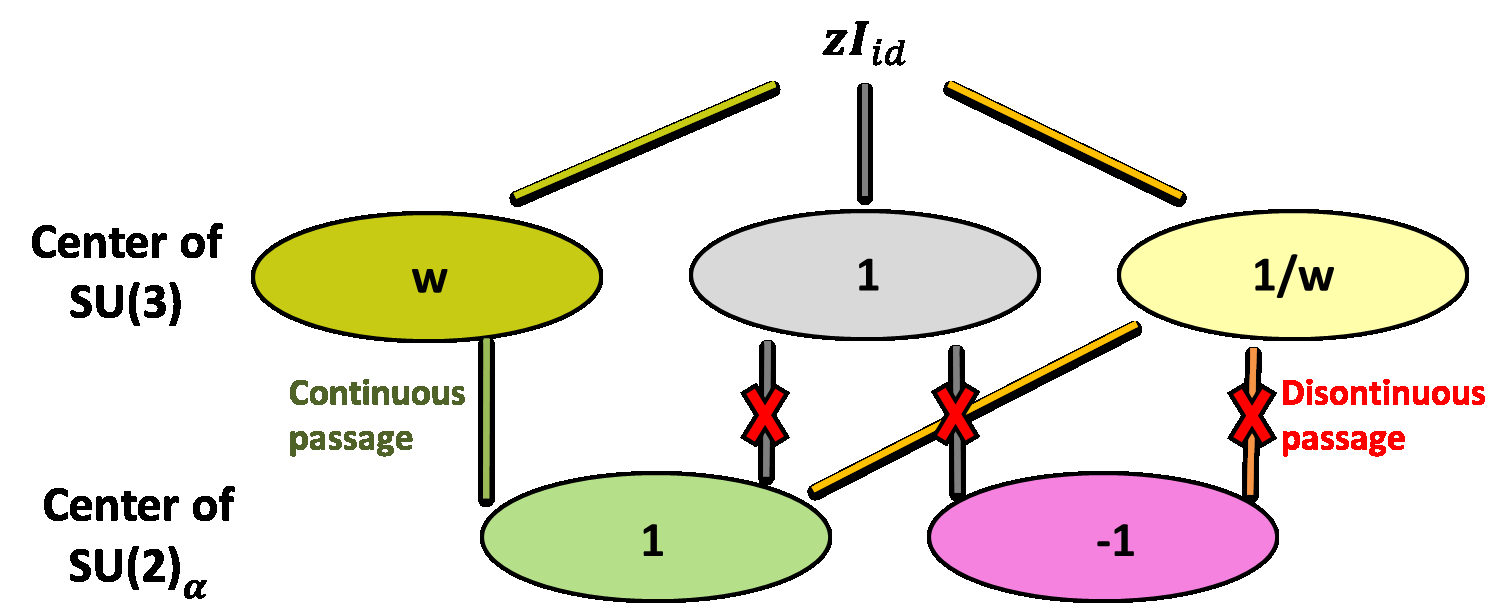}
\end{center}
\par
\vspace{0.0cm}
\caption{The different branches joining the higher spin s=3 black flowers to
the spin s=2 black holes carrying SU$\left( 2\right) $ symmetry. Except for
trivial holonomies, the descent is given by discrete transitions \textrm{%
showcasing} the fact that $\mathbb{Z}_{2}$ is not a subgroup of $\mathbb{Z}%
_{3}$.}
\label{32}
\end{figure}

\item \textbf{Case of spin s=2 gravity}\newline
Here, the generators of sl(2,$\mathbb{C}$) are in the isospin j=1/2, they
are given by the 2$\times $2 matrices (\ref{22}). The centre of sl(2,$%
\mathbb{C}$) is given by $\mathbb{Z}_{2}=\{I_{id},-I_{id}\}$\emph{\ }and is
represented by the two layers of Figure \textbf{\ref{su2}}.
\begin{figure}[h]
\begin{center}
\includegraphics[width=10cm]{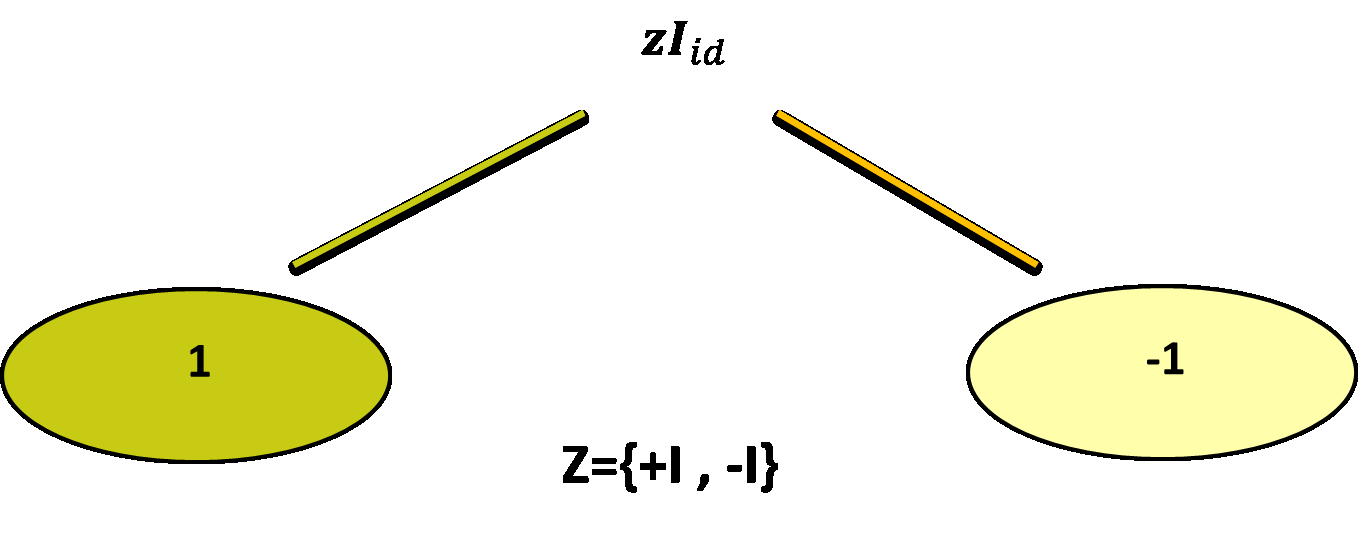}
\end{center}
\par
\vspace{-0.5cm}
\caption{Centre $\mathbb{Z}_{2}$ of the compact real form SU(2) of the spin
s=2 AdS$_{3}$ gravity. The associated chemical potential $\protect\lambda %
^{0}$ is given by $2n\protect\pi /\mathrm{\protect\beta }$ for z=1 and $%
\left( 2n+1\right) \protect\pi /\mathrm{\protect\beta }$ for $z=-1.$}
\label{su2}
\end{figure}
The regularity of the thermal holonomy reads in general as follows
\begin{equation}
\left(
\begin{array}{cc}
e^{i\mathrm{\beta }\lambda ^{0}} & 0 \\
0 & e^{-i\mathrm{\beta }\lambda ^{0}}%
\end{array}%
\right) =\left(
\begin{array}{cc}
e^{i\pi N} & 0 \\
0 & e^{-i\pi N}%
\end{array}%
\right)  \label{82}
\end{equation}%
where $N$\ is an integer solved as $\mathrm{\beta }\lambda ^{0}=\pi N$. For
even integer $N=2n$, the value $\lambda ^{0}=2n\pi /\mathrm{\beta }$
corresponds to the central identity element $I_{id}$ of the group centre $%
\mathbb{Z}_{2}$ while for odd integer $N=2n+1$, the value $\lambda
^{0}=\left( 2n+1\right) \pi /\mathrm{\beta }$\emph{\ }lead to the central
element $-I_{id}$.

\item \textbf{From SU(3) towards the three SU(2)s}\newline
Generic elements $\Theta =\zeta L_{0}+\eta W_{0}$ in the su(3) Lie algebra
with real parameters ($\zeta ,\eta $) read in the Cartan \textrm{basis} like
$\theta _{{\small 1}}H_{\alpha _{1}}+\theta _{{\small 2}}H_{\alpha _{2}}$
with $\theta _{{\small i}}$ as follows%
\begin{equation}
\theta _{{\small 1}}=\zeta +\frac{2}{3}\eta \qquad ,\qquad \theta _{{\small 2%
}}=\zeta -\frac{2}{3}\eta
\end{equation}%
we recover the three\textrm{\ SU}$\left( 2\right) _{\alpha _{i}}$\textrm{\ }%
subgroups within SU(3) by putting appropriate constraints on the parameters (%
$\zeta ,\eta $) such as%
\begin{equation}
\begin{tabular}{lll}
subgroup \ \ \  & \multicolumn{2}{l}{\ \ \ constraint eq \ \ \ } \\
SU$\left( 2\right) _{\alpha _{1}}$ & $\ \ \ \ \ \eta $ & $=+\frac{3}{2}\zeta
$ \\
SU$\left( 2\right) _{\alpha _{2}}$ & $\ \ \ \ \ \eta $ & $=-\frac{3}{2}\zeta
$ \\
SU$\left( 2\right) _{\alpha _{3}}$ & $\ \ \ \ \ \eta $ & $=0$%
\end{tabular}%
\end{equation}%
These constraint relations extend to the HS-BH entropy (\ref{max}) having
two blocks \textsc{S}$_{L}+\text{\textsc{S}}_{R}$: the first \textsc{S}$_{L}$
of the left gauge symmetry $SU(3)_{L}$ is given by $\pi |2n+m|J_{0}+\pi |%
\frac{3}{2}m+p|K_{0}$ with n, m integers and p integer mod 3$.$ The second
\textsc{S}$_{R}$ is associated with the right gauge symmetry $SU(3)_{R}$ and
is formulated as $\pi |2n+m|\tilde{J}_{0}+\pi |\frac{3}{2}m+p|\tilde{K}$ $.$
\textrm{Focusing on}\emph{\ }the left sector contribution,%
\begin{equation}
\text{\textsc{S}}_{L}=\pi |2n+m|J_{0}+\pi |\frac{3}{2}m+p|K_{0}
\end{equation}%
by imagining $J_{0}$\ and $K_{0}$\ in terms of the Cartan charges of the%
\textrm{\ }$SU(3)_{L}$\textrm{\ }symmetry and using the correspondence $%
J_{0}\leftrightarrow L_{0}$ as well as $K_{0}\leftrightarrow W_{0}$ in
addition to the similarity of \textsc{S}$_{L}$ with the Cartan charge matrix
$\Theta _{{\small sl}_{{\small 3}}}=\zeta L_{0}+\eta W_{0},$ we obtain the
following values \textrm{for} the chemical potentials%
\begin{equation}
\begin{tabular}{llllll}
subgroup \ \ \  & \multicolumn{3}{l}{\ \ \ constraint eq \ \ \ } & $\ \
\mathrm{\beta }\lambda ^{0}$ & $\ \ \mathrm{\beta }\Lambda ^{0}$ \\
\ SU$\left( 2\right) _{\alpha _{1}}$ & $3m$ & $=$ & $-2p$ & $\ 2\pi |n-\frac{%
1}{3}p|\ $ \ \ \ \  & \ 0 \\
\ SU$\left( 2\right) _{\alpha _{2}}$ & $m$ & $=$ & $-2n$ & $\ 0$ & $\ \pi
|3n-p|$ \\
\ \ SU$\left( 2\right) _{\alpha _{3}}$ & $m$ & $=$ & $4n-2p$ \ \ \ \  & $\
2\pi |3n-p|$ & $\ 2\pi |3n-p|\pi $ \ \ \ \ \
\end{tabular}%
\end{equation}%
with $p=0,1,2.$ For $p=0,$ the transition is only continuous when
considering SU$\left( 3\right) $ towards SU$\left( 2\right) _{\alpha _{1}}$
and SU$\left( 2\right) _{\alpha _{3}}$ for which the chemical potential
takes the value $\mathrm{\beta }\lambda ^{0}=2n\pi .$ The passage between
the \{$s=2,3$\} and the \{$s=2$\} black holes is illustrated by the Figure
\textbf{\ref{sl3} }
\begin{figure}[tbph]
\begin{center}
\includegraphics[width=10cm]{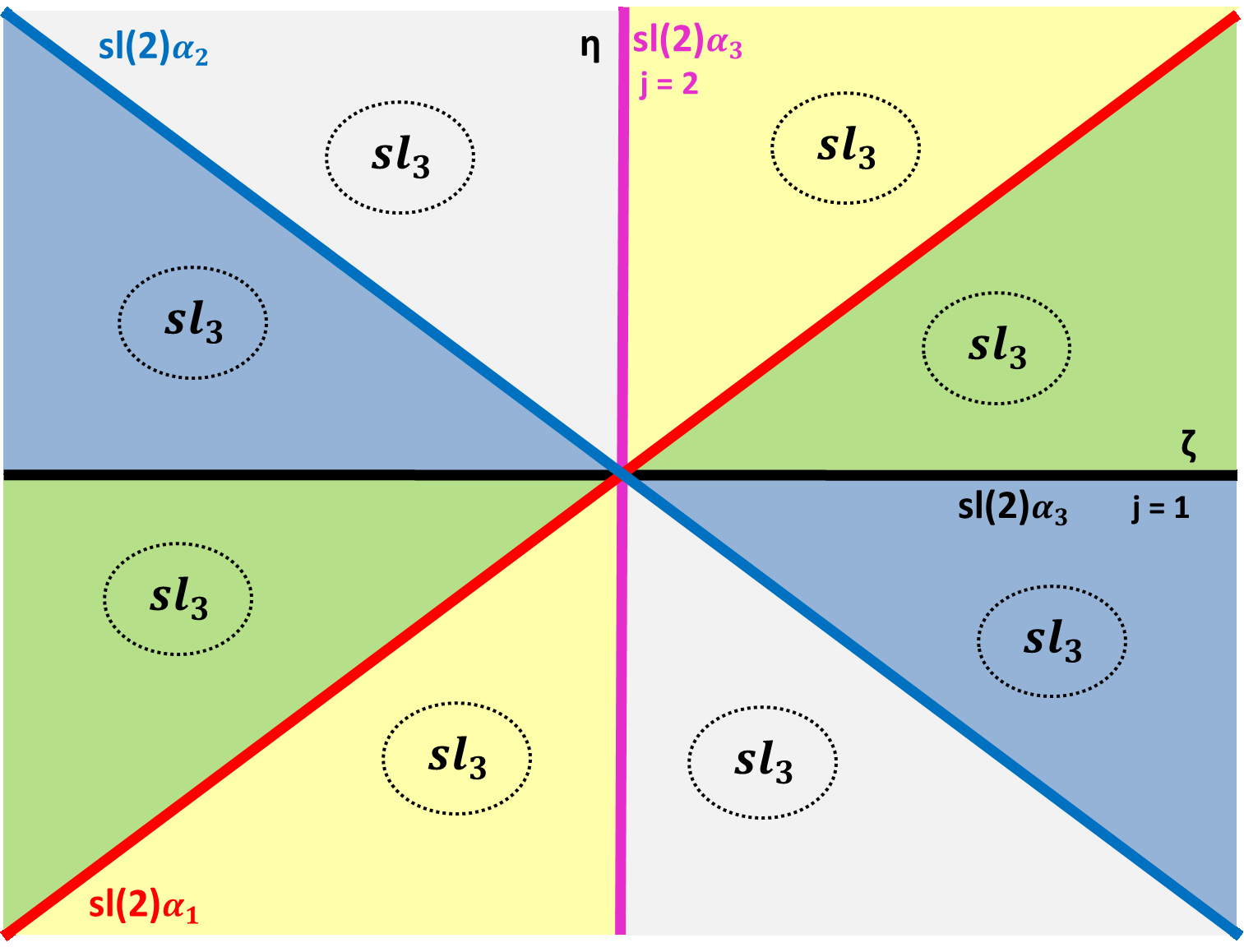}
\end{center}
\par
\vspace{0.0cm}
\caption{Transitions from HS black flowers with SU(3) symmetries to spin s=2
black holes with SU(2) subsymmetries. }
\label{sl3}
\end{figure}
where the branches of the black flowers are represented by the three SU$%
\left( 2\right) _{\mathbf{\alpha }_{1}}$, SU$\left( 2\right) _{\mathbf{%
\alpha }_{2}}$ and SU$\left( 2\right) _{\mathbf{\alpha }_{3}}$ subgroups of
the compact real form SU$\left( 3\right) $. The bridging between the three SU%
$\left( 2\right) _{\mathbf{\alpha }_{i}}$ branches is generated by\ Weyl
transformations of SU$\left( 3\right) ;$ for example, the passage from SU$%
\left( 2\right) _{\mathbf{\alpha }_{3}}$ branch (black straight line in
Figure \textbf{\ref{sl3}}) towards the SU$\left( 2\right) _{\mathbf{\alpha }%
_{1}}$ branch (red straight line in Figure \textbf{\ref{sl3}}) is given by
the discrete transformation $\sigma _{\mathbf{\alpha }_{2}}\left( \mathbf{%
\alpha }_{3}\right) =\mathbf{\alpha }_{1}$ acting as%
\begin{equation}
\sigma _{\mathbf{\alpha }_{2}}:\mathbf{\alpha }_{3}\rightarrow \mathbf{%
\alpha }_{3}-\frac{2\left( \mathbf{\alpha }_{3}.\mathbf{\alpha }_{2}\right)
}{\left( \mathbf{\alpha }_{2}.\mathbf{\alpha }_{2}\right) }\mathbf{\alpha }%
_{2}  \label{wt}
\end{equation}%
where $\mathbf{\alpha }_{3}=\mathbf{\alpha }_{1}+\mathbf{\alpha }_{2}$
\textrm{with} the inner products $\mathbf{\alpha }_{3}.\mathbf{\alpha }%
_{2}=1 $ and $\mathbf{\alpha }_{2}^{2}=2$. \textrm{Similarly, the bridge }SU$%
\left( 2\right) _{\mathbf{\alpha }_{3}}\rightarrow $ SU$\left( 2\right) _{%
\mathbf{\alpha }_{2}}$ (blue straight line in Figure \textbf{\ref{sl3}}) is
given by the Weyl transformation $\sigma _{\mathbf{\alpha }_{1}}\left(
\mathbf{\alpha }_{3}\right) =\mathbf{\alpha }_{2}$ \textrm{with }the
reflection $\sigma _{\mathbf{\alpha }_{3}}\left( \mathbf{\alpha }_{3}\right)
=-\mathbf{\alpha }_{3}.$
\end{itemize}

\subsubsection{Using primary scalars and fermions}

The third approach to work out the solutions of eq\textrm{(\ref{fem})}
exploits primary conformal field theory realisations to\textrm{\ }build
effective 2D conformal field models while assuming the conditions $\partial
_{\varphi }\Upsilon _{t}^{0}=\partial _{\varphi }\Gamma _{t}^{0}=0$ and\emph{%
\ }$\partial _{\varphi }\widetilde{\Upsilon }_{t}^{0}=\partial _{\varphi }%
\widetilde{\Gamma }_{t}^{0}=0$. Here we think about the boundary potentials
in terms of scalar and/or fermi fields with conformal stress tensor realised
in terms of oscillating fields. In this regard, notice that for the spin 2
gravity model, the affine symmetry in the AdS$_{3}$ boundary surface can be
either the compact real form SU(2)$_{\mathrm{k}}$ or the non compact SL(2,$%
\mathbb{R}$)$_{\mathrm{k}}$ depending on the considered gauge symmetry of
the CS model. These boundary symmetries are known to have different bosonic
and fermionic realisations \textrm{\cite{E1,E2,E3};} a characteristic
property shared by all real forms of the SL(N,$\mathbb{C}$)$_{\mathrm{k}}$
family.\newline
Here, we will restrict the analysis to the Cartan sectors of the real forms
of the affine symmetry $SL(3,\mathbb{C})_{\mathrm{k}}$ and consider the two
following typical conformal field representations:

\begin{description}
\item[$\left( \mathbf{a}\right) $] \textbf{Bosonic realisation} for which
the conserved currents ($\mathcal{J}_{\varphi }^{{\small 0}},\mathcal{K}%
_{\varphi }^{{\small 0}}$) and ($\mathcal{\tilde{J}}_{\varphi }^{{\small 0}},%
\mathcal{\tilde{K}}_{\varphi }^{{\small 0}}$) are given by%
\begin{equation}
\begin{tabular}{lllllll}
$\mathcal{J}_{\varphi }^{{\small 0}}$ & $\sim $ & $\partial _{\varphi }X$ & $%
\qquad ,\qquad $ & $\mathcal{\tilde{J}}_{\varphi }^{{\small 0}}$ & $\sim $ &
$\partial _{\varphi }\tilde{X}$ \\
$\mathcal{K}_{\varphi }^{{\small 0}}$ & $\sim $ & $\partial _{\varphi }Y$ & $%
\qquad ,\qquad $ & $\mathcal{\tilde{K}}_{\varphi }^{{\small 0}}$ & $\sim $ &
$\partial _{\varphi }\tilde{Y}$%
\end{tabular}
\label{B}
\end{equation}%
\textrm{where} ($X,Y$) and ($\tilde{X},\tilde{Y}$) are primary scalar fields
with periodic property $X\left( \varphi +2\pi \right) =X\left( \varphi
\right) .$ Substituting into (\ref{fem}), we obtain%
\begin{equation}
\begin{tabular}{lll}
$\partial _{t}\partial _{\varphi }X$ & $=$ & $0$ \\
$\partial _{t}\partial _{\varphi }Y$ & $=$ & $0$%
\end{tabular}%
\qquad ,\qquad
\begin{tabular}{lll}
$\partial _{t}\partial _{\varphi }\tilde{X}$ & $=$ & $0$ \\
$\partial _{t}\partial _{\varphi }\tilde{Y}$ & $=$ & $0$%
\end{tabular}%
\end{equation}%
which are just the usual field equations of motion governing the free scalar
fields.

\item[$\left( \mathbf{b}\right) $] \textbf{Fermionic representation} for
which the conserved charge potentials ($\mathcal{J}_{\varphi }^{{\small 0}},%
\mathcal{K}_{\varphi }^{{\small 0}}$) and ($\mathcal{\tilde{J}}_{\varphi }^{%
{\small 0}},\mathcal{\tilde{K}}_{\varphi }^{{\small 0}}$) are realised as
follows%
\begin{equation}
\begin{tabular}{lll}
$\mathcal{J}_{\varphi }^{{\small 0}}$ & $\sim $ & $\psi ^{1}\psi ^{2}$ \\
$\mathcal{K}_{\varphi }^{{\small 0}}$ & $\sim $ & $\chi ^{1}\chi ^{2}$%
\end{tabular}%
\qquad ,\qquad
\begin{tabular}{lll}
$\mathcal{\tilde{J}}_{\varphi }^{{\small 0}}$ & $\sim $ & $\tilde{\psi}^{1}%
\tilde{\psi}^{2}$ \\
$\mathcal{\tilde{K}}_{\varphi }^{{\small 0}}$ & $\sim $ & $\tilde{\chi}^{1}%
\tilde{\chi}^{2}$%
\end{tabular}
\label{F}
\end{equation}%
The four pairs of conformal primary fields ($\psi ^{i},\chi ^{i}$) and ($%
\tilde{\psi}^{i},\tilde{\chi}^{i}$) \textrm{are }$\left( \mathbf{i}\right) $
time independent%
\begin{equation}
\begin{tabular}{lllllll}
$\partial _{t}\psi ^{i}$ & $=$ & $0$ & $\qquad ,\qquad $ & $\partial _{t}%
\tilde{\psi}^{i}$ & $=$ & $0$ \\
$\partial _{t}\chi ^{i}$ & $=$ & $0$ & $\qquad ,\qquad $ & $\partial _{t}%
\tilde{\chi}^{i}$ & $=$ & $0$%
\end{tabular}%
\end{equation}%
thus they are only functions of\textrm{\ }the variable $\varphi $; \textrm{%
and} $\left( \mathbf{ii}\right) $ either periodic or antiperiodic Fermi
fields.
\end{description}

These realisations assume the\textrm{\ }existence of realistic quantum
scalar and/or Fermi field excitations at the boundary of the AdS$_{3}$
geometry which should incite interesting thermodynamics for our HS black
holes.\ An expanded and more elaborated discussion of this CFT$_{2}$ model\
will be reported \textrm{in section 5.}

\section{Boundary current algebras and HS-BF entropy}

\label{sec4} We first investigate the current algebras of the underlying
higher spin symmetries of the previous black flowers with generalised
boundary conditions (\ref{bb}) in connection with the real forms of the
complexified sl(N,$\mathbb{C}$). Then, we construct the quantum states
making the HS soft Heisenberg hair of these black holes. We conclude this
section with explicit entropy computations of the various higher spin black
flowers. The analysis will mainly focus on the higher spins s=2,3 black
flowers, yet it can be seamlessly generalised.

\subsection{Current algebra and sl(3,$\mathbb{C}$) real forms}

We begin by recalling that the conserved currents ($\mathcal{J}_{\varphi },%
\mathcal{K}_{\varphi }$) and\ ($\mathcal{\tilde{J}}_{\varphi },\mathcal{%
\tilde{K}}_{\varphi }$) given by the expansions (\ref{exj}) generate an
infinite dimensional symmetry on the 2D boundary gauge theory. Pertaining to
the case of real forms of the affine $\widehat{sl}(3,\mathbb{C})_{L}\times
\widehat{sl}(3,\mathbb{C})_{R}$, the 2D invariance consists of 2+2
isomorphic infinite dimensional Heisenberg symmetries namely: the 2 left
sector sub-symmetries $\hat{u}(1)_{\mathcal{J}}\times \hat{u}(1)_{\mathcal{K}%
}$ of the affine $\widehat{su}(3,\mathbb{C})_{L}$ and the 2 right sector $%
\hat{u}(1)_{\mathcal{\tilde{J}}}\times \hat{u}(1)_{\mathcal{\tilde{K}}}$
contained in the affine $\widehat{su}(3,\mathbb{C})_{R}.$

In this regard, the vanishing property of the left partial charge potential $%
da^{\text{\textsc{diag}}}=\mathcal{J}^{{\small 0}}L_{0}+\mathcal{K}^{{\small %
0}}W_{0}=0$ with periodic $\mathcal{J}^{{\small 0}}=\mathcal{J}_{\varphi }^{%
{\small 0}}d\varphi $ and $\mathcal{K}^{{\small 0}}=\mathcal{K}_{\varphi }^{%
{\small 0}}d\varphi $ shows that the currents $\mathcal{J}_{\varphi }^{%
{\small 0}}$ and $\mathcal{K}_{\varphi }^{{\small 0}}$ are conserved ($%
\partial _{t}\mathcal{J}_{\varphi }^{{\small 0}}=\partial _{t}\mathcal{K}%
_{\varphi }^{{\small 0}}=0$). And due to their periodicity property, they
expand in Fourier modes as $\sum J_{n}^{0}e^{-in\varphi }$ and $\sum
K_{n}^{0}e^{-in\varphi }$ inducing two infinite series of conserved charges
\{$J_{n}^{0}$\} and \{$K_{n}^{0}$\}. Therefore, $\mathcal{J}_{\varphi }^{%
{\small 0}}$ and $\mathcal{K}_{\varphi }^{{\small 0}}$ generate the
following current algebras (affine Kac-Moody for quantum) written for the
different real forms as%
\begin{equation}
\begin{tabular}{|c|c|}
\hline
$\widehat{sl}(3,\mathbb{C})_{L}$ real forms & {\small current algebra} \\
\hline
$\widehat{su}{\small (3)}$ & $\ \ \ \ \hat{u}(1)_{\mathcal{J}}\times \hat{u}%
(1)_{\mathcal{K}}$ $\ \ \ \ $ \\ \hline
$\widehat{su}{\small (2,1)}$ & $\ \ \ \ \hat{u}(1)_{\mathcal{J}}\times
\widehat{gl}(1)_{\mathcal{K}}$ $\ \ \ \ $ \\ \hline
$\ \ \ \ \widehat{su}{\small (1,2)}$ $\ \ \ \ $ & $\ \ \ \ \widehat{gl}(1)_{%
\mathcal{J}}\times \hat{u}(1)_{\mathcal{K}}$ $\ \ \ \ $ \\ \hline
$\widehat{sl}{\small (3,R)}$ & $\ \ \ \ \widehat{gl}(1)_{\mathcal{J}}\times
\widehat{gl}(1)_{\mathcal{K}}$ $\ \ \ \ $ \\ \hline
\end{tabular}
\label{Tab4}
\end{equation}%
\begin{equation*}
\end{equation*}%
A similar description is valid for the conserved $\mathcal{\tilde{J}}%
_{\varphi }^{{\small 0}}$ and $\mathcal{\tilde{K}}_{\varphi }^{{\small 0}}$
of the right sector, the associated\textrm{\ }current algebras are given by
\begin{equation}
\hat{u}(1)_{\mathcal{\tilde{J}}}\times \hat{u}(1)_{\mathcal{\tilde{K}}%
},\quad \widehat{gl}(1)_{\mathcal{\tilde{J}}}\times \hat{u}(1)_{\mathcal{%
\tilde{K}}},\quad \hat{u}(1)_{\mathcal{\tilde{J}}}\times \widehat{gl}(1)_{%
\mathcal{\tilde{K}}},\quad \widehat{gl}(1)_{\mathcal{\tilde{J}}}\times
\widehat{gl}(1)_{\mathcal{\tilde{K}}}
\end{equation}%
defined by Poisson brackets\textrm{\ }like
\begin{equation}
\begin{tabular}{lll}
$\left\{ \mathcal{J}_{\varphi }^{{\small 0}},\mathcal{J}_{\varphi ^{\prime
}}^{{\small 0}}\right\} $ & $=$ & $\pm \frac{4\pi }{\mathrm{k}}\partial
_{\varphi }\delta _{\varphi -\varphi ^{\prime }}$ \\
$\left\{ \mathcal{K}_{\varphi }^{{\small 0}},\mathcal{K}_{\varphi ^{\prime
}}^{{\small 0}}\right\} $ & $=$ & $\pm \frac{6\pi }{\mathrm{k}}\partial
_{\varphi }\delta _{\varphi -\varphi ^{\prime }}$%
\end{tabular}%
\qquad ,\qquad
\begin{tabular}{lll}
$\{\mathcal{\tilde{J}}_{\varphi }^{{\small 0}},\mathcal{\tilde{J}}_{\varphi
^{\prime }}^{{\small 0}}\}$ & $=$ & $\pm \frac{4\pi }{\mathrm{\tilde{k}}}%
\partial _{\varphi }\delta _{\varphi -\varphi ^{\prime }}$ \\
$\{\mathcal{\tilde{K}}_{\varphi }^{{\small 0}},\mathcal{\tilde{K}}_{\varphi
^{\prime }}^{{\small 0}}\}$ & $=$ & $\pm \frac{4\pi }{\mathrm{\tilde{k}}}%
\partial _{\varphi }\delta _{\varphi -\varphi ^{\prime }}$%
\end{tabular}%
\end{equation}%
The two sectors are \textrm{distinguished} by the sign of the CS coupling $%
\mathrm{\tilde{k}}=-\mathrm{k}$; and the $\hat{u}(1)$ and $\widehat{gl}(1)$
can be differentiated by the spectrum of $\mathrm{k}$ as described below.

By \textrm{expressing} the Fourier expansions of ($\mathcal{J}_{\varphi }^{%
{\small 0}},\mathcal{K}_{\varphi }^{{\small 0}}$) and ($\mathcal{\tilde{J}}%
_{\varphi }^{{\small 0}},\mathcal{\tilde{K}}_{\varphi }^{{\small 0}}$) in
terms of the classical modes ($J_{n}^{{\small 0}},K_{m}^{{\small 0}}$) and ($%
\tilde{J}_{n}^{{\small 0}},\tilde{K}_{n}^{{\small 0}}$), the above brackets
expand as follows%
\begin{equation}
\begin{tabular}{lll}
$i\left\{ J_{n}^{{\small 0}},J_{m}^{{\small 0}}\right\} _{\text{\textsc{pb}}%
} $ & $=$ & $\pm \frac{\mathrm{k}}{2}n\delta _{n+m,0}$ \\
$i\left\{ K_{n}^{{\small 0}},K_{m}^{{\small 0}}\right\} _{\text{\textsc{pb}}%
} $ & $=$ & $\pm \frac{2\mathrm{k}}{3}n\delta _{n+m,0}$%
\end{tabular}%
,\qquad
\begin{tabular}{lll}
$i\{\tilde{J}_{n}^{{\small 0}},\tilde{J}_{m}^{{\small 0}}\}_{\text{\textsc{pb%
}}}$ & $=$ & $\pm \frac{\mathrm{k}}{2}n\delta _{n+m,0}$ \\
$i\{\tilde{K}_{n}^{{\small 0}},\tilde{K}_{m}^{{\small 0}}\}_{\text{\textsc{pb%
}}}$ & $=$ & $\pm \frac{2\mathrm{k}}{3}n\delta _{n+m,0}$%
\end{tabular}
\label{PB}
\end{equation}%
At the quantum level, these Poisson brackets get mapped to the commutators%
\begin{equation}
\begin{tabular}{lll}
$\left[ \boldsymbol{J}_{n}^{{\small 0}},\boldsymbol{J}_{m}^{{\small 0}}%
\right] $ & $=$ & $\pm \frac{\mathrm{k}}{2}n\delta _{n+m,0}$ \\
$\left[ \boldsymbol{K}_{n}^{{\small 0}},\boldsymbol{K}_{m}^{{\small 0}}%
\right] $ & $=$ & $\pm \frac{2\mathrm{k}}{3}n\delta _{n+m,0}$%
\end{tabular}%
,\qquad
\begin{tabular}{lll}
$\lbrack \boldsymbol{\tilde{J}}_{n}^{{\small 0}},\boldsymbol{\tilde{J}}_{m}^{%
{\small 0}}]$ & $=$ & $\pm \frac{\mathrm{k}}{2}n\delta _{n+m,0}$ \\
$\lbrack \boldsymbol{\tilde{K}}_{n}^{{\small 0}},\boldsymbol{\tilde{K}}_{m}^{%
{\small 0}}]$ & $=$ & $\pm \frac{2\mathrm{k}}{3}n\delta _{n+m,0}$%
\end{tabular}
\label{KMB}
\end{equation}%
with Kac-Moody levels $\pm \frac{\mathrm{k}}{2}$, $\pm \frac{2\mathrm{k}}{3}$
\cite{1B} and adjoint conjugations $\left( \boldsymbol{J}_{n}^{{\small 0}%
}\right) ^{\dagger }=\boldsymbol{J}_{-n}^{{\small 0}},$ $\left( \boldsymbol{K%
}_{n}^{{\small 0}}\right) ^{\dagger }=\boldsymbol{K}_{-n}^{{\small 0}}$;
similar \textrm{relations are available }for the twild \textrm{sector}.

Notice that the $\left[ \boldsymbol{J}_{n}^{{\small 0}},\boldsymbol{J}_{m}^{%
{\small 0}}\right] =+\frac{\mathrm{k}}{2}n\delta _{n+m,0}$ with positive $%
\mathrm{k}$ defines the infinite dimensional \^{u}(1)$_{\mathrm{k,}\mathcal{J%
}};$ this is a subalgebra of the affine Kac-Moody $\widehat{su}(2)_{\mathrm{k%
}}\subset \widehat{\mathcal{G}}_{\mathrm{k}}$ where the container $\widehat{%
\mathcal{G}}_{\mathrm{k}}$ refers either to $\widehat{su}(3)_{\mathrm{k}}$
or to $\widehat{su}(2,1)_{\mathrm{k}}$ as in the Table (\ref{Tab4}).
However, the $\left[ \boldsymbol{J}_{n}^{{\small 0}},\boldsymbol{J}_{m}^{%
{\small 0}}\right] =-\frac{\mathrm{k}}{2}n\delta _{n+m,0}$ with the \textrm{%
minus} sign defines the infinite algebra $\widehat{gl}$(1)$_{\mathrm{k,}%
\mathcal{J}}$\emph{, }\textrm{a} subalgebra of the affine $\widehat{sl}(2,%
\mathbb{R})_{\mathrm{k}}\subset \widehat{\mathcal{G}}_{\mathrm{k}}$ with $%
\widehat{\mathcal{G}}_{\mathrm{k}}$ being either the $\widehat{sl}(3,\mathbb{%
R})_{\mathrm{k}}$ or the $\widehat{su}(2,1)_{\mathrm{k}}$. \textrm{For
instance,} the affine $\widehat{sl}(2,\mathbb{R})_{\mathrm{k}}$ is given by
the commutation relations \textrm{\cite{Dixon,Eglin,E1}},
\begin{equation}
\begin{tabular}{lll}
$\left[ \boldsymbol{J}_{n}^{{\small 0}},\boldsymbol{J}_{m}^{{\small 0}}%
\right] $ & $=$ & $-\frac{\mathrm{k}}{2}n\delta _{n+m,0}$ \\
$\left[ \boldsymbol{J}_{n}^{{\small 0}},\boldsymbol{J}_{m}^{{\small \pm }}%
\right] $ & $=$ & $\pm \boldsymbol{J}_{n+m}^{{\small \pm }}$ \\
$\left[ \boldsymbol{J}_{n}^{{\small +}},\boldsymbol{J}_{m}^{{\small -}}%
\right] $ & $=$ & $-2\boldsymbol{J}_{n+m}^{{\small 0}}+\mathrm{k}n\delta
_{n+m,0}$%
\end{tabular}%
\end{equation}%
where the first row defines precisely the subalgebra $\widehat{gl}$(1)$_{%
\mathrm{k,}\mathcal{J}}$. With these symmetries, one can determine the
charges of the conformal weights of the states of the HS soft Heisenberg
hair in the $SL(3)_{\mathtt{L}}\times $ $SL(3)_{\mathtt{R}}$ gravity model.

\subsubsection{Charges of the black flowers}

The zero modes ($\boldsymbol{J}_{{\small 0}}^{{\small 0}},\boldsymbol{K}_{%
{\small 0}}^{{\small 0}}$) and ($\boldsymbol{\tilde{J}}_{{\small 0}}^{%
{\small 0}},\boldsymbol{\tilde{K}}_{{\small 0}}^{{\small 0}}$) of the
conserved currents on the boundary surface commute with all the mode
generators of the current algebras (\ref{KMB}) namely%
\begin{equation}
\begin{tabular}{lll}
$\left[ \boldsymbol{J}_{{\small 0}}^{{\small 0}},\boldsymbol{J}_{m}^{{\small %
0}}\right] $ & $=$ & $0$ \\
$\left[ \boldsymbol{K}_{{\small 0}}^{{\small 0}},\boldsymbol{K}_{m}^{{\small %
0}}\right] $ & $=$ & $0$%
\end{tabular}%
,\qquad
\begin{tabular}{lll}
$\lbrack \boldsymbol{\tilde{J}}_{{\small 0}}^{{\small 0}},\boldsymbol{\tilde{%
J}}_{m}^{{\small 0}}]$ & $=$ & $0$ \\
$\lbrack \boldsymbol{\tilde{K}}_{{\small 0}}^{{\small 0}},\boldsymbol{\tilde{%
K}}_{m}^{{\small 0}}]$ & $=$ & $0$%
\end{tabular}
\label{cmr}
\end{equation}%
As such, they give the charges ($q_{2},q_{3}$) and ($\tilde{q}_{2},\tilde{q}%
_{3}$) labelling the quantum states of the HS theory. Denoting these
particle states as $\left\vert q_{2},q_{3}\right\rangle $ and $\left\vert
\tilde{q}_{2},\tilde{q}_{3}\right\rangle $ and $\left( \mathbf{i}\right) $
considering the example of the algebra\textrm{s} $\left[ \boldsymbol{J}_{n}^{%
{\small 0}},\boldsymbol{J}_{m}^{{\small 0}}\right] =+\frac{\mathrm{k}}{2}%
n\delta _{n+m,0}$ and $\left[ \boldsymbol{K}_{n}^{{\small 0}},\boldsymbol{K}%
_{m}^{{\small 0}}\right] =+\frac{\mathrm{k}}{2}n\delta _{n+m,0}$ with $%
\mathrm{k}>0$, then $\left( \mathbf{ii}\right) $ setting $\boldsymbol{a}_{n}=%
\sqrt{2/\left( \mathrm{k}n\right) }\boldsymbol{J}_{n}^{{\small 0}}$ for $%
n\geq 0$ and $\boldsymbol{a}_{n}^{\dagger }=\sqrt{2/\left( \mathrm{k}%
n\right) }\boldsymbol{J}_{-n}^{{\small 0}}$ as well as $\boldsymbol{b}_{n}=%
\sqrt{2/\left( \mathrm{k}n\right) }\boldsymbol{K}_{n}^{{\small 0}}$ and $%
\boldsymbol{b}_{n}^{\dagger }=\sqrt{2/\left( \mathrm{k}n\right) }\boldsymbol{%
K}_{-n}^{{\small 0}}$, the above (\ref{cmr})\ becomes%
\begin{equation}
\left[ \boldsymbol{a}_{n},\boldsymbol{a}_{m}^{{\small \dagger }}\right]
=\delta _{n+m,0}\quad ,\quad \left[ \boldsymbol{b}_{n},\boldsymbol{b}_{m}^{%
{\small \dagger }}\right] =\delta _{n+m,0}\quad ,\quad \left[ \boldsymbol{a}%
_{n},\boldsymbol{b}_{m}\right] =\left[ \boldsymbol{a}_{n},\boldsymbol{b}%
_{m}^{{\small \dagger }}\right] =0
\end{equation}%
Using these canonical commutation relations, we can build the associated HW
representations ($\boldsymbol{a}_{{\small 0}}=\boldsymbol{J}_{{\small 0}}^{%
{\small 0}}$ and $\boldsymbol{b}_{{\small 0}}=\boldsymbol{K}_{{\small 0}}^{%
{\small 0}}$) characterised by the ground states
\begin{equation}
\begin{tabular}{lllllll}
$\boldsymbol{a}_{{\small 0}}\left\vert q_{2},q_{3}\right\rangle $ & $=$ & $%
q_{2}\left\vert q_{2},q_{3}\right\rangle $ & $,\qquad $ & $\boldsymbol{a}_{%
{\small +n}}\left\vert q_{2},q_{3}\right\rangle $ & $=$ & $0$ \\
$\boldsymbol{b}_{{\small 0}}\left\vert q_{2},q_{3}\right\rangle $ & $=$ & $%
q_{3}\left\vert q_{2},q_{3}\right\rangle $ & $,\qquad $ & $\boldsymbol{b}_{%
{\small +n}}\left\vert q_{2},q_{3}\right\rangle $ & $=$ & $0$%
\end{tabular}
\label{rep}
\end{equation}%
where $q_{2}$ and $q_{3}$ are integers. The excited states generating the
two HW representations are built \textrm{per} usual as follows%
\begin{eqnarray}
\left\vert N_{0};q_{2},q_{3}\right\rangle &=&\mathcal{N}_{\text{\textsc{norm}%
}}\dprod\limits_{i=1}^{N_{\mathcal{J}}}\boldsymbol{a}_{n_{i}}^{{\small %
\dagger }}\dprod\limits_{j=1}^{N_{\mathcal{K}}}\boldsymbol{b}%
_{m_{j}}^{\dagger }\left\vert q_{2},q_{3}\right\rangle  \label{E1} \\
\left\vert \tilde{N}_{0};\tilde{q}_{2},\tilde{q}_{3}\right\rangle &=&%
\mathcal{N}_{\text{\textsc{norm}}}\dprod\limits_{i=1}^{N_{\mathcal{J}}}%
\boldsymbol{\tilde{a}}_{n_{i}}^{{\small \dagger }}\dprod\limits_{j=1}^{N_{%
\mathcal{K}}}\boldsymbol{\tilde{b}}_{m_{j}}^{\dagger }\left\vert \tilde{q}%
_{2},\tilde{q}_{3}\right\rangle  \label{E2}
\end{eqnarray}%
with normalisations $\mathcal{N}_{\text{\textsc{norm}}}$ and excitation
numbers%
\begin{equation}
N_{0}=\sum_{i=1}^{N_{\mathcal{J}}}n_{i}\qquad ,\qquad \tilde{N}%
_{0}=\sum_{j=1}^{N_{\mathcal{K}}}m_{j}
\end{equation}%
Because of the commutation relations (\ref{cmr}), all the excited states of
the above HW representations have the same charges under $\boldsymbol{a}_{%
{\small 0}}=\boldsymbol{J}_{{\small 0}}^{{\small 0}}$ and $\boldsymbol{b}_{%
{\small 0}}=\boldsymbol{K}_{{\small 0}}^{{\small 0}}$, that is
\begin{equation}
\begin{tabular}{lll}
$\boldsymbol{a}_{{\small 0}}\left\vert N_{0};q_{2},q_{3}\right\rangle $ & $=$
& $q_{2}\left\vert N_{0};q_{2},q_{3}\right\rangle $ \\
$\boldsymbol{b}_{{\small 0}}\left\vert N_{0};q_{2},q_{3}\right\rangle $ & $=$
& $q_{3}\left\vert N_{0};q_{2},q_{3}\right\rangle $%
\end{tabular}%
\end{equation}%
with mean values%
\begin{equation}
\begin{tabular}{lll}
$\left\langle N_{0}\right\vert \boldsymbol{a}_{{\small 0}}\left\vert
N_{0}\right\rangle $ & $=$ & $q_{2}$ \\
$\left\langle N_{0}\right\vert \boldsymbol{b}_{{\small 0}}\left\vert
N_{0}\right\rangle $ & $=$ & $q_{3}$%
\end{tabular}%
\qquad ,\qquad
\begin{tabular}{lll}
$\left\langle \tilde{N}_{0}\right\vert \boldsymbol{\tilde{a}}_{{\small 0}%
}\left\vert \tilde{N}_{0}\right\rangle $ & $=$ & $\tilde{q}_{2}$ \\
$\left\langle \tilde{N}_{0}\right\vert \boldsymbol{\tilde{b}}_{{\small 0}%
}\left\vert \tilde{N}_{0}\right\rangle $ & $=$ & $\tilde{q}_{3}$%
\end{tabular}
\label{mn}
\end{equation}%
These quantum states define the soft Heisenberg hair states of the HS black
holes (black flowers) \textrm{\cite{1A,1B}, }they are characterised by the
charges\textrm{\ }$q_{2},q_{3}$\textrm{\ and }$\tilde{q}_{2},\tilde{q}_{3}$
independently of $\boldsymbol{J}_{n}^{{\small 0}}$ and $\boldsymbol{K}_{n}^{%
{\small 0}}$\textrm{.}

\subsubsection{Conformal weights of black flowers}

The Fourier modes ($\boldsymbol{J}_{n}^{{\small 0}},\boldsymbol{K}_{n}^{%
{\small 0}}$) and ($\boldsymbol{\tilde{J}}_{n}^{{\small 0}},\boldsymbol{%
\tilde{K}}_{n}^{{\small 0}}$) are eigenvalues of the scale operators $%
\mathfrak{L}_{0}$ and $\mathfrak{\tilde{L}}_{0}$ of the Virasoro algebras%
\begin{equation}
\begin{tabular}{lll}
$\left[ \mathfrak{L}_{n},\mathfrak{L}_{m}\right] $ & $=$ & $\left(
n-m\right) \mathfrak{L}_{n+m}+\frac{c_{L}}{12}\left( n^{3}-n\right) \delta
_{n+m,0}$ \\
$\left[ \mathfrak{\tilde{L}}_{n},\mathfrak{\tilde{L}}_{m}\right] $ & $=$ & $%
\left( n-m\right) \mathfrak{\tilde{L}}_{n+m}+\frac{c_{R}}{12}\left(
n^{3}-n\right) \delta _{n+m,0}$%
\end{tabular}%
\end{equation}%
with $\mathfrak{L}_{n}=\int_{0}^{2\pi }e^{in\varphi }\mathcal{L}_{\varphi
\varphi }/(2\pi )$ with $\mathcal{L}_{\varphi \varphi }$ as in (\ref{L1}).
We have
\begin{equation}
\begin{tabular}{lll}
$\left[ \mathfrak{L}_{0},\boldsymbol{J}_{n}^{{\small 0}}\right] $ & $=$ & $-n%
\boldsymbol{J}_{n}^{{\small 0}}$ \\
$\left[ \mathfrak{L}_{0},\boldsymbol{K}_{n}^{{\small 0}}\right] $ & $=$ & $-n%
\boldsymbol{K}_{n}^{{\small 0}}$%
\end{tabular}%
\qquad ,\qquad
\begin{tabular}{lll}
$\lbrack \mathfrak{\tilde{L}}_{0},\boldsymbol{\tilde{J}}_{n}^{{\small 0}}]$
& $=$ & $-n\boldsymbol{\tilde{J}}_{n}^{{\small 0}}$ \\
$\lbrack \mathfrak{\tilde{L}}_{0},\boldsymbol{\tilde{K}}_{n}^{{\small 0}}]$
& $=$ & $-n\boldsymbol{\tilde{K}}_{n}^{{\small 0}}$%
\end{tabular}%
\end{equation}%
By using the change $\boldsymbol{a}_{n}=\sqrt{2/\left( \mathrm{k}n\right) }%
\boldsymbol{J}_{n}^{{\small 0}}$ and $\boldsymbol{b}_{n}=\sqrt{2/\left(
\mathrm{k}n\right) }\boldsymbol{K}_{n}^{{\small 0}}$, the above relations
read as $\left[ \mathfrak{L}_{0},\boldsymbol{a}_{n}\right] =-n\boldsymbol{a}%
_{n}$ and $\left[ \mathfrak{L}_{0},\boldsymbol{a}_{n}^{\dagger }\right] =n%
\boldsymbol{a}_{n}^{\dagger }.$ Similar formulas can be written for the
modes $(\boldsymbol{b}_{n},\boldsymbol{b}_{n}^{\dagger })$ as well as for
the twild partners. These eigenvalue equations can be solved in terms of the
oscillator numbers $\boldsymbol{a}_{n}^{\dagger }\boldsymbol{a}_{n}$, $%
\boldsymbol{b}_{n}^{\dagger }\boldsymbol{b}_{n}$ and $\boldsymbol{\tilde{a}}%
_{n}^{\dagger }\boldsymbol{\tilde{a}}_{n}$, $\boldsymbol{\tilde{b}}%
_{n}^{\dagger }\boldsymbol{\tilde{b}}_{n}$ like $\mathfrak{L}_{0}=\sum n(%
\boldsymbol{a}_{n}^{\dagger }\boldsymbol{a}_{n}+\boldsymbol{b}_{n}^{\dagger }%
\boldsymbol{b}_{n})$ and $\mathfrak{\tilde{L}}_{0}=\sum n(\boldsymbol{\tilde{%
a}}_{n}^{\dagger }\boldsymbol{\tilde{a}}_{n}+\boldsymbol{\tilde{b}}%
_{n}^{\dagger }\boldsymbol{\tilde{b}}_{n})$.

In the left sector, the quantum states $\left\vert q_{2},q_{3}\right\rangle $
given by the highest weight state representations (\ref{rep}) are also
designated by the central charge c$_{L}$ and the conformal weight $h$. The
soft Heisenberg hair states are then distinguished by their central charge
and their conformal scales as%
\begin{equation}
\mathfrak{L}_{0}\left\vert N_{0};q_{2},q_{3}\right\rangle =h\left\vert
N_{0};q_{2},q_{3}\right\rangle \qquad ,\qquad \mathfrak{L}_{0}\left\vert
\tilde{N}_{0};\tilde{q}_{2},\tilde{q}_{3}\right\rangle =\tilde{h}\left\vert
\tilde{N}_{0};\tilde{q}_{2},\tilde{q}_{3}\right\rangle
\end{equation}%
with conformal weights
\begin{equation}
h=N_{0}=\sum_{i}n_{i}\qquad ,\qquad \tilde{h}=\tilde{N}_{0}=\sum_{j}m_{j}
\end{equation}

\subsection{Entropy of black flowers}

Following \cite{1A,1B}, the entropy of the thermal HS soft Heisenberg hair
black hole is obtained as follows:

$\left( \mathbf{i}\right) $ We perform a Wick rotation $t=it_{E}$ with
Euclidean time parameterising the cycle $\left[ 0,\mathrm{\beta }\right] $
where $\mathrm{\beta }$\ is the inverse of the BH temperature. Under this
compactification, the potentials ($a_{t_{E}},a_{\varphi }$) given by eq(\ref%
{da}) become complex fields with $\tilde{a}_{\varphi /t_{E}}=\bar{a}%
_{\varphi /t_{E}}$ reading as follows%
\begin{equation}
\begin{tabular}{lll}
$a_{\varphi }^{\text{\textsc{diag}}}$ & $=$ & $\mathcal{J}_{\varphi
}^{0}L_{0}+\mathcal{K}_{\varphi }^{0}W_{0}$ \\
$a_{t}^{\text{\textsc{diag}}}$ & $=$ & $\lambda ^{0}L_{0}+\Lambda ^{0}W_{0}$%
\end{tabular}%
\qquad ,\qquad
\begin{tabular}{lll}
$\bar{a}_{\varphi }^{\text{\textsc{diag}}}$ & $=$ & $-\mathcal{\bar{J}}%
_{\varphi }^{0}L_{0}-\mathcal{\bar{K}}_{\varphi }^{0}W_{0}$ \\
$\bar{a}_{t}^{\text{\textsc{diag}}}$ & $=$ & $\bar{\lambda}^{0}L_{0}+\bar{%
\Lambda}^{0}W_{0}$%
\end{tabular}
\label{ada}
\end{equation}

$\left( \mathbf{ii}\right) $ \textrm{We} use the formula \textrm{\cite{1C}}
\begin{equation}
\text{\textsc{S}}_{\text{\textsc{hs-bh}}}=\frac{\mathrm{k}}{\epsilon _{N}}%
\dint\nolimits_{0}^{\mathrm{\beta }}dt_{E}\dint\nolimits_{0}^{2\pi }d\varphi
Tr\left[ a_{t_{E}}a_{\varphi }-\bar{a}_{t_{E}}\bar{a}_{\varphi }\right]
\end{equation}%
with SL(N) normalisation as $\epsilon _{N}=N\left( N^{2}-1\right) /12$ and $%
\epsilon _{3}=2.$ By substituting (\ref{ada}) and computing the trace, the
above \textsc{S}$_{\text{\textsc{hs-bh}}}$ reads as%
\begin{equation}
\text{\textsc{S}}_{\text{\textsc{hs-bh}}}=\frac{\mathrm{k}}{4\pi }%
\dint\nolimits_{0}^{\mathrm{\beta }}dt_{E}\dint\nolimits_{0}^{2\pi }d\varphi
Tr\left[ \lambda ^{0}\mathcal{J}_{\varphi }^{0}+\bar{\lambda}^{0}\mathcal{%
\bar{J}}_{\varphi }^{0}+\frac{4}{3}\left( \Lambda ^{0}\mathcal{K}_{\varphi
}^{0}+\bar{\Lambda}^{0}\mathcal{\bar{K}}_{\varphi }^{0}\right) \right]
\end{equation}%
For time independent potentials, the angular integration while assuming real
Lagrange parameters \textrm{gives}%
\begin{equation}
\begin{tabular}{lll}
$\text{\textsc{S}}_{\text{\textsc{hs-bh}}}$ & $=$ & $\frac{\mathrm{k\beta }}{%
2}\left[ \frac{2}{\mathrm{k}}\lambda ^{0}\left( J_{0}^{0}+\bar{J}%
_{0}^{0}\right) +\frac{4}{3}\frac{3}{2\mathrm{k}}\Lambda ^{0}\left(
K_{0}^{0}+\bar{K}_{0}^{0}\right) \right] $ \\
& $=$ & $\mathrm{\beta }\left[ \lambda ^{0}\left( J_{0}^{0}+\bar{J}%
_{0}^{0}\right) +\Lambda ^{0}\left( K_{0}^{0}+\bar{K}_{0}^{0}\right) \right]
$%
\end{tabular}%
\end{equation}%
Comparing this relation to (\ref{IPM}), we learn that \textsc{S}$_{\text{%
\textsc{hs-bh}}}$ as defined in \textrm{\cite{1B}} corresponds in our
construction to \textsc{S}$_{\text{\textsc{hs-bh}}}=2\pi \left( \mathtt{I}+%
\mathtt{\tilde{I}}\right) $ while assuming the positivity of $\mathtt{I}$
and $\mathtt{\tilde{I}}$. \textrm{U}sing (\ref{RF}),\textrm{\ the }entropy
of the black flowers in terms of the real forms of SL(3,$\mathbb{C}$) is
\begin{equation}
\begin{tabular}{|c|c|}
\hline
real form & \textsc{S}$_{\text{\textsc{hs-bh}}}$ \\ \hline\hline
$\ \ \ \ su\left( 3\right) _{L_{{\small 0}},W_{{\small 0}}}$ \ \ \ \ \  & $\
\ \ \ \ \ 2\pi \mathfrak{N}\left( J_{0}^{0}+\bar{J}_{0}^{0}\right) +3\pi
\mathfrak{M}\left( K_{0}^{0}+\bar{K}_{0}^{0}\right) $ \ \ \ \ \ \ \ \  \\
$su\left( 2\right) _{L_{{\small 0}}}$ & $2\pi N\left( J_{0}^{0}+\bar{J}%
_{0}^{0}\right) $ \\
$su\left( 2\right) _{W_{{\small 0}}}$ & $3\pi \mathfrak{M}\left( K_{0}^{0}+%
\bar{K}_{0}^{0}\right) $ \\ \hline\hline
$su(2,1)_{{\small 12}}$ & $2\pi N\left( J_{0}^{0}+\bar{J}_{0}^{0}\right) $
\\
$su(1,2)_{{\small 21}}$ & $3\pi \mathfrak{M}\left( K_{0}^{0}+\bar{K}%
_{0}^{0}\right) $ \\ \hline\hline
\end{tabular}
\label{EE}
\end{equation}%
\begin{equation*}
\end{equation*}%
where $\mathfrak{N}=N+M/2$\ and $\mathfrak{M}=M/2+p/3$ with $N$ and $M$
integers.

By using (\ref{mn}), we have%
\begin{equation}
\begin{tabular}{l|l|l}
real form & \textsc{S}$_{\text{\textsc{hs-bh}}}$ & $r_{\text{\textsc{hs-bh}}%
} $ \\ \hline
$su\left( 3\right) _{L_{{\small 0}},W_{{\small 0}}}$ & $2\pi \mathfrak{N}%
\left( q_{2}+\tilde{q}_{2}\right) +3\mathfrak{M}\left( q_{3}+\tilde{q}%
_{3}\right) $ & $\mathfrak{N}\left( q_{2}+\tilde{q}_{2}\right) +\frac{3}{2}%
\mathfrak{M}\left( q_{3}+\tilde{q}_{3}\right) $ \\ \hline
$su\left( 2\right) _{L_{{\small 0}}}$ & $2\pi N\left( q_{2}+\tilde{q}%
_{2}\right) $ & $N_{0}\left( q_{2}+\tilde{q}_{2}\right) $ \\ \hline
$su\left( 2\right) _{W_{{\small 0}}}$ & $3\pi \mathfrak{M}\left( q_{3}+%
\tilde{q}_{3}\right) $ & $\frac{3}{2}\mathfrak{M}\left( q_{3}+\tilde{q}%
_{3}\right) $ \\ \hline
$su(2,1)_{{\small 12}}$ & $2\pi N\left( q_{2}+\tilde{q}_{2}\right) $ & $%
N\left( q_{2}+\tilde{q}_{2}\right) $ \\ \hline
$su(1,2)_{{\small 21}}$ & $3\pi \mathfrak{M}\left( q_{3}+\tilde{q}%
_{3}\right) $ & $\frac{3}{2}\mathfrak{M}\left( q_{3}+\tilde{q}_{3}\right) $
\\ \hline
\end{tabular}
\label{28}
\end{equation}%
\begin{equation*}
\end{equation*}%
In what follows, we give a realisation of the HS soft Heisenberg hair states
in terms of excitations of conformal primary field operators.

\section{Scalar and Fermi fields realisations}

\label{sec5} \qquad In this section, we develop the free field realisations (%
\ref{B}-\ref{F}) by \textrm{conceiving} the boundary potentials ($a_{\varphi
}^{\text{\textsc{diag}}},a_{t}^{\text{\textsc{diag}}}$) and ($\tilde{a}%
_{\varphi }^{\text{\textsc{diag}}},\tilde{a}_{t}^{\text{\textsc{diag}}}$)
\textrm{as} conformal primary fields \textrm{with} quantum excitations.
Here, the physical charge potentials are imagined \textrm{as} realistic 2D
scalars and/or 2D Fermi fields sitting at the boundary of AdS$_{3}$; and
then at the horizon of the HS black holes. Because the boundary charge
potentials
\begin{equation}
(\mathcal{J}_{\varphi },\mathcal{K}_{\varphi })\qquad ,\qquad (\mathcal{%
\tilde{J}}_{\varphi },\mathcal{\tilde{K}}_{\varphi })
\end{equation}%
are affine Heisenberg conserved currents, they carry conformal weights and
so they can be\textrm{\ }concretely realised in terms of primary conformal
fields. \textrm{W}e first investigate the bosonic realisation (\ref{B}),
then we turn to the fermionic \textrm{one} (\ref{F}).

\subsection{Charge potentials as primary scalars}

Here, the conserved currents ($\mathcal{J}_{\varphi },\mathcal{K}_{\varphi }$%
) and ($\mathcal{\tilde{J}}_{\varphi },$ $\mathcal{\tilde{K}}_{\varphi }$)
are imagined in terms of two real 2D scalar fields $\Phi $ and $\Pi $ living
on the boundary of AdS$_{3}$ and splitting into left and right moving fields
as
\begin{equation}
\Phi =\Phi _{L}+\Phi _{R}\qquad ,\qquad \Pi =\Pi _{L}+\Pi _{R}  \label{fp}
\end{equation}%
Generally speaking, these fields are functions of the coordinate variables $%
\xi ^{\pm }=\sigma \pm t$ namely $\Phi _{L}\left( \xi ^{+}\right) $, $\Phi
_{R}\left( \xi ^{-}\right) $, $\Pi _{L}\left( \xi ^{+}\right) $ and $\Pi
_{R}\left( \xi ^{-}\right) $. Under the Wick rotation $t=it_{E},$ the
hyperbolic coordinates $\xi ^{\pm }$ become complex variables $z^{\pm
}=\sigma \pm it_{E}.$ Consequently, the left and right chiralities get
mapped into holomorphic $\Phi _{L}\left( z\right) $ (resp. $\Pi _{L}\left(
z\right) $) and antiholomorphic $\Phi _{R}\left( \bar{z}\right) $ (resp. $%
\Pi _{R}\left( \bar{z}\right) $). The non vanishing two-points correlations
of these 2D scalar fields are given by%
\begin{equation}
\begin{tabular}{lll}
$\left\langle \Phi _{L}\left( z\right) \Phi _{L}\left( w\right)
\right\rangle $ & $\sim $ & $\log \left( z-w\right) $ \\
$\left\langle \Phi _{R}\left( \bar{z}\right) \Phi _{R}\left( \bar{w}\right)
\right\rangle $ & $\sim $ & $\log \left( \bar{z}-\bar{w}\right) $%
\end{tabular}%
,\qquad
\begin{tabular}{lll}
$\left\langle \Pi _{L}\left( z\right) \Pi _{L}\left( w\right) \right\rangle $
& $\sim $ & $\log \left( z-w\right) $ \\
$\left\langle \Pi _{R}\left( \bar{z}\right) \Pi _{R}\left( \bar{w}\right)
\right\rangle $ & $\sim $ & $\log \left( \bar{z}-\bar{w}\right) $%
\end{tabular}%
\end{equation}%
from which we deduce other two-points correlations, in particular%
\begin{equation}
\begin{tabular}{lll}
$\left\langle \mathcal{J}_{L}\left( z\right) \Phi _{L}\left( w\right)
\right\rangle $ & $\sim $ & $\frac{1}{z-w}$ \\
$\left\langle \mathcal{J}_{R}\left( \bar{z}\right) \Phi _{R}\left( \bar{w}%
\right) \right\rangle $ & $\sim $ & $\frac{1}{\bar{z}-\bar{w}}$%
\end{tabular}%
\qquad ,\qquad
\begin{tabular}{lll}
$\left\langle \mathcal{K}_{L}\left( z\right) \Pi _{L}\left( w\right)
\right\rangle $ & $\sim $ & $\frac{1}{z-w}$ \\
$\left\langle \mathcal{K}_{R}\left( \bar{z}\right) \Pi _{R}\left( \bar{w}%
\right) \right\rangle $ & $\sim $ & $\frac{1}{\bar{z}-\bar{w}}$%
\end{tabular}%
\end{equation}%
where we have set
\begin{equation}
\begin{tabular}{lll}
$\mathcal{J}_{L}\left( z\right) $ & $=$ & $\frac{\partial \Phi _{L}\left(
z\right) }{\partial z}$ \\
$\mathcal{K}_{L}\left( z\right) $ & $=$ & $\frac{\partial \Pi _{L}\left(
z\right) }{\partial z}$%
\end{tabular}%
\qquad ,\qquad
\begin{tabular}{lll}
$\mathcal{J}_{R}\left( \bar{z}\right) $ & $=$ & $\frac{\partial \Phi
_{R}\left( \bar{z}\right) }{\partial \bar{z}}$ \\
$\mathcal{K}_{R}\left( \bar{z}\right) $ & $=$ & $\frac{\partial \Pi
_{R}\left( \bar{z}\right) }{\partial \bar{z}}$%
\end{tabular}
\label{cc}
\end{equation}%
Notice that the commutation relations (\ref{KMB}) can be also \textrm{f}%
ormulated as OPE with z-complex coordinate variable as follows \textrm{\cite%
{CFT}}%
\begin{equation}
\mathcal{J}_{z}\left( z_{1}\right) \mathcal{J}_{z}\left( z_{2}\right) \simeq
\frac{\mathrm{k}/2}{\left( z_{1}-z_{2}\right) ^{2}}\qquad ,\qquad \mathcal{K}%
_{z}\left( z_{1}\right) \mathcal{K}_{z}\left( z_{2}\right) \simeq \frac{2%
\mathrm{k}/3}{\left( z_{1}-z_{2}\right) ^{2}}
\end{equation}%
likewise for the right moving currents.

To make contact between (\ref{fp}) and (\ref{B}) that we recall here after ($%
\sigma =i\varphi $),
\begin{equation}
\begin{tabular}{lllllll}
$\mathcal{J}_{\varphi }$ & $=$ & $\partial _{\varphi }X$ & $\qquad ,\qquad $
& $\mathcal{\tilde{J}}_{\varphi }$ & $=$ & $\partial _{\varphi }\tilde{X}$
\\
$\mathcal{K}_{\varphi }$ & $=$ & $\partial _{\varphi }Y$ & $\qquad ,\qquad $
& $\mathcal{\tilde{K}}_{\varphi }$ & $=$ & $\partial _{\varphi }\tilde{Y}$%
\end{tabular}
\label{BB}
\end{equation}%
with $\partial _{t}\mathcal{J}_{\varphi }=\partial _{t}\mathcal{K}_{\varphi
}=0$ and $\partial _{t}\mathcal{\tilde{J}}_{\varphi }=\partial _{t}\mathcal{%
\tilde{K}}_{\varphi }=0$, we restrict the fields $\Phi $ and $\Pi $ to
depend only on the angular coordinate as
\begin{equation}
\Phi _{L}\left( \varphi \right) ,\qquad \Phi _{R}\left( \varphi \right)
,\qquad \Pi _{L}\left( \varphi \right) ,\qquad \Pi _{R}\left( \varphi \right)
\end{equation}%
then set
\begin{equation}
\Phi _{L}=X,\qquad \Pi _{L}=Y,\qquad \Phi _{R}=\tilde{X},\qquad \Pi _{R}=%
\tilde{Y}  \label{d1}
\end{equation}%
and%
\begin{equation}
\begin{tabular}{lllllll}
$\mathcal{J}_{L}$ & $=$ & $\mathcal{J}_{\varphi }$ & $,\qquad $ & $\mathcal{J%
}_{R}$ & $=$ & $\mathcal{\tilde{J}}_{\varphi }$ \\
$\mathcal{K}_{L}$ & $=$ & $\mathcal{K}_{\varphi }$ & $,\qquad $ & $\mathcal{K%
}_{R}$ & $=$ & $\mathcal{\tilde{K}}_{\varphi }$%
\end{tabular}
\label{d2}
\end{equation}%
Putting into the boundary potentials (\ref{da}), we get%
\begin{equation}
\begin{tabular}{lll}
$a_{\varphi }^{\text{\textsc{diag}}}$ & $=$ & $\left( \partial _{\varphi
}X\right) L_{0}+\left( \partial _{\varphi }Y\right) W_{0}$ \\
$a_{t}^{\text{\textsc{diag}}}$ & $=$ & $\lambda ^{0}L_{0}+\Lambda ^{0}W_{0}$%
\end{tabular}%
\quad ,\quad
\begin{tabular}{lll}
$\tilde{a}_{\varphi }^{\text{\textsc{diag}}}$ & $=$ & $-\left( \partial
_{\varphi }\tilde{X}\right) L_{0}-\left( \partial _{\varphi }\tilde{Y}%
\right) W_{0}$ \\
$\tilde{a}_{t}^{\text{\textsc{diag}}}$ & $=$ & $\tilde{\lambda}^{0}L_{0}+%
\tilde{\Lambda}^{0}W_{0}$%
\end{tabular}%
\end{equation}%
and by substituting into (\ref{L1}-\ref{L2}), we \textrm{obtain}%
\begin{equation}
\begin{tabular}{lll}
$\mathcal{L}_{\varphi \varphi }$ & $=$ & $+\frac{\mathrm{k}}{4\pi }\left[
\frac{1}{2}\left( \partial _{\varphi }X\right) ^{2}+\partial _{\varphi
}^{2}X+\frac{2}{3}\left( \partial _{\varphi }Y\right) ^{2}\right] $ \\
$\mathcal{W}_{\varphi \varphi \varphi }$ & $=$ & $-\frac{\mathrm{k}}{6\pi }%
\left[ -\frac{8}{9}\left( \partial _{\varphi }Y\right) ^{3}+\partial
_{\varphi }Y\left( 2\left( \partial _{\varphi }X\right) ^{2}+\partial
_{\varphi }^{2}X\right) +3\partial _{\varphi }X\partial _{\varphi
}^{2}Y+\partial _{\varphi }^{3}Y\right] $%
\end{tabular}%
\end{equation}%
\textrm{same expressions can be written for the twild partners.} Given the
identifications (\ref{d1}-\ref{d2}), we can determine the mode expansions of
the four (anti) chiral scalar fields $\left( X,Y\right) $ and $(\tilde{X},%
\tilde{Y})$%
\begin{equation}
\left( X,Y,\tilde{X},\tilde{Y}\right) \in \mathbb{R}^{4}
\end{equation}%
Recall that these scalars are free boundary fields related to ($\mathcal{J}%
_{\varphi },\mathcal{K}_{\varphi }$) and ($\mathcal{\tilde{J}}_{\varphi },$ $%
\mathcal{\tilde{K}}_{\varphi }$) as in eq(\ref{BB}), so we can define them
as follows
\begin{equation}
\begin{tabular}{lll}
$X-X_{0}$ & $=$ & $\dint_{0}^{\varphi }d\varphi ^{\prime }\mathcal{J}%
_{\varphi ^{\prime }}$ \\
$Y-Y_{0}$ & $=$ & $\dint_{0}^{\varphi }d\varphi ^{\prime }\mathcal{K}%
_{\varphi ^{\prime }}$%
\end{tabular}
\label{JK}
\end{equation}%
with reality properties $X^{\ast }=X$ and $Y^{\ast }=Y.$ The right $\tilde{X}
$ and $\tilde{Y}$ are given by similar integrals. By substituting the
expansions (\ref{exj}) into the above (\ref{JK}), we obtain the following
\begin{equation}
\begin{tabular}{lll}
$X-X_{0}$ & $=$ & $\frac{2}{\mathrm{k}}\dsum\limits_{n=-\infty }^{+\infty
}J_{n}\left( \dint_{0}^{\varphi }d\varphi ^{\prime }e^{-in\varphi ^{\prime
}}\right) $ \\
$Y-Y_{0}$ & $=$ & $\frac{3}{2\mathrm{k}}\dsum\limits_{n=-\infty }^{+\infty
}K_{n}\left( \dint_{0}^{\varphi }d\varphi ^{\prime }e^{-in\varphi ^{\prime
}}\right) $%
\end{tabular}%
\end{equation}%
similarly for $\tilde{X}$ and $\tilde{Y}$. The ($X_{0},Y_{0}$) and their
twild homologue ($\tilde{X}_{0},\tilde{Y}_{0}$) are integration constants to
be \textrm{determined} by demanding \textrm{periodic }scalar fields under
the mapping $\varphi \rightarrow \varphi +2\pi $. Notice that the above
relations involve the integrals $\dint_{0}^{\varphi }d\varphi ^{\prime
}e^{-in\varphi ^{\prime }}$ whose values depend on whether the integer n is
vanishing or not; we have $\left( \mathbf{i}\right) $ the result $\left(
1-e^{-in\varphi }\right) /\left( in\right) $ for $n\neq 0,$ and $\left(
\mathbf{ii}\right) $ the value $\varphi $ for $n=0.$ So, the mode expansions
of the scalar fields read as follows:

\begin{itemize}
\item \emph{left sector}%
\begin{equation}
\begin{tabular}{lll}
$X$ & $=$ & $X_{0}+\frac{2}{\mathrm{k}}J_{0}\varphi +\frac{4}{\mathrm{k}}%
\dsum\limits_{n\neq 0}\frac{\sin (\frac{n}{2}\varphi )}{n}J_{n}e^{-\frac{i}{2%
}n\varphi }$ \\
& $=$ & $\frac{2}{\mathrm{k}}\left( x_{0}+J_{0}\varphi +2\dsum\limits_{n\neq
0}\frac{\sin (\frac{n}{2}\varphi )}{n}J_{n}e^{-\frac{i}{2}n\varphi }\right) $%
\end{tabular}
\label{X1}
\end{equation}%
and
\begin{equation}
\begin{tabular}{lll}
$Y$ & $=$ & $Y_{0}+\frac{3}{2\mathrm{k}}K_{0}\varphi +\frac{6}{2\mathrm{k}}%
\dsum\limits_{n\neq 0}\frac{\sin (\frac{n}{2}\varphi )}{n}K_{n}e^{-\frac{i}{2%
}n\varphi }$ \\
& $=$ & $\frac{3}{2\mathrm{k}}\left( y_{0}+K_{0}\varphi
+2\dsum\limits_{n\neq 0}\frac{\sin (\frac{n}{2}\varphi )}{n}K_{n}e^{-\frac{i%
}{2}n\varphi }\right) $%
\end{tabular}
\label{Y1}
\end{equation}

\item \emph{Right sector}
\begin{equation}
\begin{tabular}{lll}
$\tilde{X}$ & $=$ & $\tilde{X}_{0}+\frac{2}{\mathrm{k}}\tilde{J}_{0}\varphi +%
\frac{4}{\mathrm{k}}\dsum\limits_{n\neq 0}\frac{\sin (\frac{n}{2}\varphi )}{n%
}\tilde{J}_{n}e^{-\frac{i}{2}n\varphi }$ \\
& $=$ & $\frac{2}{\mathrm{k}}\left( \tilde{x}_{0}+\tilde{J}_{0}\varphi
+2\dsum\limits_{n\neq 0}\frac{\sin (\frac{n}{2}\varphi )}{n}\tilde{J}_{n}e^{-%
\frac{i}{2}n\varphi }\right) $%
\end{tabular}
\label{X2}
\end{equation}%
and%
\begin{equation}
\begin{tabular}{lll}
$\tilde{Y}\left( \varphi \right) $ & $=$ & $\tilde{Y}_{0}+\frac{3}{2\mathrm{k%
}}K_{0}\tilde{\varphi}+\frac{6}{2\mathrm{k}}\dsum\limits_{n\neq 0}\frac{\sin
(\frac{n}{2}\varphi )}{n}\tilde{K}_{n}e^{-\frac{i}{2}n\varphi }$ \\
& $=$ & $\frac{3}{2\mathrm{k}}\left( \tilde{y}_{0}+K_{0}\tilde{\varphi}%
+2\dsum\limits_{n\neq 0}\frac{\sin (\frac{n}{2}\varphi )}{n}\tilde{K}_{n}e^{-%
\frac{i}{2}n\varphi }\right) $%
\end{tabular}
\label{Y2}
\end{equation}
\end{itemize}

\ \newline
From these equations, we learn that the fields expansion modes ($\alpha
_{n}^{{\small X}},\alpha _{n}^{{\small Y}}$) and ($\tilde{\alpha}_{n}^{%
{\small X}},\tilde{\alpha}_{n}^{{\small Y}}$) are given by%
\begin{equation}
\begin{tabular}{lll}
$\alpha _{n}^{{\small X}}$ & $=$ & $J_{n}$ \\
$\alpha _{n}^{{\small Y}}$ & $=$ & $K_{n}$%
\end{tabular}%
\qquad ,\qquad
\begin{tabular}{lll}
$\tilde{\alpha}_{n}^{{\small X}}$ & $=$ & $\tilde{J}_{n}$ \\
$\tilde{\alpha}_{n}^{{\small Y}}$ & $=$ & $\tilde{K}_{n}$%
\end{tabular}%
\end{equation}%
These relationships show that the modes ($\alpha _{n}^{{\small X}},\alpha
_{n}^{{\small Y}}$) and ($\tilde{\alpha}_{n}^{{\small X}},\alpha _{n}^{%
{\small Y}}$) satisfy the Poisson brackets (\ref{PB}), and upon quantization
they obey the infinite dimensional commutators%
\begin{equation}
\begin{tabular}{lll}
$\left[ \mathbf{\alpha }_{n}^{{\small X}},\mathbf{\alpha }_{m}^{{\small X}}%
\right] $ & $=$ & $\frac{\mathrm{k}}{2}n\delta _{n+m,0}$ \\
$\left[ \mathbf{\alpha }_{n}^{{\small Y}},\mathbf{\alpha }_{m}^{{\small Y}}%
\right] $ & $=$ & $\frac{2\mathrm{k}}{3}n\delta _{n+m,0}$%
\end{tabular}%
\qquad ,\qquad
\begin{tabular}{lll}
$\left[ \mathbf{\tilde{\alpha}}_{n}^{{\small X}},\mathbf{\tilde{\alpha}}%
_{m}^{{\small X}}\right] $ & $=$ & $\frac{\mathrm{k}}{2}n\delta _{n+m,0}$ \\
$\left[ \mathbf{\tilde{\alpha}}_{n}^{{\small Y}},\mathbf{\tilde{\alpha}}%
_{m}^{{\small Y}}\right] $ & $=$ & $\frac{2\mathrm{k}}{3}n\delta _{n+m,0}$%
\end{tabular}%
\end{equation}%
with zero modes acting on ground state as follows
\begin{equation}
\begin{tabular}{lll}
$\mathbf{\alpha }_{0}^{{\small X}}\left\vert q\right\rangle $ & $=$ & $q_{%
{\small X}}\left\vert q\right\rangle $ \\
$\mathbf{\alpha }_{0}^{{\small Y}}\left\vert q\right\rangle $ & $=$ & $q_{%
{\small Y}}\left\vert q\right\rangle $%
\end{tabular}%
\qquad ,\qquad
\begin{tabular}{lll}
$\mathbf{\tilde{\alpha}}_{0}^{{\small X}}\left\vert \tilde{q}\right\rangle $
& $=$ & $\tilde{q}_{{\small X}}\left\vert \tilde{q}\right\rangle $ \\
$\mathbf{\tilde{\alpha}}_{0}^{{\small Y}}\left\vert \tilde{q}\right\rangle $
& $=$ & $\tilde{q}_{{\small Y}}\left\vert \tilde{q}\right\rangle $%
\end{tabular}%
\end{equation}%
and $\mathbf{\alpha }_{+n}^{{\small Z}}\left\vert q\right\rangle =\mathbf{%
\tilde{\alpha}}_{+n}^{{\small Z}}\left\vert \tilde{q}\right\rangle =0$ (%
{\small Z} for {\small X} and {\small Y}). Notice also that by substituting (%
\ref{JK}), the conservation relations $\partial _{t}\mathcal{J}_{\varphi
}=\partial _{t}\mathcal{K}_{\varphi }=0$ imply $\partial _{\varphi }\left(
\partial _{t}X\right) =0$\ and $\partial _{\varphi }\left( \partial
_{t}Y\right) =0$ and \textrm{again similar formulas arise} for the twild
sector. These relations are naturally solved by
\begin{eqnarray}
\partial _{t}X &=&0\qquad ,\qquad \partial _{t}Y=0  \notag \\
\partial _{t}\tilde{X} &=&0\qquad ,\qquad \partial _{t}\tilde{Y}=0
\end{eqnarray}%
in agreement with the expansions (\ref{X1}-\ref{Y2}). Additionally, while we
have the periodicity $\mathcal{J}_{\varphi }\left( \varphi +2l\pi \right) =%
\mathcal{J}_{\varphi }\left( \varphi \right) $ and $\mathcal{K}_{\varphi
}\left( \varphi +2l\pi \right) =\mathcal{K}_{\varphi }\left( \varphi \right)
$ as well \textrm{for} their twilds homologue, the scalars ($X,Y$) and ($%
\tilde{X},\tilde{Y}$) satisfy \textrm{a} \emph{pseudo-periodicity} as follows%
\begin{equation}
\begin{tabular}{lll}
$X\left( \varphi +2l\pi \right) $ & $=$ & $X\left( \varphi \right) +\frac{%
4\pi l}{\mathrm{k}}J_{0}$ \\
$Y\left( \varphi +2l\pi \right) $ & $=$ & $Y\left( \varphi \right) +\frac{%
6\pi l}{2\mathrm{k}}K_{0}$%
\end{tabular}%
,\quad
\begin{tabular}{lll}
$\tilde{X}\left( \varphi -2l\pi \right) $ & $=$ & $\tilde{X}\left( \varphi
\right) -\frac{4\pi l}{\mathrm{k}}\tilde{J}_{0}$ \\
$\tilde{Y}\left( \varphi -2l\pi \right) $ & $=$ & $\tilde{Y}\left( \varphi
\right) -\frac{6\pi l}{2\mathrm{k}}\tilde{K}_{0}$%
\end{tabular}%
\end{equation}%
To have periodic $X\left( \varphi \right) ,$ $Y\left( \varphi \right) $ and $%
\tilde{X}\left( \varphi \right) ,$ $\tilde{Y}\left( \varphi \right) $, we
must think about the integral constants ($x_{0},y_{0}$) and ($\tilde{x}_{0},%
\tilde{y}_{0}$) \textrm{as elements of the} 4-dim lattice $\mathbf{\Lambda }%
=\{\mathbf{r}_{\boldsymbol{l}},$ $\boldsymbol{l}\in \mathbb{Z}^{4}\}$ with
site vectors $\mathbf{r}_{\boldsymbol{l}}=\left( x_{l_{1}},y_{l_{2}},\tilde{x%
}_{l_{3}},\tilde{y}_{l_{4}}\right) $ given by the discrete series%
\begin{equation}
\begin{tabular}{lllllll}
$x_{l_{1}}$ & $=$ & $\left( 2\pi J_{0}\right) l_{1}$ & \qquad $,$\qquad & $%
\tilde{x}_{l_{3}}$ & $=$ & $(2\pi \tilde{J}_{0})l_{3}$ \\
$y_{l_{2}}$ & $=$ & $\left( 2\pi K_{0}\right) l_{2}$ & \qquad $,$\qquad & $%
\tilde{y}_{l_{4}}$ & $=$ & $(2\pi \tilde{K}_{0})l_{4}$%
\end{tabular}
\label{525}
\end{equation}%
where the four $l_{i}$'s are integers that can be combined into an integer
vector $\boldsymbol{l}=(l_{1},l_{2},l_{3},l_{4}).$ Points in the lattice $%
\mathbf{\Lambda }$ obeys the additive law $\mathbf{r}_{\boldsymbol{l}+%
\boldsymbol{m}}=\mathbf{r}_{\boldsymbol{l}}+\mathbf{r}_{\boldsymbol{m}};$ it
is an abelian group isomorphic to the lattice $2\pi \mathbb{Z}^{4}$ with
parameters $\left( \alpha _{1},\alpha _{2},\alpha _{3},\alpha _{4}\right) $
equal to $(J_{0},K_{0},\tilde{J}_{0},\tilde{K}_{0}).$ Using the lattice
variables $\mathbf{r}_{\Lambda }=\left( x_{\Lambda },y_{\Lambda },\tilde{x}%
_{\Lambda },\tilde{y}_{\Lambda }\right) $, the scalars ($X,Y$) and ($\tilde{X%
},\tilde{Y}$) solving the boundary field equations of motion are given by%
\begin{equation}
\begin{tabular}{lll}
$X$ & $=$ & $\frac{2}{\mathrm{k}}x_{\Lambda }+\frac{4}{\mathrm{k}}%
\dsum\limits_{n\in \mathbb{Z}}J_{n}\frac{\sin (\frac{n}{2}\varphi )}{n}e^{-%
\frac{i}{2}n\varphi }$ \\
$Y$ & $=$ & $\frac{3}{2\mathrm{k}}y_{\Lambda }+\frac{6}{2\mathrm{k}}%
\dsum\limits_{n\in \mathbb{Z}}K_{n}\frac{\sin (\frac{n}{2}\varphi )}{n}e^{-%
\frac{i}{2}n\varphi }$%
\end{tabular}%
,\quad
\begin{tabular}{lll}
$\tilde{X}\left( \varphi \right) $ & $=$ & $\frac{2}{\mathrm{k}}\tilde{x}%
_{\Lambda }+\frac{4}{\mathrm{k}}\dsum\limits_{n\in \mathbb{Z}}\tilde{J}_{n}%
\frac{\sin (\frac{n}{2}\varphi )}{n}e^{-\frac{i}{2}n\varphi }$ \\
$\tilde{Y}\left( \varphi \right) $ & $=$ & $\frac{3}{2\mathrm{k}}\tilde{y}%
_{\Lambda }+\frac{6}{2\mathrm{k}}\dsum\limits_{n\in \mathbb{Z}}\tilde{K}_{n}%
\frac{\sin (\frac{n}{2}\varphi )}{n}e^{-\frac{i}{2}n\varphi }$%
\end{tabular}%
\end{equation}%
they satisfy the periodicities%
\begin{equation}
\begin{tabular}{lll}
$X\left( \varphi +2l_{1}\pi \right) $ & $=$ & $X\left( \varphi \right) $ \\
$Y\left( \varphi +2l_{2}\pi \right) $ & $=$ & $Y\left( \varphi \right) $%
\end{tabular}%
\qquad ,\qquad
\begin{tabular}{lll}
$\tilde{X}\left( \varphi +2l_{3}\pi \right) $ & $=$ & $\tilde{X}\left(
\varphi \right) $ \\
$\tilde{Y}\left( \varphi +2l_{4}\pi \right) $ & $=$ & $\tilde{Y}\left(
\varphi \right) $%
\end{tabular}%
\end{equation}%
\textrm{The} quantum excitations of black holes soft hair \textrm{states are}
given by
\begin{equation}
\left\vert N_{0},\tilde{N}_{0}\right\rangle =\mathcal{N}\dprod\limits_{i}%
\mathbf{\alpha }_{-n_{i}}^{{\small X}}\dprod\limits_{j}\mathbf{\alpha }%
_{-m_{j}}^{{\small Y}}\dprod\limits_{k}\mathbf{\tilde{\alpha}}_{-\tilde{n}%
_{k}}^{{\small \tilde{X}}}\dprod\limits_{l}\mathbf{\tilde{\alpha}}_{-\tilde{m%
}_{l}}^{{\small \tilde{Y}}}\left\vert q_{{\small X}},q_{{\small Y}},\tilde{q}%
_{{\small X}},\tilde{q}_{{\small Y}}\right\rangle
\end{equation}%
with $N_{0}$ and $\tilde{N}_{0}$ \textrm{being} functions of the integers ($%
n_{i},m_{j}$) and ($\tilde{n}_{k},\tilde{m}_{l}$). The HS soft hair entropy
is given by (\ref{28}), it is independent of the excitation numbers.

\subsection{Charge potentials as primary Fermions}

Here, the conserved currents ($\mathcal{J}_{\varphi },\mathcal{K}_{\varphi }$%
) and ($\mathcal{\tilde{J}}_{\varphi },$ $\mathcal{\tilde{K}}_{\varphi }$)
are \textrm{conceived as} densities of pairs of free 2D fermionic fields ($%
\psi ^{\pm },\chi ^{\pm }$) and ($\tilde{\psi}^{\pm },\tilde{\chi}^{\pm }$)
living on the boundary of AdS$_{3}$ and splitting as follows
\begin{equation}
\begin{tabular}{lll}
$\psi ^{\pm }$ & $=$ & $\psi _{L}^{\pm }+\psi _{R}^{\pm }$ \\
$\chi ^{\pm }$ & $=$ & $\chi _{L}^{\pm }+\chi _{R}^{\pm }$%
\end{tabular}%
\qquad ,\qquad
\begin{tabular}{lll}
$\tilde{\psi}^{\pm }$ & $=$ & $\tilde{\psi}_{L}^{\pm }+\tilde{\psi}_{R}^{\pm
}$ \\
$\tilde{\chi}^{\pm }$ & $=$ & $\tilde{\chi}_{L}^{\pm }+\tilde{\chi}_{R}^{\pm
}$%
\end{tabular}%
\end{equation}%
In parallel with the bosonic case,\textrm{\ }under the Wick rotation, the
left and right chiralities get mapped into holomorphic $\psi _{L}\left(
z\right) $ [resp. $\chi _{L}\left( z\right) $] and antiholomorphic $\psi
_{R}\left( \bar{z}\right) $ [resp. $\chi _{R}\left( \bar{z}\right) $]. The
non vanishing two-points correlations of these 2D Fermi fields are given by%
\begin{equation}
\begin{tabular}{lll}
$\left\langle \psi _{L}^{+}\left( z\right) \psi _{L}^{-}\left( w\right)
\right\rangle $ & $\sim $ & $\frac{1}{z-w}$ \\
$\left\langle \psi _{R}^{+}\left( \bar{z}\right) \psi _{R}^{-}\left( \bar{w}%
\right) \right\rangle $ & $\sim $ & $\frac{1}{\bar{z}-\bar{w}}$%
\end{tabular}%
\qquad ,\qquad
\begin{tabular}{lll}
$\left\langle \chi _{L}^{+}\left( z\right) \chi _{L}^{-}\left( w\right)
\right\rangle $ & $\sim $ & $\frac{1}{z-w}$ \\
$\left\langle \chi _{R}^{+}\left( \bar{z}\right) \chi _{R}^{-}\left( \bar{w}%
\right) \right\rangle $ & $\sim $ & $\frac{1}{\bar{z}-\bar{w}}$%
\end{tabular}%
\end{equation}%
from which we learn the conserved (holomorphic and antiholomorphic) currents
in terms of the fermionic densities%
\begin{equation}
\begin{tabular}{lll}
$\mathcal{J}_{L}\left( z\right) $ & $\sim $ & $\psi _{L}^{+}\psi
_{L}^{-}\left( z\right) $ \\
$\mathcal{K}_{L}\left( z\right) $ & $\sim $ & $\chi _{L}^{+}\chi
_{L}^{-}\left( z\right) $%
\end{tabular}%
\qquad ,\qquad
\begin{tabular}{lll}
$\mathcal{J}_{R}\left( \bar{z}\right) $ & $\sim $ & $\psi _{R}^{+}\psi
_{R}^{-}\left( \bar{z}\right) $ \\
$\mathcal{K}_{R}\left( \bar{z}\right) $ & $\sim $ & $\chi _{R}^{+}\chi
_{R}^{-}\left( \bar{z}\right) $%
\end{tabular}%
\end{equation}%
To establish a contact between these primary conformal fermions and the
solutions (\ref{F}) that we recall here after
\begin{equation}
\begin{tabular}{lll}
$\mathcal{J}_{\varphi }$ & $\sim $ & $\psi ^{+}\psi ^{-}$ \\
$\mathcal{K}_{\varphi }$ & $\sim $ & $\chi ^{+}\chi ^{-}$%
\end{tabular}%
\qquad ,\qquad
\begin{tabular}{lll}
$\mathcal{\tilde{J}}_{\varphi }$ & $\sim $ & $\tilde{\psi}^{+}\tilde{\psi}%
^{-}$ \\
$\mathcal{\tilde{K}}_{\varphi }$ & $\sim $ & $\tilde{\chi}^{+}\tilde{\chi}%
^{-}$%
\end{tabular}
\label{FF}
\end{equation}%
we proceed as for the scalar fields. First, because $\partial _{t}\mathcal{J}%
_{\varphi }=\partial _{t}\mathcal{K}_{\varphi }=0$ and $\partial _{t}%
\mathcal{\tilde{J}}_{\varphi }=\partial _{t}\mathcal{\tilde{K}}_{\varphi }=0$%
, the Fermi fields are functions of the angular coordinate only
\begin{equation}
\begin{tabular}{lllllll}
$\psi _{L}^{\pm }\left( \varphi \right) $ & $,\qquad $ & $\psi _{R}^{\pm
}\left( \varphi \right) $ & $,\qquad $ & $\chi _{L}^{\pm }\left( \varphi
\right) $ & $,\qquad $ & $\chi _{R}^{\pm }\left( \varphi \right) $ \\
$\tilde{\psi}_{L}^{\pm }\left( \varphi \right) $ & $,\qquad $ & $\tilde{\psi}%
_{R}^{\pm }\left( \varphi \right) $ & $,\qquad $ & $\tilde{\chi}_{L}^{\pm
}\left( \varphi \right) $ & $,\qquad $ & $\tilde{\chi}_{R}^{\pm }\left(
\varphi \right) $%
\end{tabular}
\label{dd}
\end{equation}%
Thinking about $\left( \mathbf{i}\right) $ the left and the right
chiralities of the Fermi fields like
\begin{equation}
\begin{tabular}{lll}
$\psi _{L}^{\pm }$ & $=$ & $\psi ^{\pm }$ \\
$\psi _{R}^{\pm }$ & $=$ & $\tilde{\psi}^{\pm }$%
\end{tabular}%
\qquad ,\qquad
\begin{tabular}{lll}
$\chi _{L}^{\pm }$ & $=$ & $\chi ^{\pm }$ \\
$\chi _{R}^{\pm }$ & $=$ & $\tilde{\chi}^{\pm }$%
\end{tabular}
\label{d3}
\end{equation}%
and $\left( \mathbf{ii}\right) $ the conserved currents (\ref{cc}) as follows%
\begin{equation}
\begin{tabular}{lll}
$\mathcal{J}_{L}$ & $=$ & $\mathcal{J}_{\varphi }$ \\
$\mathcal{K}_{L}$ & $=$ & $\mathcal{K}_{\varphi }$%
\end{tabular}%
\qquad ,\qquad
\begin{tabular}{lll}
$\mathcal{J}_{R}$ & $=$ & $\mathcal{\tilde{J}}_{\varphi }$ \\
$\mathcal{K}_{R}$ & $=$ & $\mathcal{\tilde{K}}_{\varphi }$%
\end{tabular}
\label{d4}
\end{equation}%
\textrm{The} boundary potentials (\ref{da}) become%
\begin{equation}
\begin{tabular}{lll}
$a_{\varphi }^{\text{\textsc{diag}}}$ & $=$ & $+\left( \psi ^{+}\psi
^{-}\right) L_{0}+\left( \chi ^{+}\chi ^{-}\right) W_{0}$ \\
$\tilde{a}_{\varphi }^{\text{\textsc{diag}}}$ & $=$ & $-\left( \tilde{\psi}%
^{+}\tilde{\psi}^{-}\right) L_{0}-\left( \tilde{\chi}^{+}\tilde{\chi}%
^{-}\right) W_{0}$%
\end{tabular}%
\quad ,\quad
\begin{tabular}{lll}
$a_{t}^{\text{\textsc{diag}}}$ & $=$ & $\lambda ^{0}L_{0}+\Lambda ^{0}W_{0}$
\\
$\tilde{a}_{t}^{\text{\textsc{diag}}}$ & $=$ & $\tilde{\lambda}^{0}L_{0}+%
\tilde{\Lambda}^{0}W_{0}$%
\end{tabular}%
\end{equation}%
From the conservation relations $\partial _{t}\mathcal{J}_{\varphi
}=\partial _{t}\mathcal{K}_{\varphi }=0$ and $\partial _{t}\mathcal{\tilde{J}%
}_{\varphi }=\partial _{t}\mathcal{\tilde{K}}_{\varphi }=0$ with ($\mathcal{J%
}_{\varphi },\mathcal{K}_{\varphi }$) and ($\mathcal{\tilde{J}}_{\varphi },%
\mathcal{\tilde{K}}_{\varphi }$) taken as in (\ref{FF}), we learn that the
above fermionic fields are free Fermi fields as they obey%
\begin{equation}
\begin{tabular}{lllllll}
$\partial _{t}\psi ^{\pm }$ & $=$ & $0$ & $\qquad ,\qquad $ & $\partial _{t}%
\tilde{\psi}^{\pm }$ & $=$ & $0$ \\
$\partial _{t}\chi ^{\pm }$ & $=$ & $0$ & $\qquad ,\qquad $ & $\partial _{t}%
\tilde{\chi}^{\pm }$ & $=$ & $0$%
\end{tabular}%
\end{equation}%
These equations (of motion) indicate that ($\psi ^{\pm },\chi ^{\pm }$) and (%
$\tilde{\psi}^{\pm },\tilde{\chi}^{\pm }$) are functions of the angular
variable $\varphi $ only as in (\ref{dd}).

Moreover, because of their quadratic forms (\ref{FF}), the periodicity
properties $\mathcal{J}_{\varphi }\left( \varphi +2\pi \right) =\mathcal{J}%
_{\varphi }\left( \varphi \right) $ and $\mathcal{K}_{\varphi }\left(
\varphi +2\pi \right) =\mathcal{K}_{\varphi }\left( \varphi \right) $ and
their twild homologue can be realized in two different ways namely by
considering periodic or antiperiodic fermions as described \textrm{%
subsequently}:

\subsubsection{Periodic fermions}

In this case, the fermionic fields ($\psi ^{\pm },\chi ^{\pm }$) and ($%
\tilde{\psi}^{\pm },\tilde{\chi}^{\pm }$) obey the periodicity property
\begin{equation}
\begin{tabular}{lll}
$\psi ^{\pm }\left( \varphi +2\pi \right) $ & $=$ & $\psi ^{\pm }\left(
\varphi \right) $ \\
$\chi ^{\pm }\left( \varphi +2\pi \right) $ & $=$ & $\chi ^{\pm }\left(
\varphi \right) $%
\end{tabular}%
,\qquad
\begin{tabular}{lll}
$\tilde{\psi}^{\pm }\left( \varphi +2\pi \right) $ & $=$ & $\tilde{\psi}%
^{\pm }\left( \varphi \right) $ \\
$\tilde{\chi}^{\pm }\left( \varphi +2\pi \right) $ & $=$ & $\tilde{\chi}%
^{\pm }\left( \varphi \right) $%
\end{tabular}%
\end{equation}%
in agreement with $\mathcal{J}\left( \varphi +2\pi \right) =\mathcal{J}%
\left( \varphi \right) $ and $\mathcal{K}\left( \varphi +2\pi \right) =%
\mathcal{K}\left( \varphi \right) $. These periodicities can be solved by
using the following expansions%
\begin{equation}
\begin{tabular}{lll}
$\psi ^{\pm }$ & $=$ & $\sqrt{\frac{2}{\mathrm{k}}}\dsum\limits_{n\in
\mathbb{Z}}\psi _{n\pm 1/2}^{\pm }e^{-in\varphi }$ \\
$\chi ^{\pm }$ & $=$ & $\sqrt{\frac{3}{2\mathrm{k}}}\dsum\limits_{m\in
\mathbb{Z}}\chi _{m\pm 1/2}^{\pm }e^{-im\varphi }$%
\end{tabular}%
\qquad ,\qquad
\begin{tabular}{lll}
$\tilde{\psi}^{\pm }$ & $=$ & $\sqrt{\frac{2}{\mathrm{k}}}\dsum\limits_{n\in
\mathbb{Z}}\tilde{\psi}_{n\pm 1/2}^{\pm }e^{+in\varphi }$ \\
$\tilde{\chi}^{\pm }$ & $=$ & $\sqrt{\frac{3}{2\mathrm{k}}}%
\dsum\limits_{m\in \mathbb{Z}}\tilde{\chi}_{m\pm 1/2}^{\pm }e^{+im\varphi }$%
\end{tabular}
\label{psi}
\end{equation}%
they \textrm{permit to} express the conserved currents as $\mathcal{J}%
_{\varphi }\sim \psi ^{+}\psi ^{-}$ and $\mathcal{K}_{\varphi }\sim \chi
^{+}\chi ^{-}$ as well as $\mathcal{\tilde{J}}_{\varphi }\sim \psi ^{+}\psi
^{-}$ and $\mathcal{\tilde{K}}_{\varphi }\sim \tilde{\chi}^{+}\tilde{\chi}%
^{-}$. \textrm{Their} Fourier modes ($J_{n},K_{n}$) and ($\tilde{J}_{n},%
\tilde{K}_{n}$) in terms of the fermionic modes ($\psi _{r}^{\pm },\chi
_{r}^{\pm }$) and ($\tilde{\psi}_{r}^{\pm },\tilde{\chi}_{r}^{\pm }$)
\textrm{are as follows}%
\begin{equation}
\begin{tabular}{lll}
$\mathcal{J}_{\varphi }$ & $=$ & $\dsum\limits_{n\in \mathbb{Z}%
}J_{n}e^{-in\varphi }$ \\
$\mathcal{K}_{\varphi }$ & $=$ & $\dsum\limits_{n\in \mathbb{Z}%
}K_{n}e^{-in\varphi }$%
\end{tabular}%
\qquad ,\qquad
\begin{tabular}{lll}
$\mathcal{\tilde{J}}_{\varphi }$ & $=$ & $\dsum\limits_{n\in \mathbb{Z}}%
\tilde{J}_{n}e^{+in\varphi }$ \\
$\mathcal{\tilde{K}}_{\varphi }$ & $=$ & $\dsum\limits_{n\in \mathbb{Z}}%
\tilde{K}_{n}e^{+in\varphi }$%
\end{tabular}%
\end{equation}%
with%
\begin{equation}
\begin{tabular}{lll}
$J_{n}$ & $=$ & $\frac{2}{\mathrm{k}}\dsum\limits_{r\in \mathbb{Z}+1/2}\psi
_{n-r}^{+}\psi _{r}^{-}$ \\
$K_{n}$ & $=$ & $\frac{3}{2\mathrm{k}}\dsum\limits_{r\in \mathbb{Z}+1/2}\chi
_{n-r}^{+}\chi _{r}^{-}$%
\end{tabular}%
\qquad ,\qquad
\begin{tabular}{lll}
$\tilde{J}_{n}$ & $=$ & $\frac{2}{\mathrm{k}}\dsum\limits_{r\in \mathbb{Z}%
+1/2}\tilde{\psi}_{n-r}^{+}\tilde{\psi}_{r}^{-}$ \\
$\tilde{K}_{n}$ & $=$ & $\frac{3}{2\mathrm{k}}\dsum\limits_{r\in \mathbb{Z}%
+1/2}\tilde{\chi}_{n-r}^{+}\tilde{\chi}_{r}^{-}$%
\end{tabular}%
\end{equation}%
\begin{equation*}
\end{equation*}%
with positive $\mathrm{k}>0$. At the quantum level, the Fourier modes ($\psi
_{r}^{\pm },\chi _{r}^{\pm }$) and ($\tilde{\psi}_{r}^{\pm },\tilde{\chi}%
_{r}^{\pm }$) are promoted to fermionic operators ($\mathbf{\psi }_{r}^{\pm
},\mathbf{\chi }_{r}^{\pm }$) and ($\mathbf{\tilde{\psi}}_{r}^{\pm },\mathbf{%
\tilde{\chi}}_{r}^{\pm }$) with adjoint conjugations $(\mathbf{\psi }%
_{r}^{-})^{\dagger }=\mathbf{\psi }_{-r}^{+}$and $(\mathbf{\chi }%
_{r}^{-})^{\dagger }=\mathbf{\chi }_{-r}^{+}$ as well as $(\mathbf{\tilde{%
\psi}}_{r}^{-})^{\dagger }=\mathbf{\tilde{\psi}}_{-r}^{+}$and $(\mathbf{%
\tilde{\chi}}_{r}^{-})^{\dagger }=\mathbf{\tilde{\chi}}_{-r}^{+}$\ with non
vanishing anticommutation relations given by%
\begin{equation}
\begin{tabular}{lll}
$\left\{ \mathbf{\psi }_{r}^{-},\mathbf{\psi }_{s}^{+}\right\} $ & $=$ & $%
\sqrt{\mathrm{k}/2}\delta _{r+s,0}$ \\
$\left\{ \mathbf{\chi }_{r}^{-},\mathbf{\chi }_{s}^{+}\right\} $ & $=$ & $%
\sqrt{2\mathrm{k}/3}\delta _{r+s,0}$%
\end{tabular}%
\qquad ,\qquad
\begin{tabular}{lll}
$\{\mathbf{\tilde{\psi}}_{r}^{-},\mathbf{\tilde{\psi}}_{s}^{+}\}$ & $=$ & $%
\sqrt{\mathrm{k}/2}\delta _{r+s,0}$ \\
$\{\mathbf{\tilde{\chi}}_{r}^{-},\mathbf{\tilde{\chi}}_{s}^{+}\}$ & $=$ & $%
\sqrt{2\mathrm{k}/3}\delta _{r+s,0}$%
\end{tabular}
\label{ps}
\end{equation}%
and normal ordered mode currents as%
\begin{equation}
\begin{tabular}{lll}
$\boldsymbol{J}_{n}$ & $=$ & $\frac{2}{\mathrm{k}}\dsum\limits_{r\in \mathbb{%
Z}+1/2}:\mathbf{\psi }_{n-r}^{+}\mathbf{\psi }_{r}^{-}:$ \\
$\boldsymbol{K}_{n}$ & $=$ & $\frac{3}{2\mathrm{k}}\dsum\limits_{r\in
\mathbb{Z}+1/2}:\mathbf{\chi }_{n-r}^{+}\mathbf{\chi }_{r}^{-}:$%
\end{tabular}%
\qquad ,\qquad
\begin{tabular}{lll}
$\boldsymbol{\tilde{J}}_{n}$ & $=$ & $\frac{2}{\mathrm{k}}\dsum\limits_{r\in
\mathbb{Z}+1/2}:\mathbf{\tilde{\psi}}_{n-r}^{+}\mathbf{\tilde{\psi}}%
_{r}^{-}: $ \\
$\boldsymbol{\tilde{K}}_{n}$ & $=$ & $\frac{3}{2\mathrm{k}}%
\dsum\limits_{r\in \mathbb{Z}+1/2}:\mathbf{\tilde{\chi}}_{n-r}^{+}\mathbf{%
\tilde{\chi}}_{r}^{-}:$%
\end{tabular}
\label{xp}
\end{equation}%
from which we learn the expression of the zero modes, namely%
\begin{equation}
\begin{tabular}{lll}
$\boldsymbol{J}_{0}$ & $=$ & $\frac{2}{\mathrm{k}}\dsum\limits_{r\in \mathbb{%
Z}+1/2}:\mathbf{\psi }_{-r}^{+}\mathbf{\psi }_{r}^{-}:$ \\
$\boldsymbol{K}_{0}$ & $=$ & $\frac{3}{2\mathrm{k}}\dsum\limits_{r\in
\mathbb{Z}+1/2}:\mathbf{\chi }_{-r}^{+}\mathbf{\chi }_{r}^{-}:$%
\end{tabular}%
\qquad ,\qquad
\begin{tabular}{lll}
$\boldsymbol{\tilde{J}}_{0}$ & $=$ & $\frac{2}{\mathrm{k}}\dsum\limits_{r\in
\mathbb{Z}+1/2}:\mathbf{\tilde{\psi}}_{-r}^{+}\mathbf{\tilde{\psi}}_{r}^{-}:$
\\
$\boldsymbol{\tilde{K}}_{0}$ & $=$ & $\frac{3}{2\mathrm{k}}%
\dsum\limits_{r\in \mathbb{Z}+1/2}:\mathbf{\tilde{\chi}}_{-r}^{+}\mathbf{%
\tilde{\chi}}_{r}^{-}:$%
\end{tabular}%
\end{equation}%
These are fermionic numbers acting on the ground state $\left\vert N_{\psi
},N_{\chi }\right\rangle \otimes |\tilde{N}_{\psi },\tilde{N}_{\chi }>$ with
left moving $N_{\psi }+N_{\chi }$ and right moving $\tilde{N}_{\psi }+\tilde{%
N}_{\chi }$ fermions constrained like
\begin{equation}
\begin{tabular}{lll}
$\boldsymbol{J}_{0}\left\vert N_{\psi },N_{\chi }\right\rangle $ & $=$ & $%
N_{\psi }\left\vert N_{\psi },N_{\chi }\right\rangle $ \\
$\boldsymbol{K}_{0}\left\vert N_{\psi },N_{\chi }\right\rangle $ & $=$ & $%
N_{\chi }\left\vert N_{\psi },N_{\chi }\right\rangle $%
\end{tabular}%
\qquad ,\qquad
\begin{tabular}{lll}
$\boldsymbol{\tilde{J}}_{0}\left\vert \tilde{N}_{\psi },\tilde{N}_{\chi
}\right\rangle $ & $=$ & $\tilde{N}_{\psi }\left\vert \tilde{N}_{\psi },%
\tilde{N}_{\chi }\right\rangle $ \\
$\boldsymbol{\tilde{K}}_{0}\left\vert \tilde{N}_{\psi },\tilde{N}_{\chi
}\right\rangle $ & $=$ & $\tilde{N}_{\chi }\left\vert \tilde{N}_{\psi },%
\tilde{N}_{\chi }\right\rangle $%
\end{tabular}%
\end{equation}%
and
\begin{equation}
\begin{tabular}{lll}
$\boldsymbol{J}_{n}\left\vert N_{\psi },N_{\chi }\right\rangle $ & $=$ & $0$
\\
$\boldsymbol{K}_{n}\left\vert N_{\psi },N_{\chi }\right\rangle $ & $=$ & $0$%
\end{tabular}%
\qquad ,\qquad
\begin{tabular}{lll}
$\boldsymbol{\tilde{J}}_{n}\left\vert \tilde{N}_{\psi },\tilde{N}_{\chi
}\right\rangle $ & $=$ & $0$ \\
$\boldsymbol{\tilde{K}}_{n}\left\vert \tilde{N}_{\psi },\tilde{N}_{\chi
}\right\rangle $ & $=$ & $0$%
\end{tabular}%
\end{equation}%
as well as excited states as follows%
\begin{equation}
\left\vert \Psi ,\tilde{\Psi}\right\rangle =\mathcal{N}\dprod\limits_{i}%
\boldsymbol{J}_{-n_{i}}\dprod\limits_{j}^{{\small Y}}\boldsymbol{J}%
_{-m_{j}}\dprod\limits_{k}\boldsymbol{\tilde{J}}_{-\tilde{n}%
_{k}}\dprod\limits_{l}\boldsymbol{\tilde{J}}_{-\tilde{m}_{l}}\left\vert
\tilde{N}_{\psi },\tilde{N}_{\chi }\right\rangle
\end{equation}%
Using the mean value relations%
\begin{equation}
\begin{tabular}{lll}
$\left\langle \Psi ,\tilde{\Psi}|\boldsymbol{J}_{0}|\Psi ,\tilde{\Psi}%
\right\rangle $ & $=$ & $N_{\psi }$ \\
$\left\langle \Psi ,\tilde{\Psi}|\boldsymbol{K}_{0}|\Psi ,\tilde{\Psi}%
\right\rangle $ & $=$ & $N_{\chi }$%
\end{tabular}%
\qquad ,\qquad
\begin{tabular}{lll}
$\left\langle \Psi ,\tilde{\Psi}|\boldsymbol{\tilde{J}}_{0}|\Psi ,\tilde{\Psi%
}\right\rangle $ & $=$ & $\tilde{N}_{\psi }$ \\
$\left\langle \Psi ,\tilde{\Psi}|\boldsymbol{\tilde{K}}_{0}|\Psi ,\tilde{\Psi%
}\right\rangle $ & $=$ & $\tilde{N}_{\chi }$%
\end{tabular}%
\end{equation}%
the entropy of the HS black flowers read as follows%
\begin{equation}
\begin{tabular}{|c|c|}
\hline
real form & $S_{\text{\textsc{hs-bh}}}$ \\ \hline
$su\left( 3\right) _{L_{{\small 0}},W_{{\small 0}}}$ & $2\pi \mathfrak{N}%
\left( N_{\psi }+\tilde{N}_{\psi }\right) +2\pi \mathfrak{M}\left( N_{\chi }+%
\tilde{N}_{\chi }\right) $ \\
$su\left( 2\right) _{L_{{\small 0}}}$ & $2\pi N\left( N_{\psi }+\tilde{N}%
_{\psi }\right) $ \\
$su\left( 2\right) _{W_{{\small 0}}}$ & $3\pi M\left( N_{\chi }+\tilde{N}%
_{\chi }\right) $ \\ \hline\hline
$su(2,1)_{{\small 12}}$ & $2\pi N\left( N_{\psi }+\tilde{N}_{\psi }\right) $
\\
$su(1,2)_{{\small 21}}$ & $3\pi M\left( N_{\chi }+\tilde{N}_{\chi }\right) $
\\ \hline
\end{tabular}
\label{NE}
\end{equation}%
where we have used $\mathfrak{N}_{0}=N+M/2$ and $\mathfrak{M}=M/2+p/3$ and
where $N$ and $M\ $are positive integers.

\subsubsection{Antiperiodic fermions}

In this sector, the fermionic fields ($\psi ^{\pm },\chi ^{\pm }$) obey the
antiperiodicity property
\begin{equation}
\begin{tabular}{lll}
$\psi ^{\pm }\left( \varphi +2\pi \right) $ & $=$ & $-\psi ^{\pm }\left(
\varphi \right) $ \\
$\chi ^{\pm }\left( \varphi +2\pi \right) $ & $=$ & $-\chi ^{\pm }\left(
\varphi \right) $%
\end{tabular}%
\qquad ,\qquad
\begin{tabular}{lll}
$\tilde{\psi}^{\pm }\left( \varphi +2\pi \right) $ & $=$ & $-\tilde{\psi}%
^{\pm }\left( \varphi \right) $ \\
$\tilde{\chi}^{\pm }\left( \varphi +2\pi \right) $ & $=$ & $-\tilde{\chi}%
^{\pm }\left( \varphi \right) $%
\end{tabular}%
\end{equation}%
preserving $\mathcal{J}\left( \varphi +2\pi \right) =\mathcal{J}\left(
\varphi \right) $ and $\mathcal{K}\left( \varphi +2\pi \right) =\mathcal{K}%
\left( \varphi \right) $. These anti-periodicities can be solved by using
the following expansions%
\begin{equation}
\begin{tabular}{lll}
$\psi ^{\pm }$ & $=$ & $\sqrt{\frac{3}{2\mathrm{k}}}\dsum\limits_{n\in
\mathbb{Z}}\psi _{n}^{\pm }e^{-i(n\pm \frac{1}{2})\varphi }$ \\
$\chi ^{\pm }$ & $=$ & $\sqrt{\frac{3}{2\mathrm{k}}}\dsum\limits_{m\in
\mathbb{Z}}\chi _{m}^{\pm }e^{-i(m\pm \frac{1}{2})\varphi }$%
\end{tabular}%
\qquad ,\qquad
\begin{tabular}{lll}
$\tilde{\psi}^{\pm }$ & $=$ & $\sqrt{\frac{3}{2\mathrm{k}}}%
\dsum\limits_{n\in \mathbb{Z}}\tilde{\psi}_{n}^{\pm }e^{-i(n\pm \frac{1}{2}%
)\varphi }$ \\
$\tilde{\chi}^{\pm }$ & $=$ & $\sqrt{\frac{3}{2\mathrm{k}}}%
\dsum\limits_{m\in \mathbb{Z}}\tilde{\chi}_{m}^{\pm }e^{-i(m\pm \frac{1}{2}%
)\varphi }$%
\end{tabular}%
\end{equation}%
\textrm{using} these modes ($\psi _{n}^{\pm },\chi _{n}^{\pm }$) and ($%
\tilde{\psi}_{n}^{\pm },\tilde{\chi}_{n}^{\pm }$), we can express \textrm{%
the }currents $\mathcal{J}_{\varphi },$ $\mathcal{K}_{\varphi }$ and $%
\mathcal{\tilde{J}}_{\varphi },$ $\mathcal{\tilde{K}}_{\varphi }$ like $%
\mathcal{J}_{\varphi }=\sum_{n\in \mathbb{Z}}J_{n}e^{-in\varphi }$ and $%
\mathcal{K}_{\varphi }=\sum K_{n}e^{-in\varphi }$\ as well as $\mathcal{%
\tilde{J}}_{\varphi }=\sum \tilde{J}_{n}e^{+in\varphi }$ and $\mathcal{%
\tilde{K}}_{\varphi }=\sum \tilde{K}_{n}e^{+in\varphi }$ with%
\begin{equation}
\begin{tabular}{lll}
$J_{n}$ & $=$ & $\frac{2}{\mathrm{k}}\dsum\limits_{m\in \mathbb{Z}}\psi
_{n-m}^{+}\psi _{m}^{-}$ \\
$K_{n}$ & $=$ & $\frac{3}{2\mathrm{k}}\dsum\limits_{m\in \mathbb{Z}}\chi
_{n-m}^{+}\chi _{m}^{-}$%
\end{tabular}%
\qquad ,\qquad
\begin{tabular}{lll}
$\tilde{J}_{n}$ & $=$ & $\frac{2}{\mathrm{k}}\dsum\limits_{m\in \mathbb{Z}}%
\tilde{\psi}_{n-m}^{+}\tilde{\psi}_{m}^{-}$ \\
$\tilde{K}_{n}$ & $=$ & $\frac{3}{2\mathrm{k}}\dsum\limits_{m\in \mathbb{Z}}%
\tilde{\chi}_{n-m}^{+}\tilde{\chi}_{m}^{-}$%
\end{tabular}%
\end{equation}%
\begin{equation*}
\end{equation*}%
At the quantum level, the Fourier modes ($\psi _{n}^{\pm },\chi _{n}^{\pm }$%
) and ($\tilde{\psi}_{n}^{\pm },\tilde{\chi}_{n}^{\pm }$)\ are promoted to
operators ($\mathbf{\psi }_{n}^{\pm },\mathbf{\chi }_{n}^{\pm }$) and $(%
\mathbf{\tilde{\psi}}_{n}^{\pm },\mathbf{\tilde{\chi}}_{n}^{\pm })$ \textrm{%
with} non vanishing anticommutation relations%
\begin{equation}
\begin{tabular}{lll}
$\left\{ \mathbf{\psi }_{n}^{-},\mathbf{\psi }_{m}^{+}\right\} $ & $=$ & $%
\sqrt{\mathrm{k}/2}\delta _{n+m,0}$ \\
$\left\{ \mathbf{\chi }_{n}^{-},\mathbf{\chi }_{m}^{+}\right\} $ & $=$ & $%
\sqrt{2\mathrm{k}/3}\delta _{n+m,0}$%
\end{tabular}%
\qquad ,\qquad
\begin{tabular}{lll}
$\left\{ \mathbf{\tilde{\psi}}_{n}^{-},\mathbf{\tilde{\psi}}_{m}^{+}\right\}
$ & $=$ & $\sqrt{\mathrm{k}/2}\delta _{n+m,0}$ \\
$\left\{ \mathbf{\tilde{\chi}}_{n}^{-},\mathbf{\tilde{\chi}}_{m}^{+}\right\}
$ & $=$ & $\sqrt{2\mathrm{k}/3}\delta _{n+m,0}$%
\end{tabular}%
\end{equation}%
each verifies an adjoint conjugation like $\left( \mathbf{\psi }%
_{n}^{-}\right) ^{\dagger }=\mathbf{\psi }_{-n}^{+}$. The corresponding
current modes are%
\begin{equation}
\begin{tabular}{lll}
$\boldsymbol{J}_{n}$ & $=$ & $\frac{2}{\mathrm{k}}\dsum\limits_{m\in \mathbb{%
Z}}:\mathbf{\psi }_{n-m}^{+}\mathbf{\psi }_{m}^{-}:$ \\
$\boldsymbol{K}_{n}$ & $=$ & $\frac{3}{2\mathrm{k}}\dsum\limits_{m\in
\mathbb{Z}}:\mathbf{\chi }_{n-m}^{+}\mathbf{\chi }_{m}^{-}:$%
\end{tabular}%
\qquad ,\qquad
\begin{tabular}{lll}
$\boldsymbol{\tilde{J}}_{n}$ & $=$ & $\frac{2}{\mathrm{k}}\dsum\limits_{m\in
\mathbb{Z}}:\mathbf{\tilde{\psi}}_{n-m}^{+}\mathbf{\tilde{\psi}}_{m}^{-}:$
\\
$\boldsymbol{\tilde{K}}_{n}$ & $=$ & $\frac{3}{2\mathrm{k}}%
\dsum\limits_{m\in \mathbb{Z}}:\mathbf{\tilde{\chi}}_{n-m}^{+}\mathbf{\tilde{%
\chi}}_{m}^{-}:$%
\end{tabular}%
\end{equation}%
from which we learn the expression of the zero modes%
\begin{equation}
\begin{tabular}{lll}
$\boldsymbol{J}_{0}$ & $=$ & $\frac{2}{\mathrm{k}}\dsum\limits_{m\in \mathbb{%
Z}}:\mathbf{\psi }_{-m}^{+}\mathbf{\psi }_{m}^{-}:$ \\
$\boldsymbol{K}_{0}$ & $=$ & $\frac{3}{2\mathrm{k}}\dsum\limits_{m\in
\mathbb{Z}}:\mathbf{\chi }_{-m}^{+}\mathbf{\chi }_{m}^{-}:$%
\end{tabular}%
\qquad ,\qquad
\begin{tabular}{lll}
$\boldsymbol{\tilde{J}}_{0}$ & $=$ & $\frac{2}{\mathrm{k}}\dsum\limits_{m\in
\mathbb{Z}}:\mathbf{\tilde{\psi}}_{-m}^{+}\mathbf{\tilde{\psi}}_{m}^{-}:$ \\
$\boldsymbol{\tilde{K}}_{0}$ & $=$ & $\frac{3}{2\mathrm{k}}%
\dsum\limits_{m\in \mathbb{Z}}:\mathbf{\tilde{\chi}}_{-m}^{+}\mathbf{\tilde{%
\chi}}_{m}^{-}:$%
\end{tabular}%
\end{equation}%
with $\left\langle N_{\psi },N_{\chi }|\boldsymbol{J}_{0}|N_{\psi },N_{\chi
}\right\rangle =N_{\psi }$ and $\left\langle N_{\psi },N_{\chi }|\boldsymbol{%
K}_{0}|N_{\psi },N_{\chi }\right\rangle =N_{\chi }$ and higher spin black
hole entropies as in eq(\ref{NE}).

\section{Conclusion and Discussions}

\label{conc}To bring this paper to a close, we summarise our main results
along with a few additional comments. Setting out with higher spin black
flowers solutions of\emph{\ }3D Anti-de Sitter gravity with $\left( \mathbf{1%
}\right) $ generalised boundary conditions as in eq(\ref{bb}) and ($\mathbf{2%
}$) gauge symmetry defined by real forms of the complexified $SL(N,\mathbb{C}%
)_{\mathtt{L}}\times SL(N,\mathbb{C})_{\mathtt{R}}$ group, we considered non
trivial holonomies $W_{\left[ \mathcal{C}_{t}\right] }=P\exp (i\int_{%
\mathcal{C}_{t}}a)$\emph{\ }taken values in the discrete group $\mathbb{Z}%
_{N}$ with boundary potentials circling the thermal cycle $\left[ \mathcal{C}%
_{t}\right] .$ We showed that these HS black flowers are in fact\emph{\ }of
various types\emph{\ }depending on the selected real form of the complex $%
SL(N,\mathbb{C})\ $whether it is the real split $SL(N,\mathbb{R}),$ the
compact $SU(N)$ or one of the several non compact $SU(N_{1},N_{2})$. We also
characterised the BH flowers by two infinite sets of conserved charges, one
left $\mathtt{I}_{r}^{\text{\textsc{l}}}$ and another right $\mathtt{I}_{r}^{%
\text{\textsc{r}}}$ ($r\in \mathbb{Z}$), obeying an affine Heisenberg- like
algebra. These (left/right) gauge invariants $\mathtt{I}_{r}$ expand like $%
\sum_{s=2}^{N}\Lambda ^{{\small (s-1)}}\mathcal{W}_{r}^{{\small (s)}}$ where
the $\Lambda ^{{\small (s-1)}}$ are real Lagrange multipliers and where $%
\mathcal{W}_{r}^{{\small (s)}}$\emph{\ }are the $(N-1)$ affine GL(1)$_{%
\mathrm{k}}$ charges of the underlying affine $SL(N)_{\mathrm{k}}$. Focusing
on the compact $SU(N)$ theory [similar discussion follows for the $%
SU(N_{1},N_{2})$] and demanding non trivial holonomies sitting in the centre
of the gauge symmetry, the conserved charges $\mathtt{I}_{r}$ as well as the
associated Lagrange multipliers $\Lambda _{n_{s},p_{N}}^{{\small (s-1)}}$
become quantized values with $n_{s}\in \mathbb{Z}$ parametrising higher
spins and an additional charges $p_{N}=0,...,N-1$ $\func{mod}N$ sweeping the
N group elements of the group centre $\mathbb{Z}_{N}$. The additional
charges index gauge invariant physical states that we have interpreted in
terms of layered HS black hole flowers. \textrm{For an in-depth discussion
on the physical significance of this new sector arising from the centre }$%
\mathbb{Z}_{N}$ \textrm{symmetry, consult \autoref{appC} for more details.}

Furthermore, by borrowing ideas from constrained Hamiltonian systems, we $%
\left( \mathbf{a}\right) $ developed an approach to address the vanishing
condition of the gauge curvature on the boundary surface of the AdS$_{3}$
space in order to compute the conserved $\mathtt{I}_{r}\left[ \Lambda \right]
;$ and $\left( \mathbf{b}\right) $ ascertained that the HS black flowers
entropy is intimately related to\textrm{\ }the invariants
\begin{equation}
\mathcal{I}_{\text{\textsc{hs-bh}}}^{+}=2\pi |\mathtt{I}_{0}^{\text{\textsc{l%
}}}+\mathtt{I}_{0}^{\text{\textsc{r}}}|\qquad ,\qquad \mathcal{I}_{\text{%
\textsc{hs-bh}}}^{-}=2\pi |\mathtt{I}_{0}^{\text{\textsc{l}}}-\mathtt{I}%
_{0}^{\text{\textsc{r}}}|  \label{up}
\end{equation}%
via the upper bound given by
\begin{equation}
\text{\textsc{S}}_{\text{\textsc{hs-bh}}}^{\text{\textsc{sl}}_{N}}=2\pi |%
\mathtt{I}_{0}^{\text{\textsc{l}}}|+2\pi |\mathtt{I}_{0}^{\text{\textsc{r}}}|
\end{equation}%
Assuming that $\Lambda _{n_{s},p}^{\text{\textsc{l}}}=\Lambda
_{n_{s},p_{N}}^{\text{\textsc{r}}}$ $\geq 0$, the entropy \textsc{S}$_{\text{%
\textsc{hs-bh}}}^{\text{\textsc{sl}}_{N}}$ can be factorised like
\begin{equation}
\begin{tabular}{lll}
$\text{\QTR{cal}{S}}_{\text{\textsc{hs-bh}}}^{\text{\textsc{sl}}_{2}}$ & $=$
& $2\pi \lambda _{n_{2},p_{2}}\left( \left\vert \mathcal{W}_{0}^{{\small (2)}%
}\right\vert +\left\vert \mathcal{\tilde{W}}_{0}^{{\small (2)}}\right\vert
\right) $ \\
&  &  \\
$\text{\QTR{cal}{S}}_{\text{\textsc{hs-bh}}}^{\text{\textsc{sl}}_{3}}$ & $=$
& $2\pi \Lambda _{n_{2},p_{2}}\left( \left\vert \mathcal{W}_{0}^{{\small (2)}%
}\right\vert +\left\vert \mathcal{\tilde{W}}_{0}^{{\small (2)}}\right\vert
\right) +3\pi \Lambda _{n_{3},p_{3}}\left( \left\vert \mathcal{W}_{0}^{%
{\small (3)}}\right\vert +\left\vert \mathcal{\tilde{W}}_{0}^{{\small (3)}%
}\right\vert \right) $%
\end{tabular}%
\end{equation}%
\begin{equation*}
\end{equation*}%
with $\lambda _{n_{2},p_{2}}=n_{2}+p_{2}/2$ for the SU$_{2}$ theory and $%
\Lambda _{n_{2},p_{2}}=n_{2}+n_{3}/2$ as well as $\Lambda
_{n_{3},p_{3}}=n_{3}/2+p_{3}/3$ for SU$_{3}$. This entropy comprises the
familiar Grumiller- Perez- Prohazka- Tempo- Troncoso (GPPTT) entropy S$_{%
\text{\textsc{hs-bh}}}^{\text{\textsc{gpptt}}}$ as a particular phase
associated with the trivial holonomy where $p_{{\small N}}=0$ mod N.

At the quantum level, the classical invariants $\mathtt{I}_{0}$ and $\mathtt{%
\tilde{I}}_{0}$ get promoted to operators denoted like $\boldsymbol{I}%
_{0}=\sum_{s=2}^{N}\Lambda _{\text{\textsc{l}}}^{{\small (s-1)}}\boldsymbol{W%
}_{0}^{{\small (s)}}$ and $\boldsymbol{\tilde{I}}_{0}=\sum_{s=2}^{N}\Lambda
_{\text{\textsc{r}}}^{{\small (s-1)}}\boldsymbol{\tilde{W}}_{0}^{{\small (s)}%
}\ $implying an interpretation of the HS-BF in terms of the vacuum
expectation value. The operator $\boldsymbol{I}_{0}$ (similarly for $%
\boldsymbol{\tilde{I}}_{0}$) is taken in the Cartan subalgebra of SL(N)$_{0}$
which is the zero mode of the affine SL(N)$_{\mathrm{k}}$. Thus, given a
highest weight representation state $\left\vert \mathbf{q}\right\rangle $ of
SL(N)$_{0}$ labelled by the $N-1$ charges like $\left\vert q_{{\small 2}%
},...,q_{{\small N}}\right\rangle $ with eigenvalue equation $\boldsymbol{W}%
_{0}^{{\small (s)}}\left\vert \mathbf{q}\right\rangle =q_{{\small s}%
}\left\vert \mathbf{q}\right\rangle $ and positive integer $q_{{\small s}},$
the invariant $\mathtt{I}_{0}$ can be considered as the mean $\left\langle
\mathbf{q}\right\vert \boldsymbol{I}_{0}\left\vert \mathbf{q}\right\rangle
=\sum_{s=2}^{N}q_{{\small s}}\Lambda ^{{\small (s-1)}}$ leading to the \emph{%
discrete }HS black flower entropy relation%
\begin{equation}
\text{\textsc{S}}_{\text{\textsc{hs-bh}}}^{\text{\textsc{sl}}_{N}}=2\pi
\sum_{s=2}^{N}\left( q_{{\small s}}+\tilde{q}_{{\small s}}\right) |\Lambda
_{n_{s},p_{N}}^{{\small (s-1)}}|  \label{qq}
\end{equation}%
For SU$_{2}$ of the spin 2 gravity model, we get $2\pi \left( q_{{\small 2}}+%
\tilde{q}_{{\small 2}}\right) \lambda _{n_{2},p_{2}}$ with $\lambda
_{n_{2},p_{2}}=n_{2}+p_{2}/2$ where $n_{2}$ is an integer and $p=0,1$ $\func{%
mod}2.$ As for the higher spin SU$_{3}$ theory, we obtain $2\pi \left( q_{%
{\small 2}}+\tilde{q}_{{\small 2}}\right) \Lambda _{n_{2},p_{2}}^{{\small (1)%
}}+2\pi \left( q_{{\small 3}}+\tilde{q}_{{\small 3}}\right) \Lambda
_{n_{3},p_{3}}^{{\small (2)}}$ where $\Lambda _{n_{2},p_{2}}^{{\small (1)}%
}=n_{2}+n_{3}/2$ and $\Lambda _{n_{3},p_{3}}^{{\small (2)}}=n_{3}/2+p_{3}/3$
with $n_{2},n_{3}$ being integers and $p=0,1,2$ $\func{mod}3.$ Within this
perspective, the generators $\boldsymbol{W}_{0}^{{\small (s)}}$ can be
interpreted in terms of number operators with eigenvalues N$_{s}^{\text{%
\textsc{particles}}}$ (bosonic N$_{s}^{\text{\textsc{bose}}}$ and/or
fermionic N$_{s}^{\text{\textsc{fermi}}}$). For instance, the ground state
with N$_{s}^{\text{\textsc{fermi}}}$ fermions gives%
\begin{equation}
\text{\textsc{S}}_{\text{\textsc{hs-bh}}}^{\text{\textsc{sl}}_{N}}=4\pi
\sum_{s=2}^{N}N_{s}^{\text{\textsc{fermi}}}|\Lambda _{n_{s},p_{N}}^{{\small %
(s-1)}}|
\end{equation}

To carry out explicit calculations, we mainly focused on the leading values $%
N=2,3,4$ of the SL(N) family; however, the construction holds for all N.
Now, the 2D boundary potentials $a^{\text{\textsc{diag}}}=a_{\varphi
}d\varphi +a_{t}dt$ and $\tilde{a}^{\text{\textsc{diag}}}=\tilde{a}_{\varphi
}d\varphi +\tilde{a}_{t}dt$ are generally valued in the Cartan subalgebra of
the gauge symmetry as exhibited by eqs(\ref{dg}-\ref{da}). For the example
of the HS $SL(3,\mathbb{R})$ model, we have $a_{\varphi /t}=a_{\varphi
/t}^{0}L_{0}+a_{\varphi /t}^{0}W_{0}$ and $\tilde{a}_{\varphi /t}=\tilde{a}%
_{\varphi /t}^{0}L_{0}+\tilde{a}_{\varphi /t}^{0}W_{0}$ with L$_{0}$ and W$%
_{0}$ being related to the usual commuting Cartan H$_{\alpha _{1}}$, H$%
_{\alpha _{2}}$ generators as in (\ref{HH}). And because $[a_{\varphi
},a_{t}]=0$ and $[\tilde{a}_{\varphi },\tilde{a}_{t}]=0,$ the flatness
conditions of the non abelian 2D gauge curvatures reduce to $da^{\text{%
\textsc{diag}}}=d\tilde{a}^{\text{\textsc{diag}}}=0$ giving the dynamical
constraints $\partial _{t}a_{\varphi }-\partial _{\varphi }a_{t}=0$ and $%
\partial _{t}\tilde{a}_{\varphi }-\partial _{\varphi }\tilde{a}_{t}=0$.
Assuming constant Lagrange multipliers ($a_{t}=cte,$ $\tilde{a}_{t}=cte$),
these relations become $\partial _{t}a_{\varphi }=0$ and $\partial _{t}%
\tilde{a}_{\varphi }=0$.

In our approach, we considered the above conditions $\partial _{t}a_{\varphi
}=0$ and $\partial _{t}\tilde{a}_{\varphi }=0$ as fields equations of motion
following from a variational principle with 2D boundary field functionals $%
\mathcal{S}_{\text{\textsc{bnd}}}^{\pm }$ interpreted in terms of total
Hamiltonian and total angular momentum. These boundary quantities are
explicitly given by eqs(\ref{eq})-(\ref{int}) and can be shortly presented
like%
\begin{equation}
\mathcal{S}_{\text{\textsc{bnd}}}^{\pm }=\frac{\mathrm{k}}{8\pi }%
\dint\nolimits_{0}^{\mathrm{\beta }}dt_{{\small E}}\dint\nolimits_{0}^{2\pi
}d\varphi Tr\left( \upsilon \partial _{t_{{\small E}}}a_{\varphi }\pm \tilde{%
\upsilon}\partial _{t_{{\small E}}}\tilde{a}_{\varphi }\right)
\end{equation}%
where the Lagrange multipliers are taken as $\upsilon =\lambda
^{0}L_{0}+\Lambda ^{0}W_{0}.$ The boundary fields $a_{\varphi }\left(
\varphi ,t_{{\small E}}\right) $\ (and equivalently $\tilde{a}_{\varphi
}\left( \varphi ,t_{{\small E}}\right) )$ are time dependent and read
explicitly like $-\frac{\mathrm{\beta }}{2}a_{\varphi }\left( \varphi
\right) $ $\times $ $\cos (\frac{\pi }{\mathrm{\beta }}t_{{\small E}})$ with
the property $a_{\varphi }\left( \varphi ,t_{{\small E}}=\mathrm{\beta }%
\right) =-a_{\varphi }\left( \varphi ,t_{{\small E}}=\mathrm{0}\right) =%
\frac{\mathrm{\beta }}{2}a_{\varphi }\left( \varphi \right) $ giving
vanishing velocities $\partial _{t_{{\small E}}}a_{\varphi }\left( \varphi
,t_{{\small E}}\right) =0$ and $\partial _{t_{{\small E}}}\tilde{a}_{\varphi
}\left( \varphi ,t_{{\small E}}\right) =0.$

With this construction, we were able to successfully replicate the
previously established results of 3D BH literature and gain further insights
on the different branches of black flowers considered in \cite{1B}. Contrary
to what was shown in \cite{1B} for trivial holonomies corresponding to $%
\omega ^{p=0}=I_{id}$, non trivial holonomies discretize the HS black holes
descent towards spin 2 black holes as there are no continuous links joining
the black flowers entropy to the spin 2's one. In consequence, to get to the
core BTZ BH, one requires discrete transitions generated by a Weyl
transformation of the compact real from of the complexified gauge symmetry
as given by (\ref{wt}).

\textrm{Additionally}, we uncovered a close-knit relationship between the
entropy calculation of HS soft hair BHs and the real forms of the
complexified gauge symmetry. To study the thermodynamical properties of
black holes, we must switch to the Euclidean framework with time coordinate $%
t_{{\small E}}=it$ which leads to a complexification of the gauge symmetry
and then to one of its real forms as described by eqs(\ref{34})-(\ref{38})
for $SL(3,\mathbb{R})$ as well as eqs(\ref{316})-(\ref{321}).

Focusing on the HS model with $SL(3,\mathbb{R})_{\mathtt{L}}\times SL(3,%
\mathbb{R})_{\mathtt{R}}$, and demanding regularity of the boundary
potentials on the thermal 1-cycle, we computed the black flower entropies
for the various compact SU(2) real forms sitting inside the complexified
gauge symmetry $SL(3,\mathbb{C})$. The values of these entropies are
collected in the tables (\ref{EE}-\ref{28}). And by using the vacuum
expectation values of the classical conserved charges $\mathtt{I}_{0}^{\text{%
\textsc{l}}}$ and $\mathtt{I}_{0}^{\text{\textsc{r}}}$ as in (\ref{qq}), the
black flowers entropies become%
\begin{equation}
\begin{tabular}{c|c}
real form & $\text{\textsc{S}}_{\text{\textsc{hs-bh}}}^{\text{\textsc{sl}}%
_{3}}$ \\ \hline\hline
\multicolumn{1}{|c|}{$\ \ \ su\left( 3\right) _{L_{{\small 0}},W_{{\small 0}%
}}$ \ \ \ \ } & \multicolumn{1}{|c|}{$\ \ \ \ 2\pi \left\vert \mathfrak{N}%
|\left( q_{2}+\tilde{q}_{2}\right) +|\mathfrak{M}\right\vert \left( q_{3}+%
\tilde{q}_{3}\right) $ \ \ \ \ \ \ \ \ } \\ \hline
\multicolumn{1}{|c|}{$su\left( 2\right) _{L_{{\small 0}}}$} &
\multicolumn{1}{|c|}{$2\pi \left\vert N\right\vert \left( q_{2}+\tilde{q}%
_{2}\right) $} \\ \hline
\multicolumn{1}{|c|}{$su\left( 2\right) _{W_{{\small 0}}}$} &
\multicolumn{1}{|c|}{$3\pi \left\vert M\right\vert \left( q_{3}+\tilde{q}%
_{3}\right) $} \\ \hline\hline
\multicolumn{1}{|c|}{$su(2,1)_{{\small 12}}$} & \multicolumn{1}{|c|}{$2\pi
\left\vert N\right\vert \left( q_{2}+\tilde{q}_{2}\right) $} \\
\multicolumn{1}{|c|}{$su(1,2)_{{\small 21}}$} & \multicolumn{1}{|c|}{$3\pi
\left\vert M\right\vert \left( q_{3}+\tilde{q}_{3}\right) $} \\ \hline\hline
\end{tabular}%
\end{equation}%
where $\mathfrak{N}=N+M/2$ and $\mathfrak{M}=M/2+p/3$ with $N$, $M$ integers
\textrm{and} $p=0,1,2$ $\func{mod}3.$ The objects ($q_{2},q_{3}$) and ($%
\tilde{q}_{2},\tilde{q}_{3}$) designate the charges of the ground state
associated to the zero modes of the four affine \^{u}(1)$_{\mathrm{k}}$
Heisenberg current algebras. \textrm{Notice that the lowest absolute value} $%
\left\vert \mathfrak{M}\right\vert $ \textrm{is equal to} $1/6$; \textrm{it
corresponds to} $\left( M,p\right) =\left( -1,1\right) .$

We concluded our study by giving $\left( \mathbf{i}\right) $ a bosonic field
realisation in terms of free left ($X,Y$) and right ($\tilde{X},\tilde{Y}$)
bosons. and $\left( \mathbf{ii}\right) $ a fermionic representation using
left ($\psi ^{\pm },\chi ^{\pm }$) and right ($\tilde{\psi}^{\pm },\tilde{%
\chi}^{\pm }$) (anti) periodic fermions. Interestingly, for a fermionic
ground state $\left\vert N_{\psi },N_{\chi }\right\rangle $, we found
amongst others that the black flower entropy is given by$\ 2\pi |\mathfrak{N}%
|(N_{\psi }+\tilde{N}_{\psi })+2\pi |\mathfrak{M}|(N_{\chi }+\tilde{N}_{\chi
})$ where ($N_{\psi },N_{\chi }$) and ($\tilde{N}_{\psi },\tilde{N}_{\chi }$%
) define the fermionic densities of the ground state of the HS soft hair
black hole.

As a future perspective, we intend to develop the free field realisation to
probe the near horizon dynamics of HS black flowers as a HS extension to the
work conducted in \cite{NH} based on the Hamiltonian reduction approach. In
\cite{NH}, the authors considered spin 2 gravity with near horizon boundary
conditions $a_{\varphi }^{\text{\textsc{diag}}}=\mathcal{J}_{{\small (2)}%
}^{0}W_{0}^{{\small (2)}}$ and $a_{t}^{\text{\textsc{diag}}}=\lambda _{%
{\small (2)}}^{0}W_{0}^{{\small (2)}}.$ And then, using the Hamiltonian
reduction method, they brought the AdS$_{3}$ gravity action to the form%
\begin{equation}
S_{NH}\left[ \Phi _{n},P_{n}\right] =\int dt\left( \sum_{n\geq 0}P_{n}%
\overset{.}{\Phi }_{n}-H_{NH}\right)  \label{nha}
\end{equation}%
with near horizon hamiltonian $H_{NH}$ given by the $J_{0}$ mode\ of the
current $\mathcal{J}_{{\small (2)}}^{0}.$ The solution to the equations of
motion of the above action obtained in \cite{NH} are similar to the primary
scalar solution (\ref{X1}) of the charge potential. In this regard, one can
gain great insights as well as substantial advantages from the free field
framework like: $\left( i\right) $ more simple dynamics as given by the
action (\ref{nha}). $\left( ii\right) $ Treatment of the soft hair states
using conformal field theory in accordance with the AdS$_{3}$/CFT$_{2}$
correspondence. $\left( iii\right) $\ Lifting the degeneracy of the soft
hair excitations via the KdV hierarchy of modified boundary conditions. $%
\left( iv\right) $\ Using fermionisation ideas, it is interesting to study
the link between (\ref{nha}) and the free fermionic realisation we gave in
subsection 5.2. Progress in these directions in relation to HS black flowers
will be reported in a later occasion.

\begin{equation*}
\end{equation*}%
\appendix

\section*{Appendices}

\qquad We give three appendices A, B and C where we report some technical
details. In appendix A, we construct HS theories having gauge symmetries
given by real forms of complex Lie algebras. In appendix B, we derive the
asymptotic algebra obeyed by the edge invariants operators. And in appendix
C, we elaborate on the physical significance of the $\mathbb{Z}_{N}$ centre
symmetry giving rise to non trivial holonomies of boundary potentials of HS
black flowers.

\section{HS theories with real forms of complex Lie algebras}

\label{appA}\textrm{Here, we} \textrm{describe} the construction of real
forms of complex Lie groups (algebras) through the leading elements of the $%
SL(N,\mathbb{C)}$ series by using graphical illustrations given by
Tits-Satake graphs. Recall that $SL(N,\mathbb{C})$ has one Dynkin diagram
with $N-1$ nodes. However it has N Tits-Satake diagrams corresponding to the
N real forms counting\textrm{\ }$\left( \mathbf{1}\right) $ the real split
form $SL(N,\mathbb{R}),$ $\left( \mathbf{2}\right) $\ the compact real form $%
SU(N)$ and $\left( \mathbf{3}\right) $ the non compact $SU(N_{1},N_{2})$s
with $N_{1}+N_{2}=N$ containing amongst others the particular subgroups%
\begin{equation}
SU(N_{1})\times SU(N_{2})\subset S\left[ U(N_{1})\times U(N_{2})\right]
\subset SU(N_{1},N_{2})
\end{equation}%
The leading member in the $SL(N,\mathbb{C)}$ family is the $SL(2,\mathbb{C)}$
group which has two real forms: $\left( \mathbf{a}\right) $ the real split
form $SL(2,\mathbb{R})\simeq SU(1,1)$ and $\left( \mathbf{b}\right) $ the
compact $SU(2).$ Using the complex variables (z$_{1}$,z$_{2}$)
parameterising $\mathbb{C}^{2}$, the global version of the groups $SU(2)$
and $SU(1,1)$\ can be defined as the set of linear unitary transformations
leaving invariant the following hermitian quadratic forms%
\begin{equation}
SU(2):z_{1}z_{1}^{\ast }+z_{2}z_{2}^{\ast }\qquad ,\qquad
SU(1,1):z_{1}z_{1}^{\ast }-z_{2}z_{2}^{\ast }
\end{equation}%
The\textrm{\ }three generators $J_{a}^{{\small sl}_{{\small 2}}}$ of the
real forms of $SL(2,\mathbb{C})$ can be formulated in terms of\emph{\ }the
usual Cartan-Weyl (Chevalley) operators $\left( H_{\alpha },E_{\pm \alpha
}\right) $ as follows%
\begin{equation}
\begin{tabular}{|c||c|c|c|c|}
\hline
generators & $J_{1}^{{\small sl}_{{\small 2}}}$ & $J_{2}^{{\small sl}_{%
{\small 2}}}$ & $J_{3}^{{\small sl}_{{\small 2}}}$ & $g_{{\small abelian}}$
\\ \hline
$sl(2,\mathbb{R})$ & $E_{+\alpha }+E_{-\alpha }$ & $E_{+\alpha }-E_{-\alpha
} $ & $H_{\alpha }$ & $e^{\theta _{\alpha }H_{\alpha }}$ \\ \hline
$su\left( 2\right) $ & $i\left( E_{+\alpha }+E_{-\alpha }\right) $ & $%
E_{+\alpha }-E_{-\alpha }$ & $iH_{\alpha }$ & $e^{i\theta _{\alpha
}H_{\alpha }}$ \\ \hline
\end{tabular}
\label{tj}
\end{equation}%
\begin{equation*}
\end{equation*}%
\textrm{with} $\alpha $ \textrm{being} the usual simple root of $sl(2,%
\mathbb{C}$)$_{\alpha }$, introduced here for later use as we aim towards a $%
sl(N,\mathbb{C}$) generalization; and where:

\begin{description}
\item[$\left( \mathbf{i}\right) $] the hermitian $g_{{\small abelian}}^{%
\mathbb{R}}=e^{\theta _{\alpha }H_{\alpha }}$ with real parameter $\theta
_{\alpha }$ \textrm{of the} non compact group $GL(1,\mathbb{R})$ has the form%
\footnote{%
\ This is an element of the real $SL(1,\mathbb{R})$ which is a subgroup of $%
GL(1,\mathbb{R}).$}%
\begin{equation}
g_{{\small abelian}}^{\mathbb{R}}=\left(
\begin{array}{cc}
e^{\theta _{\alpha }} & 0 \\
0 & e^{-\theta _{\alpha }}%
\end{array}%
\right)
\end{equation}

\item[$\left( \mathbf{ii}\right) $] the complex $g_{{\small abelian}}^{i%
\mathbb{R}}=e^{i\theta _{\alpha }^{\prime }H_{\alpha }}$ with real parameter
$\theta _{\alpha }^{\prime }$\textrm{\ of the} the compact $U(1)$ taking
values in $\left[ 0,2\pi \right] $ is as%
\begin{equation}
g_{{\small abelian}}^{i\mathbb{R}}=\left(
\begin{array}{cc}
e^{i\theta _{\alpha }^{\prime }} & 0 \\
0 & e^{-i\theta _{\alpha }^{\prime }}%
\end{array}%
\right)
\end{equation}
\end{description}

\ \ \newline
Notice that the composition of these two group elements can be combined into
the complex $g_{{\small abelian}}^{\mathbb{C}}=e^{z_{\alpha }H_{\alpha }}\in
GL(1,\mathbb{C})_{\alpha }$ with complex parameter $z_{\alpha }=\theta
_{\alpha }+i\theta _{\alpha }^{\prime }$ such that
\begin{equation}
g_{{\small abelian}}^{\mathbb{C}}=\left(
\begin{array}{cc}
e^{z_{\alpha }} & 0 \\
0 & e^{-z_{\alpha }}%
\end{array}%
\right)
\end{equation}%
Notice also that the $SL(2,\mathbb{R})$ and the $SU(2)$ can be nicely
discriminated via the Cartan involution $\vartheta $ as it operates\emph{\ }%
differently on their Cartan generators \textrm{\cite{ST, ST2},}%
\begin{equation}
\begin{tabular}{|c|cc||c|c|c|}
\hline
\multicolumn{2}{|c}{Cartan involution $\vartheta $} &  & $\alpha $ & $%
H_{\alpha }$ & $E_{+\alpha }$ \\ \hline
compact $SU(2)_{\alpha }$ & $\vartheta =+1$ &  & $+\alpha $ & $H_{\alpha }$
& $E_{-\alpha }$ \\ \hline
real split $SL(2,\mathbb{R})_{\alpha }$ & $\vartheta =-1$ &  & $-\alpha $ & $%
-H_{\alpha }$ & $-E_{-\alpha }$ \\ \hline
\end{tabular}%
\end{equation}%
\textrm{\ }thus permitting to draw different characteristic graphs known as
Tits-Satake diagrams. The Tits-Satake diagrams of the real split form\textrm{%
\ }$sl\left( 2,\mathbb{R}\right) $\textrm{\ }and the complex\textrm{\ }$%
su\left( 2\right) $\textrm{\ }have one (un)painted node as depicted in the
last column of the following Table%
\begin{equation}
\begin{tabular}{|c|c|c|c|}
\hline
$sl(2,C)$ real forms & Cartan subalgebra & \multicolumn{2}{|c|}{Tits-Satake
diagram} \\ \hline
$su\left( 2\right) \simeq so\left( 3\right) $ & $u\left( 1\right) \simeq
so\left( 2\right) $ & black node & \includegraphics[width=0.7cm]{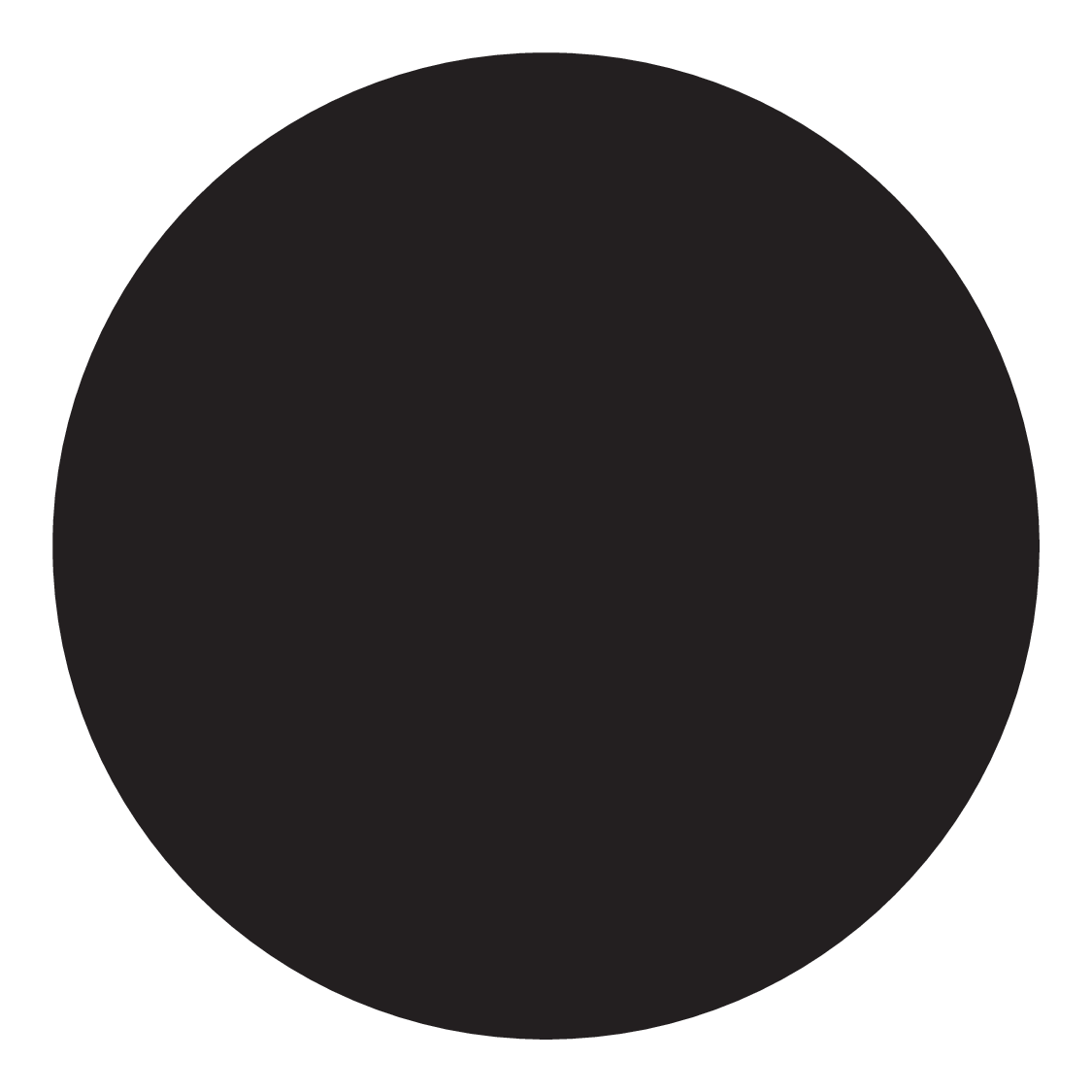} \\
\hline
$sl\left( 2,\mathbb{R}\right) \simeq so\left( 1,2\right) $ & $gl\left( 1,%
\mathbb{R}\right) \simeq so\left( 1,1\right) $ & white node & %
\includegraphics[width=0.7cm]{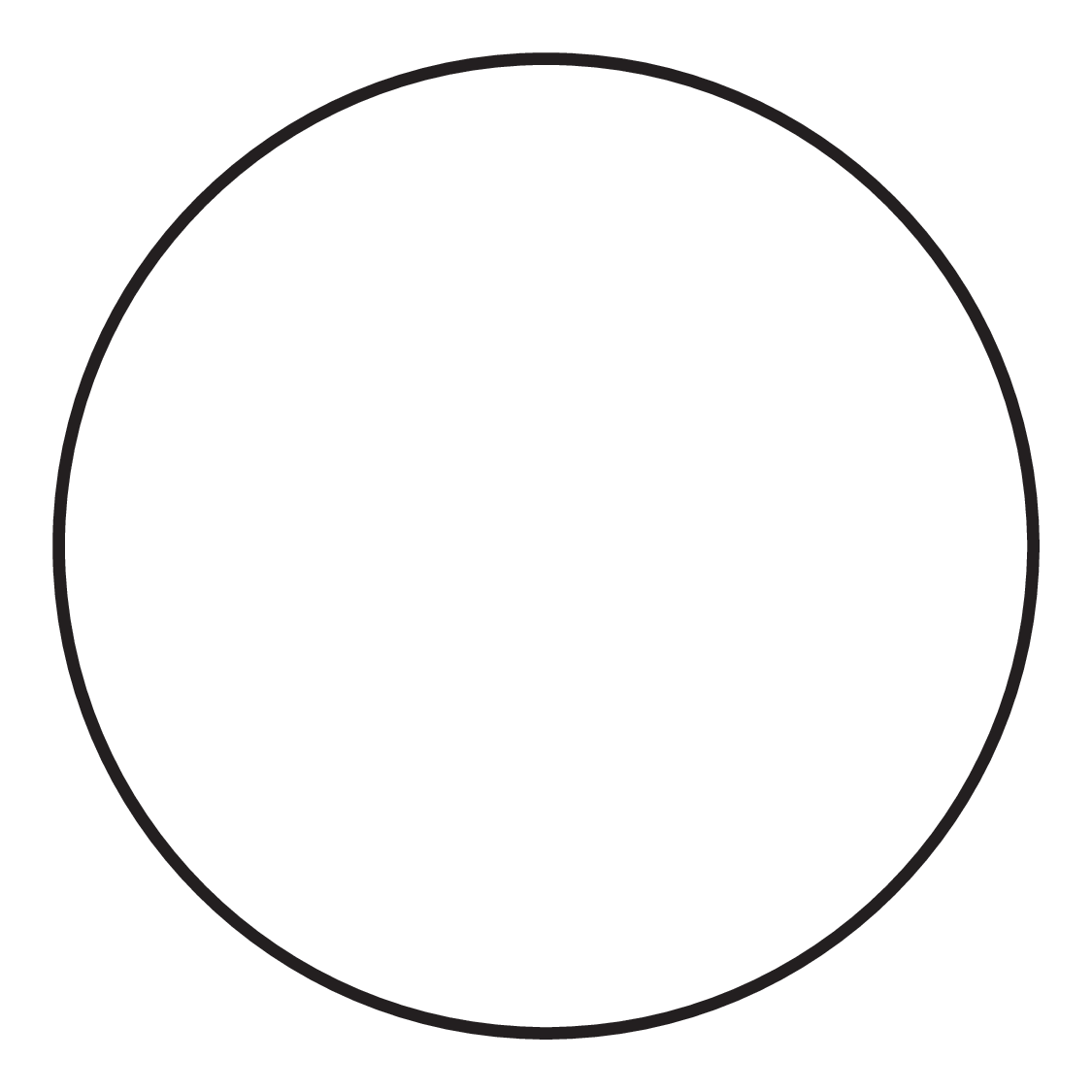} \\ \hline
\end{tabular}%
\end{equation}%
\begin{equation*}
\end{equation*}%
From this \textrm{graphical} representation,\emph{\ }\textrm{it is evident
that}\emph{\ }a black node refers to an $SU(2)$ compact group while a white
node designates an $SL\left( 2,\mathbb{R}\right) \simeq SU(1,1)$ non compact
group.

\subsection*{Case of SL(3) model}

\begin{figure}[tbph]
\begin{center}
\includegraphics[width=10cm]{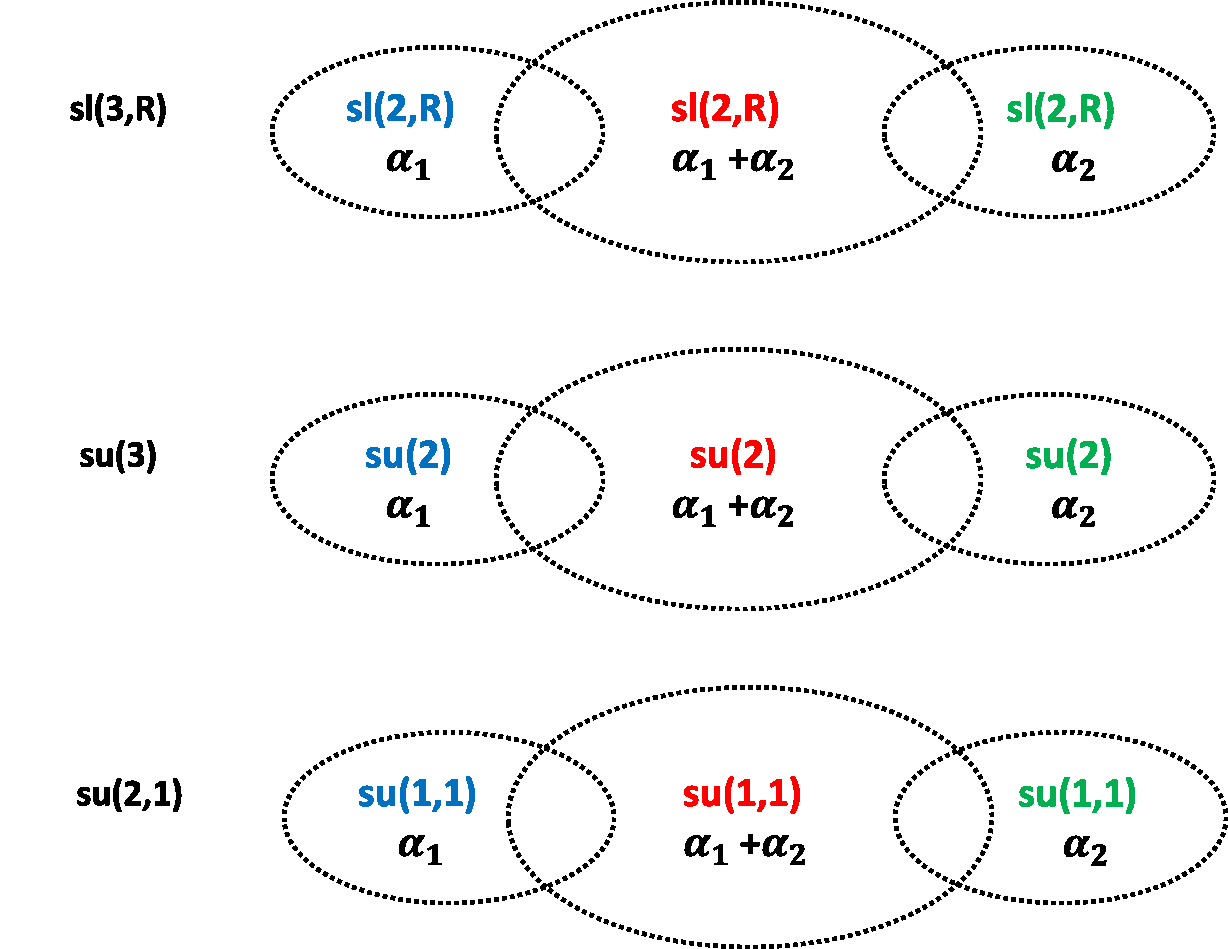}
\end{center}
\par
\vspace{0.0cm}
\caption{A sketch of the three real forms of $sl(3,\mathbb{C})$ containing: $%
(i)$ the real split form $sl(3,\mathbb{R}),$ $(ii)$ the real compact $su(3,%
\mathbb{C})$ and ($iii$) the $su(2,1)\simeq su(1,2).$}
\label{subgr}
\end{figure}

\begin{figure}[tbp]
\begin{center}
\includegraphics[width=10cm]{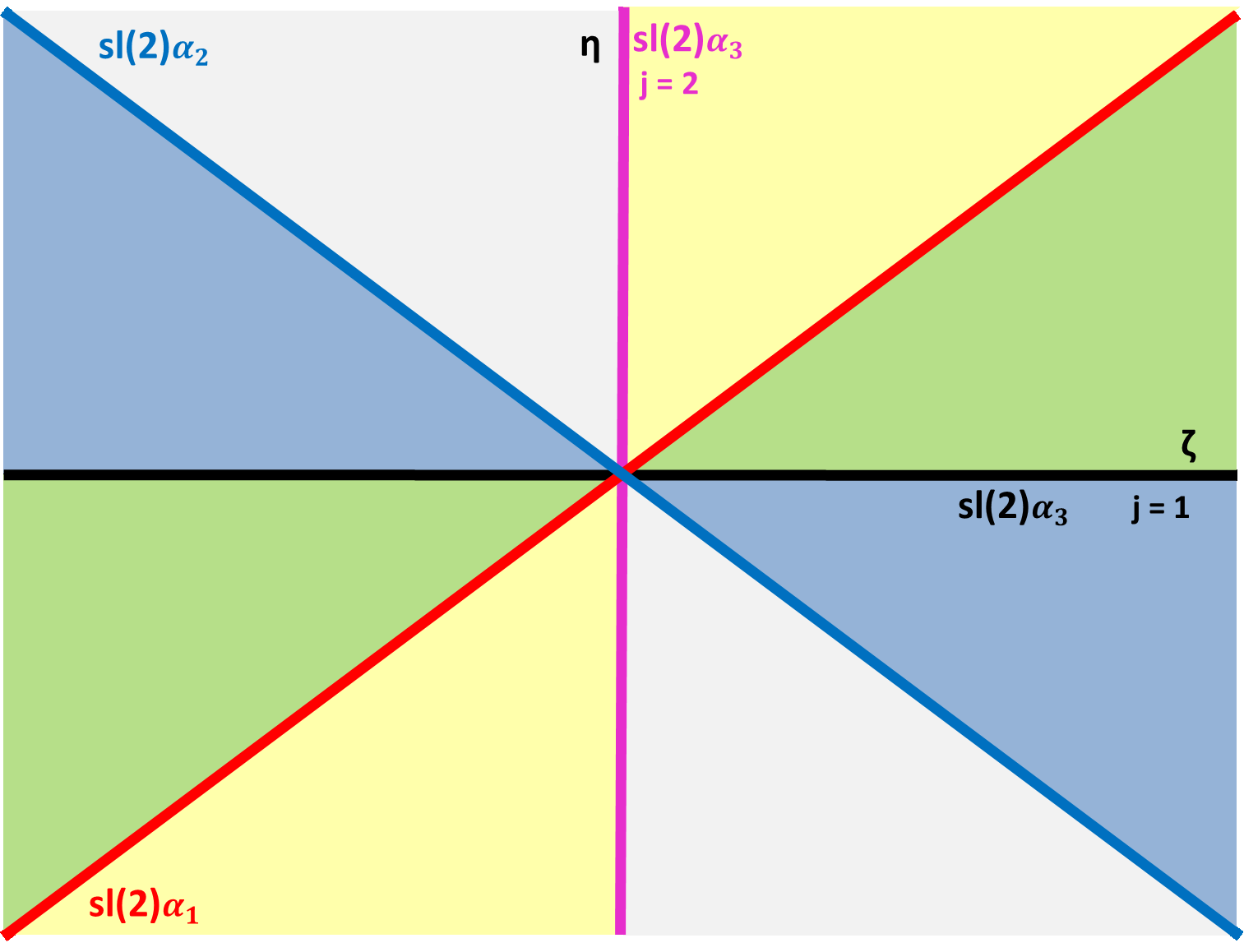}
\end{center}
\par
\vspace{0.0cm}
\caption{Plane of the Cartan subalgebra of sl(3,$\mathbb{R}$) with the
orthogonal generators L$_{0}$ (horizontal) and W$_{0}$ (vertical). Points in
this plane are labelled as $\Theta =\protect\zeta ^{0}L_{0}+\protect\eta %
^{0}W_{0}$ with $\protect\zeta ^{0},\protect\eta ^{0}$ real and Tr$\left(
L_{0}W_{0}\right) =0.$ Coloured regions represent 2D chambers in the Cartan
plane of SL(3,$\mathbb{R}$). And transitions from one chamber to a
neighbouring one require crossing the SL(2,$\mathbb{R}$)$_{\protect\alpha %
_{i}}$ lines.}
\label{sl2}
\end{figure}

The previous construction holds also for higher dimensional Lie algebras
beyond SL(2). For instance, the complex Lie algebra $sl(3,\mathbb{C)}$ has
three real forms: the real split $sl(3,\mathbb{R}),$ the compact real form $%
su(3)$ and the non compact $su(2,1)\simeq su(1,2).$ Using the complex
variables ($z_{1},z_{2},z_{3}$) parameterising $\mathbb{C}^{3}$, the global $%
SU(2,1)$ and $SU(1,2)$\ can be respectively defined as the sets of linear
unitary transformations leaving invariant the hermitian quadratic forms
\begin{equation}
z_{1}z_{1}^{\ast }+z_{2}z_{2}^{\ast }-z_{3}z_{3}^{\ast }\qquad ,\qquad
z_{1}z_{1}^{\ast }-z_{2}z_{2}^{\ast }-z_{3}z_{3}^{\ast }
\end{equation}%
The three real forms of $sl(3,\mathbb{C)}$ are sketched in the Figure
\textbf{\ref{subgr}} where we display, for each real form, the subgroup
associated to each positive root\textrm{;} see also Figure \textbf{\ref{sl2}}%
.
The Cartan subalgebra of these real forms of $sl(3,\mathbb{C)}$ has two
diagonal generators ($H_{\alpha _{1}},H_{\alpha _{2}}$) labelled by the two
simple roots ($\alpha _{1},\alpha _{2}$) of the complex $sl(3,\mathbb{C)}$,
\textrm{they can be realised as}%
\begin{equation}
H_{\alpha _{1}}=\left(
\begin{array}{ccc}
1 & 0 & 0 \\
0 & -1 & 0 \\
0 & 0 & 0%
\end{array}%
\right) \quad ,\quad H_{\alpha _{2}}=\left(
\begin{array}{ccc}
0 & 0 & 0 \\
0 & 1 & 0 \\
0 & 0 & -1%
\end{array}%
\right)
\end{equation}%
showing that
\begin{equation}
\begin{tabular}{lll}
$L_{0}$ & $=$ & $H_{\alpha _{1}}+H_{\alpha _{2}}$ \\
$W_{0}$ & $=$ & $\frac{2}{3}\left( H_{\alpha _{1}}-H_{\alpha _{2}}\right) $%
\end{tabular}%
\quad ,\quad
\begin{tabular}{lll}
$H_{\alpha _{1}}$ & $=$ & $\frac{1}{4}\left( 2L_{0}+3W_{0}\right) $ \\
$H_{\alpha _{2}}$ & $=$ & $\frac{1}{4}\left( 2L_{0}-3W_{0}\right) $%
\end{tabular}
\label{HH}
\end{equation}%
with $\left( W_{n}\right) ^{\ast }=W_{n}$ and $\left( W_{n}\right) ^{\dagger
}=\left( -\right) ^{n}W_{-n}$ as well as
\begin{equation}
Tr\left( W_{0}W_{0}\right) =\frac{8}{3},\qquad Tr\left( W_{-}W_{+}\right)
=-4,\qquad Tr\left( W_{-2}W_{+2}\right) =16
\end{equation}%
Accordingly, the ($L_{{\small 0}},W_{{\small 0}}$) are the generators of the
algebra $gl\left( 1,\mathbb{R}\right) _{+}\oplus gl\left( 1,\mathbb{R}%
\right) _{-}$ also denoted like\textrm{\ }$gl\left( 1\right) _{{\small L}_{%
{\small 0}}}\oplus gl\left( 1\right) _{{\small W}_{{\small 0}}}.$

In the next part, we develop a HS basis\emph{\ }using simple roots of Lie
algebra with illustrative examples. We shall first distinguish between the
usual HS basis built o\textrm{n} $SL_{2}$ representations\emph{\ }and the HS
basis \textrm{f}ounded on simple roots of Lie algebras (\textrm{i.e}
Cartan-Weyl basis). Evidently,\ the two\ HS bases are\ connected\ via bridge
transformations.

$\left( \mathbf{1}\right) $ \textbf{HS basis using}\emph{\ }$SL_{2}$\emph{\ }%
\textbf{representations}\newline
The matrix representation of the higher spin generators $W_{m_{s}}^{{\small %
(s)}}$ of the sl$_{N}$ Lie algebra relies on decomposing the total group
dimension as $\dim \left( sl_{N}\right) =\sum_{s=2}^{N}\dim \left(
\boldsymbol{R}_{s}\right) $ with $SL_{2}$ representation dimensions given by
$\dim \left( \boldsymbol{R}_{s}\right) =2s-1$ with $s=j+1.$ It is
constructed as follows:

$(\mathbf{i})$ single out the $L_{0,\pm }=W_{m_{2}}^{{\small (2)}}$
expanding like $\sum_{j,k}\left\vert j\right\rangle (L_{0,\pm
})_{jk}\left\langle k\right\vert $ with representative matrices as
\begin{equation}
\begin{tabular}{ccc}
$\left( L_{+}\right) _{jk}$ & $=$ & $-\sqrt{j\left( N-j\right) }\delta
_{j+1,k}$ \\
$\left( L_{-}\right) _{jk}$ & $=$ & $+\sqrt{k\left( N-k\right) }\delta
_{j,k+1}$ \\
$\left( L_{0}\right) _{jk}$ & $=$ & $+\frac{1}{2}\left( N+1-2j\right) \delta
_{j,k}$ \\
$Tr\left( L_{0}L_{0}\right) $ & $=$ & $+\frac{N}{12}\left( N^{2}-1\right)
\equiv \epsilon _{N}$%
\end{tabular}%
\   \label{sn}
\end{equation}%
then compute the powers $\left( L_{\pm }\right) ^{s-1}$ to get%
\begin{eqnarray}
\left( L_{+}\right) ^{s-1} &=&\left( -\right) ^{s-1}\sum_{l=1}^{N-s-1}\left(
\dprod\limits_{n=0}^{s-2}\sqrt{\left( l+n\right) \left( N-l-n\right) }%
\right) \left\vert l\right\rangle \left\langle l+s-1\right\vert \\
\left( L_{-}\right) ^{s-1} &=&\sum_{l=1}^{N-s-1}\left(
\dprod\limits_{n=0}^{s-2}\sqrt{\left( l+n\right) \left( N-l-n\right) }%
\right) \left\vert l+s-1\right\rangle \left\langle l\right\vert
\end{eqnarray}%
obeying the nilpotence property $\left( L_{\pm }\right) ^{N}=0.$

$\left( \mathbf{ii}\right) $ consider the higher spin $W_{m_{s}}^{{\small (s)%
}}$ \textrm{as} isospins $j=s-1$ representations. They are accordingly
obtained by performing successive actions of\textrm{\ }$ad_{L_{-}}$on the
highest weight state $\left( L_{+}\right) ^{s-1}$\textrm{\ }as follows%
\begin{equation}
W_{m_{s}}^{{\small (s)}}=2\left( -\right) ^{s-m_{s}-1}\frac{\left(
s-m_{s}-1\right) !}{\left( 2s-2\right) !}\left( ad_{L_{-}}\right)
^{s-m_{s}-1}\left( L_{+}\right) ^{s-1}  \label{sm}
\end{equation}%
with the property%
\begin{equation}
Tr(W_{0}^{{\small (s)}}W_{0}^{{\small (s)}})=\frac{48\epsilon _{N}\left[
\left( s-1\right) !\right] ^{4}}{\left( 2s-1\right) !\left( 2s-2\right) !}%
\dprod\limits_{j=2}^{s-1}\left( N^{2}-j^{2}\right) :=\epsilon _{N}\varpi _{s}
\end{equation}%
The \textrm{full} structure of the HW representations \textrm{is as}
sketched here after
\begin{equation}
\begin{tabular}{cccccc}
$j=1$ \ \ \  &  & $j=2$ &  & $\ldots $ & $\ \ \ \ j=N-1$ \ \ \ \  \\ \hline
\multicolumn{1}{|c}{$L_{+}$} & $\rightarrow _{L_{+}}$ & $\left( L_{+}\right)
^{2}$ & $\rightarrow _{L_{+}}$ & $\ldots $ & \multicolumn{1}{c|}{$\left(
L_{+}\right) ^{N-1}$} \\
\multicolumn{1}{|c}{$\ \ \downarrow ad_{L_{-}}^{m_{2}}$} &  & $\ \downarrow
ad_{L_{-}}^{m_{3}}$ &  &  & \multicolumn{1}{c|}{$\ \downarrow ad_{L_{-}}^{m_{%
{\small N}}}$} \\
\multicolumn{1}{|c}{$L_{-}$} & $\rightarrow _{L_{-}}$ & $\left( L_{-}\right)
^{2}$ & $\rightarrow _{L_{-}}$ & $\ldots $ & \multicolumn{1}{c|}{$\left(
L_{-}\right) ^{N-1}$} \\ \hline
\end{tabular}%
\end{equation}%
with labels taking the values $1-s\leq m_{{\small s}}\leq s-1.$

$\left( \mathbf{2}\right) $ \textbf{HS basis using roots system}\newline
This HS basis can be derived from the above construction by using the
projectors $\boldsymbol{P}_{l}=\left\vert l\right\rangle \left\langle
l\right\vert $, labelled by $l=1,...,N$, and the basic step operators $%
\boldsymbol{E}_{l}=\left\vert l\right\rangle \left\langle l+1\right\vert $
as well as $\boldsymbol{F}_{l}=\left\vert l+1\right\rangle \left\langle
l\right\vert $. These operators satisfy the properties $\boldsymbol{E}_{l}%
\boldsymbol{F}_{l}=\boldsymbol{P}_{l}$ and $\boldsymbol{F}_{l}\boldsymbol{E}%
_{l}=\boldsymbol{P}_{l+1}$. As such, the previous realisation of the $sl_{2}$
generators L$_{0,\pm }$\emph{\ }can be expanded in terms of ($\boldsymbol{P}%
_{l},\boldsymbol{E}_{l},\boldsymbol{F}_{l}$) as follows
\begin{eqnarray}
L_{+} &=&-\sum_{l=1}^{N-1}\sqrt{l\left( N-l\right) }\boldsymbol{E}_{l}
\notag \\
L_{-} &=&\sum_{l=1}^{N-1}\sqrt{l\left( N-l\right) }\boldsymbol{F}_{l} \\
L_{0} &=&\frac{1}{2}\sum_{l=1}^{N}\left( N+1-2l\right) \boldsymbol{P}_{l}
\notag
\end{eqnarray}%
where the charge operator $L_{0}$ can be recast as%
\begin{equation}
L_{0}=\frac{1}{2}\sum_{l=1}^{\left[ N/2\right] }\left( N+1-2l\right) \left(
\boldsymbol{P}_{{\small l}}-\boldsymbol{P}_{{\small N+1-l}}\right)
\end{equation}%
For $N=2,$ the $L_{0}$ charge matrix reads as $\frac{1}{2}\left( \boldsymbol{%
P}_{1}-\boldsymbol{P}_{2}\right) $ and $L_{+}=-\boldsymbol{E}_{1}$ as well
as $L_{-}=+\boldsymbol{F}_{1}.$\emph{\ }As for $N=3,$ we have
\begin{equation}
L_{0}=\boldsymbol{P}_{1}-\boldsymbol{P}_{3},\quad L_{+}=-\sqrt{2}\left(
\boldsymbol{E}_{1}+\boldsymbol{E}_{2}\right) ,\quad L_{-}=+\sqrt{2}\left(
\boldsymbol{F}_{1}+\boldsymbol{F}_{2}\right)
\end{equation}%
with the properties $L_{+}^{2}=2\left( \boldsymbol{E}_{1}\boldsymbol{E}%
_{2}\right) $ and $L_{+}^{3}=0$. We can also take on the case with $N=4,$
\textrm{where}
\begin{eqnarray}
L_{+} &=&-\sqrt{3}\left( \boldsymbol{E}_{1}+\boldsymbol{E}_{3}\right) -\sqrt{%
4}\boldsymbol{E}_{2}  \notag \\
L_{-} &=&+\sqrt{3}\left( \boldsymbol{F}_{1}+\boldsymbol{F}_{3}\right) +\sqrt{%
4}\boldsymbol{F}_{2} \\
L_{0} &=&+\frac{3}{2}\left( \boldsymbol{P}_{1}-\boldsymbol{P}_{4}\right) +%
\frac{1}{2}\left( \boldsymbol{P}_{2}-\boldsymbol{P}_{3}\right)  \notag
\end{eqnarray}%
Moreover, using $\boldsymbol{H}_{i}=\boldsymbol{P}_{i}-\boldsymbol{P}_{i+1},$
the $L_{0}$ is given by a linear combination of the diagonal $\boldsymbol{H}%
_{i}$'s as follows%
\begin{equation}
\begin{tabular}{lll}
$L_{0}$ & $=$ & $\frac{3}{2}\left( \boldsymbol{H}_{1}+\boldsymbol{H}_{2}+%
\boldsymbol{H}_{3}\right) +\frac{1}{2}\boldsymbol{H}_{2}$ \\
& $=$ & $\frac{3}{2}\left( \boldsymbol{H}_{1}+\boldsymbol{H}_{3}\right) +2%
\boldsymbol{H}_{2}$%
\end{tabular}%
\end{equation}%
For generic $N,$ we \textrm{have}%
\begin{equation}
L_{0}=\frac{1}{2}\dsum\limits_{l=1}^{\left[ N/2\right] }\left( N+1-2l\right)
\left( \boldsymbol{H}_{l}+\boldsymbol{H}_{l+1}+...+\boldsymbol{H}%
_{N-l}\right)
\end{equation}%
In terms of the simple roots $\mathbf{\alpha }_{i}$ of the $sl_{N}$ Lie
algebra, the above matrix realisations of the principal sl$_{2}$\ can be
rewritten in terms of the generators $\left( \boldsymbol{H}_{\alpha _{l}},%
\boldsymbol{E}_{\pm \alpha _{l}}\right) $ of sl$_{N}$ as follows%
\begin{eqnarray}
L_{+} &=&-\sum_{l=1}^{N-1}\sqrt{l\left( N-l\right) }\boldsymbol{E}_{+\alpha
_{l}}  \notag \\
L_{-} &=&+\sum_{l=1}^{N-1}\sqrt{l\left( N-l\right) }\boldsymbol{E}_{-\alpha
_{l}} \\
L_{0} &=&+\frac{1}{2}\dsum\limits_{l=1}^{\left[ N/2\right] }\left(
N+1-2l\right) \boldsymbol{h}_{l}  \notag
\end{eqnarray}%
with%
\begin{equation}
\boldsymbol{h}_{l}=\dsum\limits_{n=0}^{N-2l}\boldsymbol{H}_{\alpha _{l+n}}
\end{equation}

$\left( \mathbf{3}\right) $ \textbf{Abelian subgroups of}\emph{\ }$SL(3,%
\mathbb{C})$\newline
The two group elements $g_{1}$ and $g_{2}$ generating the Cartan subgroup of
the real forms of\emph{\ }$SL(3,\mathbb{C})$ are given by the following
exponentiations\textrm{\ }%
\begin{equation}
\begin{tabular}{|c|c|c|c|c|c|c|}
\hline
real forms & $g_{1}$ & $g_{2}$ & {\small Cartan subgroup} &
\multicolumn{3}{|c|}{\small Cartan geometry} \\ \hline
$SU(3)$ & $e^{i\theta _{\alpha _{{\small 1}}}H_{\alpha _{{\small 1}}}}$ & $%
e^{i\theta _{\alpha _{{\small 2}}}H_{\alpha _{{\small 2}}}}$ & ${\small %
U(1)\times U(1)}$ & $\mathbb{S}^{1}\times \mathbb{S}^{1}$ & $\simeq $ &
{\small 2-Torus} \\ \hline
$SU(2,1)$ & $e^{\theta _{\alpha _{{\small 1}}}H_{\alpha _{{\small 1}}}}$ & $%
e^{i\theta _{\alpha _{{\small 2}}}H_{\alpha _{{\small 2}}}}$ & ${\small %
GL(1)\times U(1)}$ & $\mathbb{R}\times \mathbb{S}^{1}$ & $\simeq $ & {\small %
cylindre I} \\ \hline
$SU(1,2)$ & $e^{i\theta _{\alpha _{{\small 1}}}H_{\alpha _{{\small 1}}}}$ & $%
e^{\theta _{\alpha _{{\small 2}}}H_{\alpha _{{\small 2}}}}$ & ${\small %
U(1)\times GL(1)}$ & $\mathbb{S}^{1}\times \mathbb{R}$ & $\simeq $ & {\small %
cylindre II} \\ \hline
$SL(3,\mathbb{R})$ & $e^{\theta _{\alpha _{{\small 1}}}H_{\alpha _{{\small 1}%
}}}$ & $e^{\theta _{\alpha _{{\small 2}}}H_{\alpha _{{\small 2}}}}$ & $%
{\small GL(1)\times GL(1)}$ & $\mathbb{R}\times \mathbb{R}$ & $\simeq $ &
{\small real plane} \\ \hline
\end{tabular}
\label{geo}
\end{equation}%
\begin{equation*}
\end{equation*}%
where the real $\theta _{\alpha _{{\small 1}}},\theta _{\alpha _{{\small 2}%
}} $ are the group parameters. From these expressions, we see that these
abelian Lie subgroups have different geometrical interpretations leading\ to
different values of entropy which showcases the dependency of the
calculations on the type of real form of the complexified gauge symmetry.

The Tits-Satake diagrams of the\textrm{\ above} real forms of\emph{\ }$SL(3,%
\mathbb{C})$ have two (un)painted nodes, they \textrm{are depicted in the
next} Table (\ref{T1})%
\begin{equation}
\begin{tabular}{|c|c|c|c|c|c|}
\hline
{\small real forms} & \multicolumn{2}{|c|}{\small Cartan subalgebra} &
\multicolumn{3}{|c|}{\small Tits-Satake diagram} \\ \hline
$su\left( 3\right) $ & $u{\small (1)}_{\alpha _{{\small 1}}}$ & $u{\small (1)%
}_{\alpha _{{\small 2}}}$ & {\small 2 black node} & %
\includegraphics[width=0.7cm]{black} & \includegraphics[width=0.7cm]{black}
\\ \hline
$sl\left( 3,\mathbb{R}\right) $ & $gl{\small (1)}_{\alpha _{{\small 1}}}$ & $%
gl{\small (1)}_{\alpha _{{\small 2}}}$ & {\small 2 white node} & %
\includegraphics[width=0.7cm]{blank} & \includegraphics[width=0.7cm]{blank}
\\ \hline
$su(2,1)$ & $gl{\small (1)}_{\alpha _{{\small 1}}}$ & $u{\small (1)}_{\alpha
_{{\small 2}}}$ & {\small black }${\small \leftrightarrow }$ {\small white}
& \includegraphics[width=0.7cm]{black} & \includegraphics[width=0.7cm]{blank}
\\ \hline
$su(1,2)$ & $u{\small (1)}_{\alpha _{{\small 1}}}$ & $gl{\small (1)}_{\alpha
_{{\small 2}}}$ & {\small white }${\small \leftrightarrow }$ {\small black}
& \includegraphics[width=0.7cm]{blank} & \includegraphics[width=0.7cm]{black}
\\ \hline
\end{tabular}
\label{T1}
\end{equation}%
\begin{equation*}
\text{ \ }
\end{equation*}%
where the diagrams of $su(2,1)$\ and $su(1,2)$\ are exchangeable under the
permutation $\alpha _{1}\leftrightarrow \alpha _{2}$. From this graphic
representation, we learn interesting features; in particular the following:

\paragraph{A) Characterising the real forms:}

\ \newline
Because each node $\mathfrak{N}_{i}$ of the Dynkin (and Tits-Satake)
diagrams is associated with a simple root $\alpha _{i}$ and a\textrm{\ }$%
SL\left( 2,\mathbb{C}\right) _{\alpha _{i}}$ algebra, several other
properties can be derived\textrm{:}

\textbf{A.1}) \textbf{Real split form} $SL\left( 3,\mathbb{R}\right) $%
\textbf{:} \newline
The $SL\left( 3,\mathbb{R}\right) $ has three \emph{particular} non compact
3d subgroups $SL\left( 2,\mathbb{R}\right) ,$ each one is associated with
one of the positive roots\textrm{\ }$\alpha _{1}$\textrm{, }$\alpha _{2}$%
\textrm{\ }and\textrm{\ }$\alpha _{3}=\alpha _{1}+\alpha _{2}$; they are
given by\textrm{\ }$SL\left( 2,\mathbb{R}\right) _{\alpha _{1}}$\textrm{, }$%
SL\left( 2,\mathbb{R}\right) _{\alpha _{2}}$ and the subgroup\textrm{\ }$%
SL\left( 2,\mathbb{R}\right) _{\alpha _{3}}.$ These groups will be
subsequently visualised as straight lines in the real ($\zeta ^{0},\eta ^{0}$%
) plane defined as
\begin{equation}
\text{\textsc{Cart}}_{sl_{3}}=\left\{ \Theta _{{\small sl}_{{\small 3}%
}}=\zeta ^{0}L_{0}+\eta ^{0}W_{0}|\quad \zeta ^{0},\eta ^{0}\in \mathbb{R}%
\right\} \qquad ,\qquad \text{\textsc{Cart}}_{sl_{3}}\simeq \mathbb{R}^{2}
\end{equation}%
and depicted in Figure \textbf{\ref{sl2}} with directions given by the
Cartan generators L$_{0}$ (horizontal) and W$_{0}$ (vertical). Notice that
the $SL\left( 3,\mathbb{R}\right) $ is non compact and has no compact SU$%
\left( 2\right) $ subgroup$\mathrm{.}$ \newline
The hermitian matrix elements $\Theta _{{\small sl}_{{\small 3}}}$ of the
Cartan plane of $sl\left( 3,\mathbb{R}\right) $ expand like $\zeta
^{0}L_{0}+\eta ^{0}W_{0}$ with real parameters $\zeta ^{0},$ $\eta ^{0}$ and
trace $\xi ^{2}=Tr(\Theta _{{\small sl}_{{\small 3}}}\Theta _{{\small sl}_{%
{\small 3}}}^{\dagger })/2$ given by $\left( \zeta ^{0}\right) ^{2}+\frac{4}{%
3}\left( \eta ^{0}\right) ^{2}.$ By substituting $L_{0}=H_{\alpha
_{1}}+H_{\alpha _{2}}$ and $W_{0}=2\left( H_{\alpha _{1}}-H_{\alpha
_{2}}\right) /3$, we can put the expansion of $\Theta _{{\small sl}_{{\small %
3}}}$ as follows $\theta _{\alpha _{1}}H_{\alpha _{1}}+\theta _{\alpha
_{2}}H_{\alpha _{2}}$ with%
\begin{equation}
\theta _{\alpha _{1}}=\zeta ^{0}+\frac{2}{3}\eta ^{0}\qquad ,\qquad \theta
_{\alpha _{2}}=\zeta ^{0}-\frac{2}{3}\eta ^{0}
\end{equation}%
Using the Cartan matrix $K_{ij}=\mathbf{\alpha }_{i}\mathbf{.\alpha }_{j},$
we obtain $\xi ^{2}=(\theta _{i}K_{ij}\theta _{j})/2$. From these relations,
one can establish the three sl(2,$\mathbb{R}$)$_{\alpha _{i}}$ sub-Cartans
within the Cartan of sl(3,$\mathbb{R}$). They are represented by straight
lines in the real plane $\left( \zeta ,\eta \right) $ \textrm{such that}:

\begin{itemize}
\item the Cartan of sl(2,$\mathbb{R}$)$_{\alpha _{1}}$ is given by the
\textrm{trajectory} $\eta ^{0}=+3\zeta ^{0}/2$.

\item the Cartan of sl(2,$\mathbb{R}$)$_{\alpha _{2}}$ is represented by the
straightline $\eta ^{0}=-3\zeta ^{0}/2.$

\item the Cartan of sl(2,$\mathbb{R}$)$_{\alpha _{3}}$ is characterised by
the horizontal line $\eta ^{0}=0$ \textrm{and} designates\emph{\ }the
isospin j=1 representation.

\item the Cartan of sl(2,$\mathbb{R}$)$_{\alpha _{3}}$ in the isospin j=2 is
given by the vertical line $\zeta ^{0}=0$.
\end{itemize}

\textbf{A.2)} \textbf{Real compact form }$SU\left( 3\right) $\textbf{: }%
\newline
The compact $SU\left( 3\right) $ has three particular $SU(2)$ subgroups,
each one is linked to a positive root as follows
\begin{equation}
SU\left( 2\right) _{\alpha _{1}}\qquad ,\qquad SU\left( 2\right) _{\alpha
_{2}}\qquad ,\qquad SU\left( 2\right) _{\alpha _{1}+\alpha _{2}}
\end{equation}%
These subgroups are compact and will be considered below with more details
especially when studying the distinct branches of HS black flowers and their
reduction towards the spin 2 black holes. Here, the homologue of Figure
\textbf{\ref{sl2}} is a real 2-torus; i.e:%
\begin{equation}
\text{\textsc{Cart}}_{su_{3}}=\left\{ \Theta _{{\small su}_{{\small 3}%
}}=\zeta ^{0}L_{0}+\eta ^{0}W_{0}|\quad \zeta ^{0},\eta ^{0}\in \left[
0,2\pi \right] \right\} \qquad ,\qquad \text{\textsc{Cart}}_{sl_{3}}\simeq
\mathbb{T}^{2}
\end{equation}

\textbf{A.3) Real non compact }$SU\left( 2,1\right) :$ \newline
It has two particular subgroups, a compact\textrm{\ }$SU\left( 2\right) $%
\textrm{\ }and a non compact\textrm{\ }$SU\left( 1,1\right) ,\ $in a one to
one correspondence with the simple roots\textrm{\ }$\alpha _{1}$\textrm{\
and }$\alpha _{2}$\textrm{.} \textrm{Hence the two} \textrm{following }%
pictures:

\begin{itemize}
\item the compact subgroup $SU\left( 2\right) _{\alpha _{1}}$\ is engendered
by $\alpha _{1}$\ while the non compact $SU\left( 1,1\right) _{\alpha _{2}}$%
\ is generated by\textrm{\ }$\alpha _{2}.$ Elements $\Theta _{su_{2,1}}$ in
the Cartan subalgebra of $SU\left( 2,1\right) $ are complex, they have the
structure $\theta _{\alpha _{1}}H_{\alpha _{1}}+i\theta _{\alpha
_{2}}H_{\alpha _{2}}$ splitting like $\zeta ^{0}L_{0}+\eta ^{0}W_{0}$ with
complex parameters $\zeta ^{0}$ and $\eta ^{0}$ \textrm{taken} as follows%
\begin{equation}
\zeta ^{0}=w\qquad ,\qquad \eta ^{0}=\frac{3}{2}\bar{w}\qquad ,\qquad w=%
\frac{1}{2}\left( \theta _{\alpha _{1}}+i\theta _{\alpha _{2}}\right)
\end{equation}

\item \textrm{the inverse picture describes the case where the non compact }$%
SU\left( 1,1\right) _{\alpha _{1}}$\textrm{\ subgroup is generated by }$%
\alpha _{1}$\textrm{\ while the compact }$SU\left( 2\right) _{\alpha _{2}}$%
\textrm{\ is the one associated with }$\alpha _{2}.$\newline
Below, we refer to these two possibilities as follows%
\begin{equation}
\begin{tabular}{|l|l|l|}
\hline
\ Group & \multicolumn{2}{|l|}{\ \ \ particular subgroups} \\ \hline
$\ \ SU\left( 2,1\right) _{12}$ \ \  & $\ \ \ SU\left( 1,1\right) _{\alpha
_{1}}$ & $SU\left( 2\right) _{\alpha _{2}}$ \ \ \  \\ \hline
$\ \ \ SU\left( 2,1\right) _{21}$ & $\ \ \ SU\left( 1,1\right) _{\alpha
_{2}} $ & $SU\left( 2\right) _{\alpha _{1}}$ \\ \hline
\end{tabular}%
\end{equation}%
\textrm{And} the homologue of the Figure \textbf{\ref{sl2}} is a real
cylinder; i.e:%
\begin{equation*}
\begin{tabular}{lll}
$\text{\textsc{Cart}}_{su_{2,1}}$ & $=$ & $\left\{ \Theta _{su_{2,1}}=\zeta
^{0}L_{0}+\eta ^{0}W_{0}|\quad \zeta ^{0}\in \left[ 0,2\pi \right] ,\eta
^{0}\in \mathbb{R}\right\} $ \\
& $\simeq $ & $\mathbb{S}^{1}\times \mathbb{R}$%
\end{tabular}%
\end{equation*}
\end{itemize}

\paragraph{B) Cartan subgroups:}

\ \newline
By using (\ref{HH}), we have%
\begin{equation}
\begin{tabular}{|c|c|c|c|}
\hline
real forms & $\exp (\zeta ^{{\small 0}}L_{0})$ & $\exp (\eta ^{{\small 0}%
}W_{0})$ & {\small Cartan subgroup} \\ \hline
$SU(3)$ & $\ \ e^{i\zeta ^{{\small 0}}(H_{\alpha _{{\small 1}}}+H_{\alpha _{%
{\small 2}}})}$ \ \  & $\ \ e^{\frac{2i}{3}\eta ^{{\small 0}}(H_{\alpha _{%
{\small 1}}}-H_{\alpha _{{\small 2}}})}$ \ \  & ${\small U(1)}_{+}{\small %
\times U(1)}_{-}$ \\ \hline
$SU(2,1)_{12}$ & $e^{\zeta ^{{\small 0}}(H_{\alpha _{{\small 1}}}+iH_{\alpha
_{{\small 2}}})}$ & $e^{\frac{2}{3}\eta ^{{\small 0}}(H_{\alpha _{{\small 1}%
}}-iH_{\alpha _{{\small 2}}})}$ & ${\small GL(1,\mathbb{C})\times }\overline{%
{\small GL(1,\mathbb{C})}}$ \\ \hline
$SU(1,2)_{21}$ & $e^{\zeta ^{{\small 0}}(iH_{\alpha _{{\small 1}}}+H_{\alpha
_{{\small 2}}})}$ & $e^{\frac{2}{3}\eta ^{{\small 0}}(iH_{\alpha _{{\small 1}%
}}-H_{\alpha _{{\small 2}}})}$ & ${\small GL(1,\mathbb{C})\times }\overline{%
{\small GL(1,\mathbb{C})}}$ \\ \hline
$SL(3,\mathbb{R})$ & $e^{\zeta ^{{\small 0}}(H_{\alpha _{{\small 1}%
}}+H_{\alpha _{{\small 2}}})}$ & $e^{\frac{2}{3}\eta ^{{\small 0}}(H_{\alpha
_{{\small 1}}}-H_{\alpha _{{\small 2}}})}$ & $\ {\small GL(1,}\mathbb{R}%
{\small )_{+}\times GL(1,\mathbb{R})}_{-}$ \  \\ \hline
\end{tabular}
\label{316}
\end{equation}%
\begin{equation*}
\end{equation*}%
where the parameters $\zeta ^{{\small 0}},$ $\eta ^{{\small 0}}$ are as\emph{%
\ }previously provided. As we will see, they can be constrained by demanding
\textrm{the} regularity of the gauge potentials around the thermal cycle.
\textrm{Note} that the complex group $GL(1,\mathbb{C})$ is isomorphic to $%
GL(1,\mathbb{R})\times U(1){\small .}$\newline
The exponentiation of the circulation $\int_{\mathcal{C}_{t}}dta_{t}^{\text{%
\textsc{diag}}}$ around some line $\mathcal{C}_{t}$ \textrm{has}\emph{\ }the
geometric interpretations%
\begin{equation}
\begin{tabular}{|c|c|c|}
\hline
$SL(3,\mathbb{C})${\small \ real forms} & $\ \ \ \exp (\int_{\mathcal{C}%
_{t}}dta_{t}^{\text{\textsc{diag}}})$ \ \ \ \ \  & \ \ {\small Cartan}
{\small geometry \ \ \ \ } \\ \hline
$SU\left( 3\right) $ & $e^{i\Upsilon ^{0}L_{0}}e^{i\Gamma ^{0}W_{0}}$ & $%
\mathbb{S}_{+}^{1}\times \mathbb{S}_{-}^{1}$ \\ \hline
$SU(2,1)_{12}$ & $e^{\Upsilon ^{0}L_{0}}e^{i\Gamma ^{0}W_{0}}$ & $\mathbb{R}%
\times \mathbb{S}_{-}^{1}$ \\ \hline
$SU(1,2)_{21}$ & $e^{i\Upsilon ^{0}L_{0}}e^{\Gamma ^{0}W_{0}}$ & $\mathbb{S}%
_{+}^{1}\times \mathbb{R}$ \\ \hline
$SL\left( 3,\mathbb{R}\right) $ & $e^{\Upsilon ^{0}L_{0}}e^{\Gamma
^{0}W_{0}} $ & $\mathbb{R}^{\ast }\times \mathbb{R}$ \\ \hline
\end{tabular}%
\end{equation}%
\begin{equation*}
\end{equation*}%
for the left sector. As for the right sector, we have%
\begin{equation}
\begin{tabular}{|c|c|c|}
\hline
$\widetilde{SL}(3,\mathbb{C})${\small \ real forms } & $\ \ \ \ \exp (\tilde{%
a}_{t}^{\text{\textsc{diag}}})$ \ \ \ \ \  & \ \ \ \ {\small geometry \ \ \
\ } \\ \hline
$\widetilde{SU}\left( 3\right) $ & $e^{-i\tilde{\Upsilon}^{0}L_{0}}e^{-i%
\tilde{\Gamma}^{0}W_{0}}$ & $\mathbb{\tilde{S}}_{+}^{1}\times \mathbb{\tilde{%
S}}_{-}^{1}$ \\ \hline
$\widetilde{SU}(2,1)_{12}$ & $e^{\tilde{\Upsilon}^{0}L_{0}}e^{-i\tilde{\Gamma%
}^{0}W_{0}}$ & $\mathbb{\tilde{R}}\times \mathbb{\tilde{S}}_{-}^{1}$ \\
\hline
$\widetilde{SU}(1,2)_{21}$ & $e^{-i\tilde{\Upsilon}^{0}L_{0}}e^{\tilde{\Gamma%
}^{0}W_{0}}$ & $\mathbb{S}_{+}^{1}\times \mathbb{\tilde{R}}$ \\ \hline
$\widetilde{SL}\left( 3,\mathbb{R}\right) $ & $e^{\tilde{\Upsilon}%
^{0}L_{0}}e^{i\tilde{\Gamma}^{0}W_{0}}$ & $\mathbb{\tilde{R}}\times \mathbb{%
\tilde{R}}$ \\ \hline
\end{tabular}
\label{321}
\end{equation}

\subsection*{Beyond the SL(3) symmetry}

The above sl(3) construction extends straightforwardly to encompass other
real forms of the $sl\left( N,\mathbb{C}\right) $ family as well as other
Lie algebras of the Cartan classification. Building upon the SL(2) and SL(3)
models, we give the HS generators $W_{m_{2}}^{(2)},$ $W_{m_{3}}^{(3)},$\ $%
W_{m_{4}}^{(4)}$ of SL$_{4}$ in terms of the simple roots \{$\mathbf{\alpha }%
_{1},\mathbf{\alpha }_{2},\mathbf{\alpha }_{3}$\}.

\paragraph{\textbf{I. HS spin basis using simple roots}:}

\ \ \newline
The generators $L_{0,\pm }$ of the principal $sl_{2}$ in terms of the
Cartan-Weyl generators $\boldsymbol{H}_{\alpha _{i}},\boldsymbol{E}_{\pm
\alpha _{i}}$ of the Lie algebra $sl_{4}$ are given by%
\begin{eqnarray}
L_{+} &=&-\sqrt{3}\boldsymbol{E}_{\alpha _{1}}-2\boldsymbol{E}_{\alpha _{2}}-%
\sqrt{3}\boldsymbol{E}_{\alpha _{3}}  \notag \\
L_{0} &=&+\frac{3}{2}\boldsymbol{H}_{\alpha _{1}}+2\boldsymbol{H}_{\alpha
_{2}}+\frac{3}{2}\boldsymbol{H}_{\alpha _{3}} \\
L_{-} &=&+\sqrt{3}\boldsymbol{E}_{-\alpha _{1}}+2\boldsymbol{E}_{-\alpha
_{2}}+\sqrt{3}\boldsymbol{E}_{-\alpha _{3}}  \notag
\end{eqnarray}%
with $\left[ \boldsymbol{E}_{\alpha _{i}},\boldsymbol{E}_{-\alpha _{i}}%
\right] =\boldsymbol{H}_{\alpha _{i}}.$ These expressions are invariant
under the discrete \textrm{mapping} $(\mathbf{\alpha }_{1},\mathbf{\alpha }%
_{2},\mathbf{\alpha }_{3})$ $\leftrightarrow $ $(\mathbf{\alpha }_{3},%
\mathbf{\alpha }_{2},\mathbf{\alpha }_{1})$ exchanging $\mathbf{\alpha }_{1}$
and $\mathbf{\alpha }_{3}$ while preserving $\mathbf{\alpha }_{2}.$ This
discrete $\mathbb{Z}_{2}$ symmetry is nothing other than the outer
automorphism of the Dynkin diagram of sl(4,$\mathbb{C}$). The corresponding
matrix realisation can be written as%
\begin{equation}
L_{+}=\left(
\begin{array}{cccc}
{\small 0} & {\small -}\sqrt{{\small 3}} & {\small 0} & {\small 0} \\
{\small 0} & {\small 0} & {\small -2} & {\small 0} \\
{\small 0} & {\small 0} & {\small 0} & {\small -}\sqrt{{\small 3}} \\
{\small 0} & {\small 0} & {\small 0} & {\small 0}%
\end{array}%
\right) ,\text{ \ }L_{-}=\left(
\begin{array}{cccc}
{\small 0} & {\small 0} & {\small 0} & {\small 0} \\
\sqrt{{\small 3}} & {\small 0} & {\small 0} & {\small 0} \\
{\small 0} & {\small 2} & {\small 0} & {\small 0} \\
{\small 0} & {\small 0} & \sqrt{{\small 3}} & {\small 0}%
\end{array}%
\right) ,\text{ \ }L_{0}=\left(
\begin{array}{cccc}
\frac{3}{2} & {\small 0} & {\small 0} & {\small 0} \\
{\small 0} & \frac{1}{2} & {\small 0} & {\small 0} \\
{\small 0} & {\small 0} & {\small -}\frac{1}{2} & {\small 0} \\
{\small 0} & {\small 0} & {\small 0} & {\small -}\frac{3}{2}%
\end{array}%
\right)
\end{equation}%
From these relations, one can deduce the highest weight states $L_{+}^{2}$
and\ $L_{+}^{3}$ of the isospin j=2 and j=3 representations namely%
\begin{equation}
\begin{tabular}{lll}
$L_{+}^{2}$ & $=$ & $+\sqrt{12}\left( \boldsymbol{E}_{\alpha _{1}+\alpha
_{2}}+\boldsymbol{E}_{\alpha _{2}+\alpha _{3}}\right) $ \\
$L_{+}^{3}$ & $=$ & $-6\boldsymbol{E}_{\alpha _{1}+\alpha _{2}+\alpha _{3}}$%
\end{tabular}%
\end{equation}%
with $\boldsymbol{E}_{\alpha +\beta }=\varepsilon _{\alpha \beta }\left[
\boldsymbol{E}_{\alpha },\boldsymbol{E}_{\beta }\right] .$\emph{\ }The
associated\ matrix realisations are as%
\begin{equation}
L_{+}^{2}=\left(
\begin{array}{cccc}
0 & 0 & \sqrt{12} & 0 \\
0 & 0 & 0 & \sqrt{12} \\
0 & 0 & 0 & 0 \\
0 & 0 & 0 & 0%
\end{array}%
\right) \qquad ,\qquad W_{0}^{{\small (4)}}=\left(
\begin{array}{cccc}
0 & 0 & 0 & -6 \\
0 & 0 & 0 & 0 \\
0 & 0 & 0 & 0 \\
0 & 0 & 0 & 0%
\end{array}%
\right)
\end{equation}%
The rest of the generators\textrm{\ }$W_{m_{2}}^{{\small (2)}}$\textrm{\ and
}$W_{m_{3}}^{{\small (3)}}$\textrm{\ }are respectively given by successive
actions o\textrm{f }$ad_{L_{-}}$\textrm{\ }on the isospin\textrm{\ }$j=2$%
\textrm{\ and }$j=3$\textrm{\ }representations with highest weight states%
\textrm{\ }$\left( L_{+}\right) ^{j}$\textrm{\ as follows}%
\begin{equation}
\begin{tabular}{lll}
$W_{m_{3}}^{{\small (3)}}$ & $=$ & $+\left( -\right) ^{m_{3}}\frac{\left(
2-m_{3}\right) !}{12}\left( ad_{L_{-}}\right) ^{2-m_{3}}\left( L_{+}\right)
^{2}$ \\
$W_{m_{4}}^{{\small (4)}}$ & $=$ & $-\left( -\right) ^{m_{4}}\frac{\left(
3-m_{4}\right) !}{360}\left( ad_{L_{-}}\right) ^{3-m_{4}}\left( L_{+}\right)
^{3}$%
\end{tabular}%
\end{equation}%
\textrm{Besides} the charge operator $L_{0},$ the other diagonal generators
of $sl_{4}$ \textrm{are}%
\begin{equation}
\begin{tabular}{lll}
$W_{0}^{{\small (3)}}$ & $=$ & $+\frac{1}{6}\left( ad_{L_{-}}\right)
^{2}\left( L_{+}\right) ^{2}$ \\
$W_{0}^{{\small (4)}}$ & $=$ & $-\frac{1}{60}\left( ad_{L_{-}}\right)
^{3}\left( L_{+}\right) ^{3}$%
\end{tabular}%
\end{equation}%
\textrm{they can be expanded} like%
\begin{equation}
\begin{tabular}{lll}
$W_{0}^{{\small (3)}}$ & $=$ & $+\frac{1}{3}\left( \left\{
L_{-},L_{+}\right\} +4L_{0}L_{0}\right) $ \\
$W_{0}^{{\small (4)}}$ & $=$ & $-\frac{1}{5}\left( 4L_{0}^{3}+L_{0}\left\{
L_{+},L_{-}\right\} +L_{-}\left\{ L_{0},L_{+}\right\} +L_{+}\left\{
L_{-},L_{0}\right\} \right) $%
\end{tabular}%
\end{equation}%
\textrm{With} \textrm{the} matrix representations%
\begin{equation}
W_{0}^{{\small (3)}}=\left(
\begin{array}{cccc}
2 & 0 & 0 & 0 \\
0 & -2 & 0 & 0 \\
0 & 0 & -2 & 0 \\
0 & 0 & 0 & 2%
\end{array}%
\right) \qquad ,\qquad W_{0}^{{\small (4)}}=-\frac{3}{5}\left(
\begin{array}{cccc}
1 & 0 & 0 & 0 \\
0 & -3 & 0 & 0 \\
0 & 0 & 3 & 0 \\
0 & 0 & 0 & -1%
\end{array}%
\right)
\end{equation}%
\begin{equation*}
\end{equation*}%
indicating that $Tr(W_{0}^{{\small (3)}}W_{0}^{{\small (3)}})=16$ and $%
Tr(W_{0}^{{\small (4)}}W_{0}^{{\small (4)}})=\frac{36}{5}.$ Moreover, with
the formula $\boldsymbol{H}_{i}=\boldsymbol{P}_{i}-\boldsymbol{P}_{i+1},$
the diagonal generators can be recast as linear combinations of the Cartans
as follows%
\begin{equation}
\begin{tabular}{lll}
$W_{0}^{{\small (2)}}$ & $=$ & $+\frac{3}{2}\boldsymbol{H}_{\alpha _{1}}+2%
\boldsymbol{H}_{\alpha _{2}}+\frac{3}{2}\boldsymbol{H}_{\alpha _{3}}$ \\
$W_{0}^{{\small (3)}}$ & $=$ & $+2\left( \boldsymbol{H}_{\alpha _{1}}-%
\boldsymbol{H}_{\alpha _{3}}\right) $ \\
$W_{0}^{{\small (4)}}$ & $=$ & $-\frac{3}{5}\left( \boldsymbol{H}_{\alpha
_{1}}-2\boldsymbol{H}_{\alpha _{2}}+\boldsymbol{H}_{\alpha _{3}}\right) $%
\end{tabular}
\label{ww}
\end{equation}%
They form an orthogonal set such that $Tr(L_{0}W_{0}^{{\small (s)}%
})=Tr(W_{0}^{{\small (3)}}W_{0}^{{\small (4)}})=0.$ The five isospin\textrm{%
\ }$j=2$\textrm{\ }states with HW\textrm{\ }$6L_{+}^{2}$\textrm{\ }can be
also read in terms of the simple roots as follows
\begin{equation}
\begin{tabular}{lll}
$W_{++}^{{\small (3)}}$ & $=$ & $+\sqrt{3}\left( \boldsymbol{E}_{\alpha
_{1}+\alpha _{2}}+\boldsymbol{E}_{\alpha _{2}+\alpha _{3}}\right) $ \\
$W_{+}^{{\small (3)}}$ & $=$ & $-8\sqrt{3}\left( \boldsymbol{E}_{\alpha
_{1}}-\boldsymbol{E}_{\alpha _{3}}\right) $ \\
$W_{0}^{{\small (3)}}$ & $=$ & $+2\left( \boldsymbol{H}_{\alpha _{1}}-%
\boldsymbol{H}_{\alpha _{3}}\right) $ \\
$W_{-}^{{\small (3)}}$ & $=$ & $-72\sqrt{3}\left( \boldsymbol{E}_{-\alpha
_{1}}-\boldsymbol{E}_{-\alpha _{3}}\right) $ \\
$W_{--}^{{\small (3)}}$ & $=$ & $+576\sqrt{3}\left( \boldsymbol{E}_{-\alpha
_{1}-\alpha _{2}}-\boldsymbol{E}_{-\alpha _{2}-\alpha _{3}}\right) $%
\end{tabular}%
\end{equation}%
Similar relationships can be written down for the\emph{\ 7 }states \textrm{%
of }$W_{m_{4}}^{{\small (4)}}$ with HW $W_{+++}^{{\small (4)}}=-6\boldsymbol{%
E}_{\alpha _{1}+\alpha _{2}+\alpha _{3}}.$

\paragraph{\textbf{II. the simple weight vectors\ }$\mathbf{\protect\gamma }%
_{i}$\textbf{\ of SL}$_{\mathbf{4}}:$}

\ \ \newline
Using (\ref{ww}), we can introduce three weight vectors ($\mathbf{\gamma }%
_{1},\mathbf{\gamma }_{2},\mathbf{\gamma }_{3}$) sitting \textrm{in }$%
\mathbb{R}^{4}$ \textrm{via} the correspondence:
\begin{equation}
\begin{tabular}{lll}
$Tr(W_{0}^{{\small (2)}}W_{0}^{{\small (2)}})$ & $=$ & $\mathbf{\gamma }_{1}.%
\mathbf{\gamma }_{1}$ \\
$Tr(W_{0}^{{\small (3)}}W_{0}^{{\small (3)}})$ & $=$ & $\mathbf{\gamma }_{2}.%
\mathbf{\gamma }_{2}$ \\
$Tr(W_{0}^{{\small (4)}}W_{0}^{{\small (4)}})$ & $=$ & $\mathbf{\gamma }_{3}.%
\mathbf{\gamma }_{3}$ \\
$Tr(W_{0}^{{\small (i+1)}}W_{0}^{{\small (j+1)}})$ & $=$ & $\mathbf{\gamma }%
_{i}.\mathbf{\gamma }_{j}=0$%
\end{tabular}%
\end{equation}%
These weight vectors are related to the simple roots ($\mathbf{\alpha }_{1},%
\mathbf{\alpha }_{2},\mathbf{\alpha }_{3}$) of SL$_{\mathbf{4}}$ as follows
\begin{equation}
\begin{tabular}{lll}
$\mathbf{\gamma }_{1}$ & $=$ & $+\frac{3}{2}\mathbf{\alpha }_{1}+2\mathbf{%
\alpha }_{2}+\frac{3}{2}\mathbf{\alpha }_{3}$ \\
$\mathbf{\gamma }_{2}$ & $=$ & $+2\mathbf{\alpha }_{1}-2\mathbf{\alpha }_{3}$
\\
$\mathbf{\gamma }_{3}$ & $=$ & $-\frac{3}{5}\mathbf{\alpha }_{1}+\frac{6}{5}%
\mathbf{\alpha }_{2}-\frac{3}{5}\mathbf{\alpha }_{3}$%
\end{tabular}%
\qquad ,\qquad
\begin{tabular}{lll}
$\mathbf{\alpha }_{1}$ & $=$ & $\frac{1}{5}\mathbf{\gamma }_{1}+\frac{1}{4}%
\mathbf{\gamma }_{2}-\frac{1}{3}\mathbf{\gamma }_{3}$ \\
$\mathbf{\alpha }_{2}$ & $=$ & $\frac{1}{5}\mathbf{\gamma }_{1}+\frac{1}{2}%
\mathbf{\gamma }_{3}$ \\
$\mathbf{\alpha }_{3}$ & $=$ & $\frac{1}{5}\mathbf{\gamma }_{1}-\frac{1}{4}%
\mathbf{\gamma }_{2}-\frac{1}{3}\mathbf{\gamma }_{3}$%
\end{tabular}%
\end{equation}%
These are orthogonal vectors with lengths $\mathbf{\gamma }_{1}^{2}=5,$ $%
\mathbf{\gamma }_{2}^{2}=16\ $and$\ \mathbf{\gamma }_{3}^{2}=\frac{36}{5}.$
Their expression in terms of the canonical weight vectors $\mathbf{e}_{i}$
read as follows
\begin{equation}
\begin{tabular}{lll}
$\mathbf{\gamma }_{1}$ & $=$ & $+\frac{3}{2}\mathbf{e}_{1}+\frac{1}{2}%
\mathbf{e}_{2}-\frac{1}{2}\mathbf{e}_{3}-\frac{3}{2}\mathbf{e}_{4}$ \\
$\mathbf{\gamma }_{2}$ & $=$ & $+2\mathbf{e}_{1}-2\mathbf{e}_{2}-2\mathbf{e}%
_{3}+2\mathbf{e}_{4}$ \\
$\mathbf{\gamma }_{3}$ & $=$ & $-\frac{3}{5}\mathbf{e}_{1}+\frac{9}{5}%
\mathbf{e}_{2}-\frac{9}{5}\mathbf{e}_{3}+\frac{3}{5}\mathbf{e}_{4}$%
\end{tabular}%
\end{equation}

\paragraph{\textbf{III. Commuting Cartan subalgebra in sl(4,}$\mathbb{R}$)%
\textbf{:}}

\ \ \newline
The homologue of \textbf{Figure} \textbf{\ref{sl2}} describing the SL(4,$%
\mathbb{R}$) Cartan geometry \textsc{Cart}$_{sl_{4}}$ is given by the
3-dimensional space ($\dim $\textsc{C}$\text{\textsc{art}}_{sl_{4}}=3$) with
planes, straight lines and chambers as given below:

\begin{itemize}
\item 2-dim Cartan planes \textsc{C}$\text{\textsc{art}}_{sl_{3}}$
associated with the sl(3,$\mathbb{R}$)$_{\alpha _{i}\text{-}\alpha _{j}}$
Lie subalgebras separating 3-dim chambers.

\item various straight lines \textsc{C}$\text{\textsc{art}}_{sl_{2}}$
\textrm{of} the sl(2,$\mathbb{R}$)$_{\alpha _{i}}$ Lie subalgebras carving
up 2-dim chambers in the sl(3,$\mathbb{R}$)$_{\alpha _{i}\text{-}\alpha
_{j}} $ planes.
\end{itemize}

The typical diagonal matrix elements $\Theta _{sl_{4}}$ of the Cartan
subspace \textsc{C}$\text{\textsc{art}}_{sl_{4}}$ expand like
\begin{equation}
\Theta _{sl_{4}}=\zeta _{(2)}^{0}W_{0}^{(2)}+\zeta
_{(3)}^{0}W_{0}^{(3)}+\zeta _{(4)}^{0}W_{0}^{(4)}
\end{equation}%
with three real parameters $\zeta _{{\small s-1}}^{0}$ and associated HS
generators $W_{0}^{(s)}$. By using the alternative (\ref{ww}), we get instead%
\begin{equation}
\begin{tabular}{lll}
$\Theta _{sl_{4}}$ & $=$ & $(\frac{3}{2}\zeta _{(2)}^{0}+2\zeta _{(3)}^{0}-%
\frac{3}{5}\zeta _{(4)}^{0})\boldsymbol{H}_{\alpha _{1}}$ \\
&  & $+(2\zeta _{(2)}^{0}+\frac{6}{5}\zeta _{(4)}^{0})\boldsymbol{H}_{\alpha
_{2}}$ \\
&  & $+(\frac{3}{2}\zeta _{(2)}^{0}-2\zeta _{(3)}^{0}-\frac{3}{5}\zeta
_{(4)}^{0})\boldsymbol{H}_{\alpha _{3}}$%
\end{tabular}
\label{ac}
\end{equation}%
\begin{equation*}
\end{equation*}%
The two dimensional subspaces \textsc{C}$\text{\textsc{art}}_{sl_{3}}$
sitting in (\ref{ac}) are obtained upon imposing a constraint relation on
these parameters,\ for instance the condition%
\begin{equation}
\zeta _{(4)}^{0}=\frac{5}{2}\zeta _{(2)}^{0}-\frac{10}{3}\zeta _{(3)}^{0}
\end{equation}%
restricts the above relation to%
\begin{equation}
\Theta _{sl_{3}}=4\zeta _{(3)}^{0}\boldsymbol{H}_{\alpha _{1}}+(5\zeta
_{(2)}^{0}-4\zeta _{(3)}^{0})\boldsymbol{H}_{\alpha _{2}}
\end{equation}%
If we moreover set $5\zeta _{(2)}^{0}-4\zeta _{(3)}^{0}$, we reach the%
\textrm{\ }\textsc{Cart}$_{sl_{2}}$ having the elements $\Theta
_{sl_{2}}=5\zeta _{(2)}^{0}\boldsymbol{H}_{\alpha _{1}}.$

We conclude this\ description by noting that the complexified gauge symmetry
group $SL(4,\mathbb{C})$ has four real forms given by%
\begin{equation}
SL(4,\mathbb{R})\quad ,\quad SU(4)\quad ,\quad SU(2,2)\quad ,\quad
SU(3,1)\simeq SU(1,3)
\end{equation}%
where the compact real form $SU(4)$ has $3+3$ compact SU(2)$_{\alpha }$
subgroups labelled by the 6 positive roots%
\begin{equation}
\alpha _{{\small 1}},\quad \alpha _{{\small 2}},\quad \alpha _{{\small 3}%
},\quad \alpha _{{\small 1}}+\alpha _{{\small 2}},\quad \alpha _{{\small 2}%
}+\alpha _{{\small 3}},\quad \alpha _{{\small 1}}+\alpha _{{\small 2}%
}+\alpha _{{\small 3}}
\end{equation}%
and typical group elements \textrm{of} the Cartan subgroup \textrm{are}
given by $e^{i\Theta _{sl_{{\small N}}}}$ with angular matrix $\Theta _{sl_{%
{\small N}}}=\sum \zeta _{{\small s-1}}^{0}W_{0}^{(s)}$ expanding as in (\ref%
{ac}).
\begin{equation*}
\end{equation*}

\section{More on the edge invariants}

\label{appB}In this appendix, we derive the algebra governing the asymptotic
symmetry of the edge invariants operators.

\paragraph{\textbf{1)} \textbf{Fourier modes of the conserved currents}}

\ \ \newline
The periodicity properties of the angular variable\textrm{\ }$\varphi $%
\textrm{\ }reading as%
\begin{eqnarray}
\mathcal{J}_{\varphi }^{{\small 0}}\left( \varphi +2\pi \right) &=&\mathcal{J%
}_{\varphi }^{{\small 0}}\left( \varphi \right) \qquad ,\qquad \mathcal{K}%
_{\varphi }^{{\small 0}}\left( \varphi +2\pi \right) =\mathcal{K}_{\varphi
}^{{\small 0}}\left( \varphi \right) \\
\mathcal{\tilde{J}}_{\varphi }^{{\small 0}}\left( \varphi +2\pi \right) &=&%
\mathcal{\tilde{J}}_{\varphi }^{{\small 0}}\left( \varphi \right) \qquad
,\qquad \mathcal{\tilde{K}}_{\varphi }^{{\small 0}}\left( \varphi +2\pi
\right) =\mathcal{\tilde{K}}_{\varphi }^{{\small 0}}\left( \varphi \right)
\end{eqnarray}%
allow to expand the charge potentials in Fourier series as follows%
\begin{equation}
\begin{tabular}{lll}
$\mathcal{J}_{\varphi }^{{\small 0}}$ & $=$ & $\frac{2}{\mathrm{k}}%
\dsum\limits_{n=-\infty }^{+\infty }e^{-in\varphi }J_{n}^{{\small 0}}$ \\
$\mathcal{K}_{\varphi }^{{\small 0}}$ & $=$ & $\frac{3}{2\mathrm{k}}%
\dsum\limits_{n=-\infty }^{+\infty }e^{-in\varphi }K_{n}^{{\small 0}}$%
\end{tabular}%
\qquad ,\qquad
\begin{tabular}{lll}
$\mathcal{\tilde{J}}_{\varphi }^{{\small 0}}$ & $=$ & $\frac{2}{\mathrm{k}}%
\dsum\limits_{n=-\infty }^{+\infty }e^{+in\varphi }\tilde{J}_{n}^{{\small 0}%
} $ \\
$\mathcal{\tilde{K}}_{\varphi }^{{\small 0}}$ & $=$ & $\frac{3}{2\mathrm{k}}%
\dsum\limits_{n=-\infty }^{+\infty }e^{+in\varphi }\tilde{K}_{n}^{{\small 0}%
} $%
\end{tabular}
\label{exj}
\end{equation}%
where ($J_{n}^{{\small 0}},K_{n}^{{\small 0}}$) and ($\tilde{J}_{n}^{{\small %
0}},\tilde{K}_{n}^{{\small 0}}$) are Fourier modes. \textrm{The}
normalisation factors $\frac{2}{\mathrm{k}}$ and $\frac{3}{2\mathrm{k}}$
insure appropriate constant structures for the W$_{3}$-algebra. We have
\begin{equation}
\begin{tabular}{lll}
$\frac{2}{\mathrm{k}}J_{n}^{{\small 0}}$ & $=$ & $\frac{1}{2\pi }%
\dint\nolimits_{0}^{2\pi }d\varphi e^{+in\varphi }\mathcal{J}_{\varphi }^{%
{\small 0}}$ \\
$\frac{3}{2\mathrm{k}}K_{n}^{{\small 0}}$ & $=$ & $\frac{1}{2\pi }%
\dint\nolimits_{0}^{2\pi }d\varphi e^{+in\varphi }\mathcal{K}_{\varphi }^{%
{\small 0}}$%
\end{tabular}%
\qquad ,\qquad
\begin{tabular}{lll}
$\frac{2}{\mathrm{k}}\tilde{J}_{n}^{{\small 0}}$ & $=$ & $\frac{1}{2\pi }%
\dint\nolimits_{0}^{2\pi }d\varphi e^{-in\varphi }\mathcal{\tilde{J}}%
_{\varphi }^{{\small 0}}$ \\
$\frac{3}{2\mathrm{k}}\tilde{K}_{n}^{{\small 0}}$ & $=$ & $\frac{1}{2\pi }%
\dint\nolimits_{0}^{2\pi }d\varphi e^{-in\varphi }\mathcal{\tilde{K}}%
_{\varphi }^{{\small 0}}$%
\end{tabular}%
\end{equation}

\paragraph{\textbf{2)} \textbf{Edge invariants}}

\ \ \newline
The conserved charges $(\mathtt{I},\mathtt{\tilde{I}})$ given by (\ref{II})
are\textrm{, in fact, }the zero mode $(\mathtt{I}_{{\small 0}},\mathtt{%
\tilde{I}}_{{\small 0}})$ of \textrm{the} two infinite families $(\mathtt{I}%
_{{\small n}},\mathtt{\tilde{I}}_{{\small n}})$ defined as follows
\begin{eqnarray}
2\pi \mathtt{I}_{{\small n}} &=&\frac{\mathrm{k\beta }}{4}\int_{0}^{2\pi
}d\varphi e^{in\varphi }Tr\left[ \lambda \mathcal{J}_{\varphi }+\Lambda
\mathcal{K}_{\varphi }\right]  \notag \\
2\pi \mathtt{\tilde{I}}_{{\small n}} &=&\frac{\mathrm{k\beta }}{4}%
\int_{0}^{2\pi }d\varphi e^{-in\varphi }Tr\left[ \tilde{\lambda}\mathcal{%
\tilde{J}}_{\varphi }+\tilde{\Lambda}\mathcal{\tilde{K}}_{\varphi }\right]
\end{eqnarray}%
By integration, we obtain%
\begin{eqnarray}
\mathtt{I}_{{\small n}} &=&\mathrm{\beta }\left[ \lambda
^{0}J_{n}^{0}+\Lambda ^{0}K_{n}^{0}\right]  \notag \\
\mathtt{\tilde{I}}_{{\small n}} &=&\mathrm{\beta }\left[ \tilde{\lambda}^{0}%
\tilde{J}_{n}^{0}+\tilde{\Lambda}^{0}\tilde{K}_{n}^{0}\right]
\end{eqnarray}%
Being a linear combination of ($J_{n}^{0},K_{n}^{0}$) and ($\tilde{J}%
_{n}^{0},\tilde{K}_{n}^{0}$), they satisfy the algebra%
\begin{equation}
\begin{tabular}{lll}
$i\left\{ \mathtt{I}_{n},\mathtt{I}_{m}\right\} _{\text{\textsc{pb}}}$ & $=$
& $\pm \frac{\mathrm{\beta k}}{2}\left( \lambda ^{0}+\frac{4}{3}\Lambda
^{0}\right) n\delta _{n+m,0}$ \\
$i\left\{ \mathtt{\tilde{I}}_{n},\mathtt{\tilde{I}}_{m}\right\} _{\text{%
\textsc{pb}}}$ & $=$ & $\pm \frac{\mathrm{\beta k}}{2}\left( \tilde{\lambda}%
^{0}+\frac{4}{3}\tilde{\Lambda}^{0}\right) n\delta _{n+m,0}$%
\end{tabular}%
\end{equation}%
At the quantum level, the above Poisson brackets get mapped to the
commutators%
\begin{equation}
\begin{tabular}{lll}
$\left[ \boldsymbol{J}_{n}^{{\small 0}},\boldsymbol{J}_{m}^{{\small 0}}%
\right] $ & $=$ & $\pm \frac{\mathrm{k}}{2}n\delta _{n+m,0}$ \\
$\left[ \boldsymbol{K}_{n}^{{\small 0}},\boldsymbol{K}_{m}^{{\small 0}}%
\right] $ & $=$ & $\pm \frac{2\mathrm{k}}{3}n\delta _{n+m,0}$%
\end{tabular}%
,\qquad
\begin{tabular}{lll}
$\lbrack \boldsymbol{\tilde{J}}_{n}^{{\small 0}},\boldsymbol{\tilde{J}}_{m}^{%
{\small 0}}]$ & $=$ & $\pm \frac{\mathrm{k}}{2}n\delta _{n+m,0}$ \\
$\lbrack \boldsymbol{\tilde{K}}_{n}^{{\small 0}},\boldsymbol{\tilde{K}}_{m}^{%
{\small 0}}]$ & $=$ & $\pm \frac{2\mathrm{k}}{3}n\delta _{n+m,0}$%
\end{tabular}%
\end{equation}%
and%
\begin{equation}
\begin{tabular}{lll}
$\left[ \boldsymbol{I}_{n},\boldsymbol{I}_{m}\right] $ & $=$ & $\pm \frac{%
\mathrm{k}}{2}\mathrm{\beta }\left( \lambda ^{{\small 0}}+\frac{4}{3}\Lambda
^{{\small 0}}\right) n\delta _{n+m,0}$ \\
$\left[ \boldsymbol{\tilde{I}}_{n},\boldsymbol{\tilde{I}}_{m}\right] $ & $=$
& $\pm \frac{\mathrm{k}}{2}\mathrm{\beta }\left( \tilde{\lambda}^{{\small 0}%
}+\frac{4}{3}\tilde{\Lambda}^{{\small 0}}\right) n\delta _{n+m,0}$%
\end{tabular}%
\end{equation}%
Substituting (\ref{RF}), we \textrm{obtain}:%
\begin{equation}
\begin{tabular}{|c|c|}
\hline
{\small real forms} & $\mathtt{I}_{n}$ \\ \hline\hline
$su\left( 3\right) $ & $\ \ \ \ \ 2\pi \mathfrak{N}J_{n}^{0}+3\pi \mathfrak{M%
}K_{n}^{0}$\ \ \ \ \ \ \  \\
$su\left( 2\right) _{L_{{\small 0}}}$ & $2\pi NJ_{n}^{0}$ \\
$su\left( 2\right) _{W_{{\small 0}}}$ & $3\pi \mathfrak{M}K_{n}^{0}$ \\
\hline\hline
$\ \ \ \ \ \ su(2,1)_{{\small 12}}$ \ \ \ \ \ \  & $\left( 2\pi N\right)
J_{n}^{0}$ \\
$su(1,2)_{{\small 21}}$ & $\left( 3\pi \mathfrak{M}\right) K_{n}^{0}$ \\
\hline
\end{tabular}%
\end{equation}%
\begin{equation*}
\end{equation*}%
where we set $\mathfrak{N}_{0}=N+M/2$ and $\mathfrak{M}=M/2+p/3$. Similar
relations hold also for the twild sector $\mathtt{\tilde{I}}_{n}$.%
\begin{equation*}
\end{equation*}

\section{The $\mathbb{Z}_{N}$ centre symmetry}

\label{appC}In this appendix, we elaborate on the centre symmetry of the
higher spin black flowers with gauge field holonomy sitting in $\mathbb{Z}%
_{N}$. This finite abelian discrete group divides the gauge symmetry $%
\mathcal{G}$ of the HS Chern-Simons model for the black flowers into N
subsets $\mathcal{G}_{p}$ like
\begin{equation}
\mathcal{G}=\cup _{p=0}^{N-1}\mathcal{G}_{p}  \label{grad}
\end{equation}%
with zero mode $\mathcal{G}_{0}\simeq SU(N)$. Because of this gradation of
the gauge group $\mathcal{G}$, the centre symmetry gives rise to holonomy
states $\left\vert \psi \left( \mathcal{G}_{p}\right) \right\rangle $
labelled by the N elements of the discrete symmetry
\begin{equation}
\omega ^{p}\quad ;\quad 0\leq p<N
\end{equation}%
generated by $\omega =e^{i2\pi /N}.$ In what follows, we first give useful
generalities. Then, we describe the physical states in Yang-Mills theory at
finite temperature with centre symmetry $\mathbb{Z}_{N}$\emph{. }After that,
we consider the holonomy states $\left\vert \psi \left( \mathcal{G}%
_{p}\right) \right\rangle $\emph{\ }in HS black flowers with discrete\emph{\
}$\mathbb{Z}_{N}$.

\subsection*{Generalities}

As a front matter of this description, we start by recalling two useful
features and turn after to give technical details.

First, recall that at classical level, a given gauge-field configuration $A$
in the holonomy (\ref{1W}) and in $SU(N)$ YM theory in general can be seen
as representing a certain state $\left\vert \psi \left( A\right)
\right\rangle $ of the gauge model (HS black flowers). However, since $%
\mathcal{G}_{0}\simeq SU(N)$ is the group of gauge transformations $U_{0}$,
the potential $A$ is not uniquely defined and consequently all the gauge
configurations
\begin{equation}
A^{U_{0}}=U_{0}AU_{0}^{-1}+iU_{0}dU_{0}^{-1}
\end{equation}%
should represent the same physical state in the sense that all the $%
\left\vert \psi \left( A^{U_{0}}\right) \right\rangle $'s describe the same
physical state as $\left\vert \psi \left( A\right) \right\rangle .$ Also,
this means that the physical $\left\vert \psi \left( A\right) \right\rangle $
is in fact in 1:1 correspondence with the $\mathcal{G}_{0}$-orbit of the
gauge configurations defined as%
\begin{equation}
\mathcal{A}=\left\{ A^{U_{0}}\quad |\quad U_{0}\in \mathcal{G}_{0}\right\}
\qquad ,\qquad \mathcal{G}_{0}/\mathcal{G}_{0}\simeq I_{id}  \label{orb}
\end{equation}
So, it is natural to refer to the physical state in this family of gauge
configurations just as $\left\vert \psi \left( \mathcal{G}_{0}\right)
\right\rangle $ instead of $\left\vert \psi \left( A\right) \right\rangle .$

Second, from a general view, the above state $\left\vert \psi \left(
\mathcal{G}_{0}\right) \right\rangle $ is in fact a particular state as it
is labelled by $\mathcal{G}_{0}$ which is a subgroup of $\mathcal{G}.$ By
considering the full gauge symmetry $\mathcal{G}=\cup _{p=0}^{N-1}\mathcal{G}%
_{p}$ having subsets $\mathcal{G}_{p}$ demanded by the non trivial actions
of the centre $\mathbb{Z}_{N}$, one then ends up with N physical states
labelled as $\left\vert \psi \left( \mathcal{G}_{p}\right) \right\rangle $.
The relationship of these states with the centre $\mathbb{Z}_{N}$ is given
by the correspondence between $\mathcal{G}_{p}/\mathcal{G}_{0}$ and the
group elements $\omega ^{p}$ (i.e: $\mathcal{G}_{p}/\mathcal{G}%
_{0}\leftrightarrow \omega ^{p}$) indicating in turn that $\mathcal{G}_{p}$
can be imagined as given by $\omega ^{p}\mathcal{G}_{0}.$ For convenience,
we refer below to these (holonomy) states $\left\vert \psi \left( \mathcal{G}%
_{p}\right) \right\rangle $ simply as
\begin{equation*}
\left\vert p\right\rangle \quad ,\quad 0\leq p<N
\end{equation*}%
they are individually invariant under the zero mode $\mathcal{G}_{0}$ but
not under the extended $\mathcal{G}$.

To get more insight into the properties of the physical states $\left\vert
p\right\rangle $ in HS\ black flower, we propose to first discuss the
physical significance of the discrete symmetry\textrm{\ }$\mathbb{Z}_{N}$%
\textrm{\ }in Yang-Mills theories at finite temperature that is invariant
under SU(N). Then, we comment on the importance of the gauge invariant
states indexed by the group elements\textrm{\ }$\omega ^{p}$ due to the
centre symmetry $\mathbb{Z}_{N}$. After that, we turn to\textrm{\ }our main
concern regarding the states $\left\vert p\right\rangle $ descending from
the regularity of the holonomy (\ref{1W}) namely%
\begin{equation}
W\left[ a_{\beta }\right] =\exp \left( i\int_{0}^{\beta }a_{t_{{\small E}}}^{%
\text{\textsc{diag}}}dt_{{\small E}}\right)  \label{WT}
\end{equation}%
sitting in the centre $\mathbb{Z}_{N}$ in the HS Chern-Simons theory for
black flowers. The algebraic setting of the states $\left\vert
p\right\rangle $ obtained in our paper can be put in correspondence with
those physical states investigated in \cite{M1} for YM at finite temperature
with which we begin this appendix.

\subsection*{Discrete\emph{\ }$\mathbb{Z}_{N}$\emph{\ }in YM theory at
finite temperature}

Following \cite{M1}, the centre $\mathbb{Z}_{N}$ is a discrete symmetry that
is present in SU(N) Yang-Mills theories at finite temperature. In this
theory, the gauge potential\textrm{\ A}$_{\mu }\left( \tau ,\mathbf{x}%
\right) $\textrm{\ }is valued in the Lie algebra of the SU(N) gauge
invariance\textrm{\ (A}$_{\mu }=\sum t_{a}$\textrm{A}$_{\mu }^{a}$\textrm{)}%
; and it is a function of the euclidian time $\tau $ and the spatial
variables\textrm{\ }$\mathbf{x}$\textrm{\ \cite{M1,M2,M3}}. Contrary to the
usual gauge transformations of \textrm{SU(N) which are not physical,} it
happens that the transformations under $\mathbb{Z}_{N}$ action are physical
in the sense that they permit to jump between gauge invariant states $%
\left\vert p\right\rangle $ (observables $\mathcal{O}_{p}$) of the YM theory
labelled by the group elements $\omega ^{p}$. This feature can be nicely
illustrated by thinking about the $\left\vert p\right\rangle $ (observables $%
\mathcal{O}_{p})$ in terms of the factorisation
\begin{equation}
\left\vert p\right\rangle =\left\vert 0,\omega ^{p}\right\rangle \qquad
;\qquad \mathcal{O}_{p}=\omega ^{p}\mathcal{O}_{0}  \label{om}
\end{equation}%
where $\left\vert 0\right\rangle $ ($\mathcal{O}_{0}$) is a basic state
corresponding to the neutral group element $\omega ^{0}=I_{id}\in \mathbb{Z}%
_{N}$. An interesting example of such observables is given by the so-called
Polyakov loop $\mathfrak{l}=\left\langle \Phi \left[ A\right] \right\rangle $
with gauge functional $\Phi \left[ A\right] $ as%
\begin{equation}
\Phi \left[ A\right] =\frac{1}{N}tr\left[ \mathcal{P}\exp \left(
ig\dint\nolimits_{0}^{\mathrm{\beta }}d\tau A_{0}\left( \tau ,\mathbf{x}%
\right) \right) \right]  \label{fa}
\end{equation}%
where $\left[ 0,\mathrm{\beta }\right] \equiv \mathcal{C}_{\tau }$
designates the thermal cycle with temperature $T=1/\mathrm{\beta }$ and
where g is the gauge coupling constant. Below, we show that, modulo
unphysical \textrm{SU(N)} gauge transformations, the discrete $\mathbb{Z}%
_{N} $ acts trivially on gauge fields \textrm{A}$_{\mu }\left( \tau ,\mathbf{%
x}\right) $ but not \textrm{on} the functional $\Phi \left[ A\right] $.
Here, we draw the main lines of the description towards (\ref{om}) and refer
to the work of \textrm{\cite{M1}} for further explicit details.

\textrm{The action of the discrete }$\mathbb{Z}_{N}$ on the potential
\textrm{A}$_{\mu }$ and the functional $\Phi \left[ A\right] $ arises from
the observation that at finite temperature the usual $SU\left( N\right) $
gauge potential transformations%
\begin{equation}
\mathrm{A}_{\mu }^{U}=U\mathrm{A}_{\mu }U^{-1}+\frac{i}{g}U\partial _{\mu
}U^{-1}\qquad ,\qquad U=U\left( \tau ,\mathbf{x}\right)  \label{Tr}
\end{equation}%
that leave invariant the classical field action S$_{YM}\left[ A\right] $ do
not all qualify as exact symmetries under the functional integrals type (\ref%
{fa}). This remarkable feature is \textrm{due to} the periodic boundary
condition of the potential \textrm{A}$_{\mu }\left( \tau ,\mathbf{x}\right) $
at finite temperature which restricts the gauge field to obey the constraint%
\begin{equation}
A_{\mu }\left( \tau +\mathrm{\beta },\mathbf{x}\right) =A_{\mu }\left( \tau ,%
\mathbf{x}\right) \qquad ,\qquad \tau \in \mathcal{C}_{\tau }  \label{t1}
\end{equation}%
\textrm{but} allows for \textrm{larger} gauge transformations going beyond
the usual SU(N) as explicitly showed in \textrm{\cite{M1}}. For
concreteness, we denote these extended gauge transformations like
\begin{equation}
V\left( \tau ,\mathbf{x};\omega ^{p}\right) =V_{(p)}\left( \tau ,\mathbf{x}%
\right)
\end{equation}%
with $V_{(0)}\left( \tau ,\mathbf{x}\right) $ equals to the transformation $%
U\left( \tau ,\mathbf{x}\right) $ in (\ref{Tr}) that read in terms of the
group $\lambda \left( \tau ,\mathbf{x}\right) =\sum t_{a}\lambda ^{a}\left(
\tau ,\mathbf{x}\right) $ and the identity $I_{id}$ as follows
\begin{equation}
U\left( \tau ,\mathbf{x}\right) =e^{i\lambda \left( \tau ,\mathbf{x}\right)
}=I_{id}+i\lambda \left( \tau ,\mathbf{x}\right) +...
\end{equation}%
The $V_{(p)}$'s go beyond SU(N), they belong to an extended symmetry $%
\mathcal{G}$ with gradation $\cup _{p=1}^{N}\mathcal{G}_{p}$. This extended
group contains N subsets $\mathcal{G}_{p}$ including: $\left( \mathbf{i}%
\right) $ the subset $\mathcal{G}_{0}$ having a group structure isomorphic
to SU(N); it is a normal subgroup of the extended $\mathcal{G}$. $\left(
\mathbf{ii}\right) $ The other N-1 subsets $\mathcal{G}_{p}$ with $p\neq 0$ $%
\func{mod}N$ are not subgroups of $\mathcal{G}$; they obey the composition
law%
\begin{equation}
\mathcal{G}_{p}\mathcal{G}_{q}\simeq \mathcal{G}_{p+q}\qquad ,\qquad
p,q=0,...,N-1
\end{equation}%
required by the $\mathbb{Z}_{N}$ gradation. Actually, these commutative
compositions define the group of physical centre transformations isomorphic
to $\mathbb{Z}_{N}$.

So, given a gradation level p, the gauge transformation $V_{(p)}\left( \tau ,%
\mathbf{x}\right) $\ sits in the subset $\mathcal{G}_{p}$; and is related to
the basic SU(N) transformation $U\left( \tau ,\mathbf{x}\right) $ belonging
to $\mathcal{G}_{0}$ as follows%
\begin{equation}
\begin{tabular}{lll}
$U\left( \tau +\mathrm{\beta },\mathbf{x}\right) $ & $=$ & $V_{(p)}\left(
\tau ,\mathbf{x}\right) $ \\
$V_{(p)}\left( \tau ,\mathbf{x}\right) $ & $=$ & $e^{i\frac{2\pi p}{N}%
}U\left( \tau ,\mathbf{x}\right) $%
\end{tabular}
\label{t2}
\end{equation}%
These relations are remarkable; they indicate that the boundary condition of
$U\left( \tau ,\mathbf{x}\right) $ holds only up to a $\mathbb{Z}_{N}$
transformation $U^{(h)}=hU\left( \tau ,\mathbf{x}\right) h^{-1}$ with $h\in
\mathbb{Z}_{N}$. From eqs(\ref{t1}-\ref{t2}), one gets the following results:

\begin{description}
\item[$\left( \mathbf{i}\right) $] the potential $A_{\mu }\left( \tau ,%
\mathbf{x}\right) $ and the gauge transformation\ $U\left( \tau ,\mathbf{x}%
\right) $\ have the same boundary conditions only for the subgroup $\mathcal{%
G}_{0}\simeq SU(N)$ of the extended $\mathcal{G}$. This state corresponds to
trivial holonomy in HS black flowers.

\item[$\left( \mathbf{ii}\right) $] only the changes of $\mathcal{G}_{0}$
\textrm{s}hould be considered as genuine gauge transformations in the sense
of unphysical transformations which do not alter the physical states $%
\left\vert p\right\rangle $ and observables $\mathcal{O}_{p}$ of the $%
\mathbb{Z}_{N}$- system. Formally, we have $\mathcal{G}_{0}\left\vert
p\right\rangle =\left\vert p\right\rangle $ and $\mathcal{G}_{0}.\mathcal{O}%
_{0}=\mathcal{O}_{0}$.

\item[$\left( \mathbf{iii}\right) $] any transformation belonging to the
complement $\mathcal{K}=\mathcal{G}\backslash \mathcal{G}_{0}$ should be
seen as a physical transformation that permits to jump between the states $%
\oplus _{p=1}^{N}\left\vert p\right\rangle $ and observables $\mathcal{O}%
=\oplus _{p=1}^{N}\mathcal{O}_{p}$ of the system. These transformations
allow to leap from a gauge invariant state $\left\vert p\right\rangle $ to a
different gauge invariant state $\left\vert q\right\rangle $ with $p\neq q.$
The same for $\mathcal{O}_{p}$ towards $\mathcal{O}_{q}.$

\item[$\left( \mathbf{iv}\right) $] for the case of the Polyakov loop, the
functional $\Phi \left[ A\right] $ changes as for the \emph{YM} observables $%
\mathcal{O}_{p}$ (\ref{om}). So, under the gauge transformation $%
A\rightarrow A^{U}$, the Polyakov loop transforms like
\begin{equation}
\Phi \left[ A^{U}\right] =e^{i2\pi p/N}\Phi \left[ A\right]
\end{equation}%
where $e^{i2\pi p/N}=\omega ^{p}$ is an element of the $\mathbb{Z}_{N}$
centre \textrm{of SU(N)}. For the explicit derivation of the above relation,
we refer to the sections II and IV in the work \cite{M1}; see also eq(32)
there.
\end{description}

\subsection*{Discrete\emph{\ }$\mathbb{Z}_{N}$\emph{\ }in HS black flowers}

The previous description done for generic YM theory at finite temperature is
also valid for HS black flowers considered in this paper. Here, the centre
symmetry $\mathbb{Z}_{N}$ arises in AdS$_{3}$ higher spin theories with
SU(N) due to the Wick transformation $t=it_{E}$ necessary for the study and
the computation of the thermodynamical properties like the black hole's
entropy. For HS black flowers, we have
\begin{equation*}
\tau =t_{E}\qquad ,\qquad \mathbf{x}=\varphi
\end{equation*}%
\textrm{\ }while the role of YM potential\textrm{\ A}$_{\mu }\left( \tau ,%
\mathbf{x}\right) $\textrm{\ }is now played by the boundary potential\textrm{%
\ }$a_{t}\left( t_{E},\varphi \right) $\textrm{\ }valued in the Cartan
subalgebra of the unitary gauge symmetry SU(N).

Given this bridging,\ the holonomy of the boundary potential $W\left[
a_{\beta }\right] $ (\ref{WT}) can be put in correspondence with the
Polyakov loop (\ref{fa}). The properties of the transformation $W\left[
a_{\beta }\right] $ under the $\mathbb{Z}_{N}$ centre symmetry behaves as
for $\Phi \left[ A\right] $; and as such the action of the $\mathbb{Z}_{N}$
is \textrm{a} physical symmetry. This means that even though the set of
transformations given by%
\begin{equation}
a_{t}^{U}=Ua_{t}U^{-1}+iU\partial _{t}U^{-1}\qquad ,\qquad U=U\left(
t_{E},\varphi \right)
\end{equation}%
leave the HS CS Lagrangian invariant, only some of them are considered as
symmetries of $W\left[ a_{\beta }\right] $. This is due to the periodicity
property acting differently on the SU(N) elements $U\left( t_{E},\varphi
\right) $ in comparison to the gauge fields $a_{t}\left( t_{E},\varphi
\right) $. In fact, the periodicity property restricts the boundary gauge
field $a_{t}\left( t_{E},\varphi \right) $\ as:%
\begin{equation}
a_{t}\left( t_{E}+\beta ,\varphi \right) =a_{t}\left( t_{E},\varphi \right)
\end{equation}%
where $\beta $ designates the inverse of temperature. However, it allows for
a much larger symmetry for the gauge transformations because of the property%
\begin{equation}
U\left( t_{E}+\beta ,\varphi \right) =e^{2i\pi p/N}U\left( t_{E},\varphi
\right)
\end{equation}%
and%
\begin{equation}
W\left[ a_{\beta }^{U}\right] =e^{2i\pi p/N}W\left[ a_{\beta }\right]
\end{equation}%
indicating \textrm{that} the $W\left[ a_{\beta }^{U}\right] $ has a $\mathbb{%
Z}_{N}$ gradation like $\oplus _{p=1}^{N}W_{\left( p\right) }\left[ a_{\beta
}\right] $. The new states $W_{\left( p\right) }\left[ a_{\beta }\right] $
with $p\neq 0$ are demanded by the periodicity condition of boundary gauge
potential $a_{t}\left( t_{E},\varphi \right) $ requiring in turn an extended
gauge symmetry (\ref{grad})\textrm{\ }with $\mathcal{G}_{p}$ behaving like $%
e^{2i\pi p/N}\mathcal{G}_{0}.$

At the end of this appendix and in the light of the centre symmetry\textrm{\
}$\mathbb{Z}_{N}$\textrm{, }one learns that a HS black flower has N physical
layers\textrm{\ }$\mathfrak{L}_{0},...,\mathfrak{L}_{N-1}$\textrm{\ }indexed
by the group elements\textrm{\ }$\omega ^{p}$\textrm{. }The transition
between the layers is given by discrete leaps. This property was not
apparent when considering unphysical transformations via trivial holonomies.
Additionally, we were able to establish a link between the BH entropy and
compact subgroups of the real forms of the complexified gauge symmetry and
give an algebraic interpretation to the HS BH transitions: for example, the
discontinuous passage between the HS 3 black holes and the spin 2 BH is
merely due to the fact that the centre $\mathbb{Z}_{2}$ is not a subgroup of
the HS symmetry centre $\mathbb{Z}_{3}.$%
\begin{equation*}
\end{equation*}

\end{document}